\documentclass[a4paper,11pt]{article}
\pdfoutput=1 

\usepackage{amsmath,amssymb,bm,bbm,mathtools}
\usepackage{ascmac}
\usepackage{cancel}
\usepackage{xcolor} 
\usepackage{graphicx} 
\usepackage[whole]{bxcjkjatype} 
\usepackage{CJKutf8}
\usepackage{indentfirst}
\usepackage{multirow}
\usepackage{multicol}
\usepackage{marvosym,wasysym}
\usepackage{arydshln}
\usepackage{tikz}
\usepackage{tikz-3dplot} 
\usepackage{pgfplots}
\usetikzlibrary{arrows, calc, decorations, decorations.markings, intersections, positioning, snakes, 3d, matrix, patterns}
\newdimen\tbaselineshift 
\usepackage[unicode,bookmarks=true,bookmarksnumbered=true,bookmarkstype=toc,linktocpage=true]{hyperref} 
\hypersetup{
pdfauthor={Tetsuji KIMURA, Shin Sasaki and Kenta Shiozawa},
pdftitle={Semi-doubled GLSM for 5-branes of codim-2},
colorlinks={true},
linkcolor={blue},
urlcolor={blue},
filecolor={brown},
citecolor={blue}
}
\definecolor{refkey}{rgb}{0.40, 0.55, 0.55}
\definecolor{labelkey}{rgb}{0.90, 0.55, 0.55}
\usepackage{ulem}

 \font\f=cmr10
 \pdffontexpand\f 30 20 10 autoexpand
\usepackage[a4paper,top=33truemm,textheight=227truemm,left=20truemm,right=20truemm]{geometry}
\def\baselinestretch{1.05}
\makeatletter

 \@addtoreset{equation}{section}
\makeatother

\makeatletter
\def\tbcaption{\def\@captype{table}\caption}
\def\figcaption{\def\@captype{figure}\caption}
\makeatother

\newcounter{Enumerate}

\DeclareFontFamily{U}{rsf}{}
\DeclareFontShape{U}{rsf}{m}{n}{
  <5> <6> rsfs5 <7> <8> <9> rsfs7 <10-> rsfs10}{}
\DeclareMathAlphabet\Scr{U}{rsf}{m}{n}

\usepackage[mathscr]{eucal}

\newcommand{\del}{\partial}

\newcommand{\half}{\frac{1}{2}}

\newcommand{\LS}{\ \ \ \ \ \ \ \ \ \ }
\newcommand{\ls}{\ \ \ \ \ }
\newcommand{\wt}{\widetilde}
\newcommand{\wh}{\widehat}
\newcommand{\ve}{\varepsilon}
\newcommand{\ol}{\overline}

\newcommand{\bsubeq}{\begin{subequations}}
\newcommand{\esubeq}{\end{subequations}}

\newcommand{\noi}{\noindent}
\newcommand{\mr}{\mathring}

\newcommand{\eps}{\epsilon}

\newcommand{\nn}{\nonumber}

\renewcommand{\I}{{\rm i}}
\newcommand{\N}{\mathcal{N}}

\renewcommand{\d}{{\rm d}}
\newcommand{\e}{{\rm e}}
\renewcommand{\l}{\ell}
\newcommand{\w}{\wedge}

\newcommand{\slb}{\scalebox}

\def\+{{+\!\!\!+}} 

\begin{document}
\allowdisplaybreaks{
\thispagestyle{empty}


\begin{flushright}
October 2018 \\
\end{flushright}

\vspace{20mm}

\begin{center}
\noi
\slb{2.1}{Semi-doubled Gauged Linear Sigma Model}

\vspace{3mm}

\noi
\slb{2.1}{for Five-branes of Codimension Two}

\vspace{15mm}

\slb{1.2}{Tetsuji {\sc Kimura}$^{\,{\sf a} \ast}$}, \
\slb{1.2}{Shin {\sc Sasaki}$^{\,{\sf b} \dagger}$} 
\ and \ 
\slb{1.2}{Kenta {\sc Shiozawa}$^{\,{\sf b} \dagger\dagger}$}

\vspace{5mm}

\slb{.9}{
\begin{tabular}{l}
{\renewcommand{\arraystretch}{1.1}
\begin{tabular}{rl}
${\sf a}$ & 
{\sl Research Institute of Science and Technology,} 
\\
& {\sl College of Science and Technology, Nihon University} 
\\
& {\sl Kanda Surugadai 1-8-14, Chiyoda-ku, Tokyo 101-8308, JAPAN}
\\
& ${\ast}$ {\tt kimura.tetsuji \_at\_ nihon-u.ac.jp}
\end{tabular}
}
\\
\vphantom{A}
\\
{\renewcommand{\arraystretch}{1.1}
\begin{tabular}{rl}
${\sf b}$ & {\sl Department of Physics, Kitasato University}
\\
& {\sl Kitasato 1-15-1, Minami-ku, Sagamihara 252-0373, JAPAN}
\\
& ${\dagger}$ {\tt shin-s \_at\_ kitasato-u.ac.jp}, \ \ 
${\dagger\dagger}$ {\tt k.shiozawa \_at\_ sci.kitasato-u.ac.jp}
\end{tabular}
}
\end{tabular}
}

\end{center}

\vfill
\renewcommand{\abstractname}{\sc Abstract}
\begin{abstract}
We establish a double dualization in two-dimensional supersymmetric gauge theory.
We construct a gauged linear sigma model (GLSM) which contains a complex
 twisted linear superfield coupled to two sets of Abelian vector superfields.
In the IR regime, the GLSM provides a string sigma model whose target spaces are a defect NS5-brane, a Kaluza-Klein vortex and an exotic $5^2_2$-brane.
All of them are five-branes of codimension two and are related by T-duality.
This model is a natural extension of the GLSM proposed by Tong which gives a sigma model for an H-monopole, i.e., a smeared NS5-brane of codimension three.
This is also regarded as an alternative system of the GLSM for exotic five-branes proposed by the present authors.
In this analysis, we confirm that the T-duality transformation procedure
 in terms of the complex twisted linear superfield is applicable to dualize both the real and imaginary parts of the twisted chiral superfield even at the UV level, beyond the IR limit.
This indicates that the T-duality transformations at finite gauge couplings can be performed in terms of reducible superfields in the same way as irreducible (twisted) chiral superfields.
Furthermore, we study quantum vortex corrections to the GLSM at the UV level.
In the IR limit, these corrections are mapped to string worldsheet instanton corrections to the five-branes of codimension two.
The result completely agrees with those in double field theory analysis.
\end{abstract}

\newpage
{
\renewcommand{\baselinestretch}{.1}
\setcounter{tocdepth}{2}
\tableofcontents
}

\newpage
\section{Introduction}
\label{S:introduction}

Exotic brane is a mysterious object in string theory.
Exotic brane is a spacetime extended object of codimension two, one, or zero.
Each is also called a defect brane (or a vortex, in the soliton physics framework), a domain wall, and a space-filling brane.
D7-branes, D8-branes, and D9-branes are typical examples in D-brane physics \cite{Polchinski:1995mt}.
All of them play a central role in development of string theory.
Indeed, D7-branes provide us F-theory \cite{Vafa:1996xn} beyond small string coupling constant.
D8-branes yield the Romans mass \cite{Romans:1985tz} in type IIA theory.
This is one of the simplest deformation of ten-dimensional supergravity.
D9-branes in type IIB string theory \cite{Hull:1997kt} lead us to type I string theory, where closed strings and open strings coexist.
After the discovery of exotic branes 
\cite{Elitzur:1997zn, Blau:1997du, Obers:1997kk, Obers:1998fb, Eyras:1999at, LozanoTellechea:2000mc},
they have been investigated and applied to various configurations.
In particular, in black hole quantum mechanics, they play a significant role 
\cite{Bena:2011uw}.

Exotic branes and Kaluza-Klein (KK) monopoles reveal stringy corrections to our spacetime.
It is known that their geometries receive string winding mode corrections that cannot be traced in the framework of supergravity \cite{Gregory:1997te,Harvey:2005ab,Kimura:2013zva}.
Beyond the investigation of the stringy corrections to the KK-monopole, 
people developed many formulations in various fields such as  
supergravity and superstrings \cite{Siegel, deBoer:2010ud, deBoer:2012ma, Kimura:2016xzd},
worldvolume theory \cite{Chatzistavrakidis:2013jqa,Kimura:2014upa,Kimura:2016anf,Blair:2017hhy},
(non)geometric fluxes \cite{Hassler:2013wsa,Sakatani:2014hba,Andriot:2014uda}, 
double field theory (DFT) \cite{Hull:2009mi,Hohm:2010jy,Hohm:2010pp,Berman:2014jsa,Bakhmatov:2016kfn}, 
$\N=(4,4)$ gauged linear sigma model (GLSM) \cite{Tong:2002rq, Harvey:2005ab, Okuyama:2005gx, Kimura:2013fda}, and many others.
Quite recently, all of the exotic branes of codimension two, one and zero are completely classified by virtue of string dualities \cite{Fernandez-Melgarejo:2018yxq, Berman:2018okd}.
Even though their physical feature is still unclear,
it is natural to think that exotic branes provide us new physics beyond perturbative property of string theory and M-theory.

Now, we focus on an exotic $5^2_2$-brane.
This is an NS-NS type brane object (called the NS-brane, for short) coupled to B-field in string theory.
Indeed, this appears when two of four transverse directions of a single NS5-brane is T-dualized.
Since the KK-monopole is also obtained via T-duality of the NS5-brane,
the $5^2_2$-brane is an interesting object to research.
This configuration can be described in the framework of string worldsheet sigma model when F-string is utilized as a probe.
This is a reason why NS-branes have been analyzed in terms of nonlinear sigma model (NLSM) and its UV completion, i.e., GLSM \cite{Tong:2002rq, Harvey:2005ab, Okuyama:2005gx, Kimura:2013fda}.
GLSM was discussed by Witten \cite{Witten:1993yc} to study string worldsheet sigma model on Calabi-Yau manifold and its corresponding Landau-Ginzburg CFT. 
However, since the formulation is quite general, the GLSM can be applied to various situations. 
Worldsheet instanton corrections to the geometry of the H-monopole (equivalently, the smeared NS5-brane in a compact circle $S^1$) is one of the typical applications via brane configurations \cite{Tong:2002rq}.
It is the most important that, in the GLSM framework,
the string worldsheet instanton corrections can be captured by the gauge theory vortex corrections \cite{Witten:1993yc}.
By virtue of this, the vortex corrections to the H-monopole geometry has been computed.
The essential ingredient in the GLSM is the topological term:
\begin{align}
\e^{- S_{\text{E}}}
\ &\sim \ 
\exp \Big\{ - \frac{\I}{2 \pi} \int \d^2 x \, \vartheta F_{12} \Big\}
\, ,
\label{eq:Euclidean_topological_term}
\end{align}
where $S_{\text{E}}$ is the Euclidean action associated with the two-dimensional GLSM and $F_{12}$ is the field strength of the Abelian gauge field\footnote{In the main part of this paper except here and in section \ref{S:instantons}, the analysis will be discussed in two-dimensional spacetime of the Lorentz signature. These two frameworks can be exchanged via the Wick rotation $x^0 = - \I x^2$, $F_{01} = + \I F_{21} = - \I F_{12}$.}.
The scalar field $\vartheta$ represents the coordinate of the compact circle in the four transverse directions of the NS5-brane. 
One immediately finds that when the gauge field takes value in a non-trivial topological sector classified by the first Chern number $n = - \frac{1}{2\pi} \int \! \d^2 x F_{12}$, 
the isometry of the geometry along the $\vartheta$-direction is broken.
This substantially leads to the localization of the NS5-brane in $S^1$.
A remarkable consequence of this fact can be seen in the T-dualized KK-monopole picture. 
Since the coordinate $\vartheta$ is the Fourier dual of the KK-modes associated with the compact circle, the corresponding corrections to the KK-monopole is realized as the dual coordinate associated with the string {\it winding} modes (T-dual of the KK-modes).
Indeed, this is confirmed by the direct calculations of instanton effects to the Taub-NUT geometry based on the GLSM \cite{Harvey:2005ab}.

Supersymmetry is a key ingredient to uncover the instanton effects. 
Since the H- and KK-monopoles are half-BPS solutions to ten-dimensional supergravity with 32 supersymmetry, the corresponding NLSM and GLSM possess the two-dimensional $\N = (4,4)$ supersymmetry.
This $\N = (4,4)$ supersymmetry is conveniently realized by the $\N = (2,2)$ superfield formalism.
The topological term appeared in (\ref{eq:Euclidean_topological_term}) can be expressed as a twisted superpotential in two-dimensional $\N=(2,2)$ supersymmetric theory:
\begin{align}
\vartheta F_{12} 
\ &\sim \ 
\int \d^2 \wt{\theta} \, \Theta \Sigma + \text{(h.c.)}
\, , \label{topological-F}
\end{align}
where $\Theta$ is a twisted chiral superfield whose imaginary part is the scalar field $\vartheta$, and $\Sigma$ is another twisted chiral superfield which involves the gauge field strength (for details, see section \ref{S:twistedlinear}).
Superfield formalism is a useful tool to study supersymmetry comprehensively.
Indeed this formalism is very powerful to study the corrections to the NS5-brane and the KK-monopole.
The discussion is generalized to exotic branes.
The exotic $5^2_2$-brane has a two-torus fibration in the ten-dimensional spacetime viewpoint,
i.e., two compact circles, which come from the T-duality of NS5-brane.
We constructed a GLSM for the exotic $5^2_2$-brane \cite{Kimura:2013fda} in which the two compact directions are assigned by $(r^2, \vartheta)$.
The model naturally incorporated the topological term (\ref{topological-F}) including $\vartheta$.
In order to evaluate the string worldsheet instanton corrections along $r^2$, we expect the study of a topological term
\begin{align}
\exp \Big\{ - \frac{\I}{2 \pi} \int \d^2 x \, r^2 \wh{F}_{12} \Big\}
\, ,
\end{align}
where $\wh{F}_{12}$ is a new gauge field strength. 
Unfortunately, however, this form cannot be realized in \cite{Kimura:2013fda}.
This is because $r^2$ belongs to the chiral superfield $\Psi$ which never couples to a twisted chiral superfield as in (\ref{topological-F}).
Instead of this, the present authors tried to find an alternative formulation in which the corrections can be captured \cite{Kimura:2013zva, Kimura:2013khz, Kimura:2014aja, Kimura:2015yla, Kimura:2015cza}, but this approach has not been completed.

Recently, we explicitly obtained the string worldsheet instanton
corrections along the two-torus in the DFT framework \cite{Kimura:2018hph}.
In there we studied the corrections to the $5^2_2$-brane geometry that come from the string winding modes in the two-torus $T^2$.
As anticipated from the discussion of the KK-monopole, the corrections break the isometry along the two dual coordinates associated with the winding modes.
It is now indispensable to study the microscopic origin of the winding corrections to the exotic branes. Namely, we examine GLSM that captures the geometry with two-torus fibration correctly.
This implies that we have to find a new description in which we can compute the instanton corrections along two isometry directions in $T^2$ and their duals.
The only one alternative in our hand is $r^3$, i.e., the real part of the twisted chiral $\Theta$.
Precisely speaking, we have to prepare two topological terms
\bsubeq \label{2-top}
\begin{align}
\int \d^2 x \, \vartheta F_{12} + \int \d^2 x \, r^3 \wh{F}_{12} 
\, ,
\end{align}
and its superfield expression 
\begin{align}
\int \d^2 \wt{\theta} \, \Theta \big( \Sigma + \I \wh{\Sigma} \big)
+ \text{(h.c.)}
\, .
\end{align}
\esubeq
By using these new terms, we will study the simultaneous compactifications along both the real and imaginary parts of $\Theta$, and develop a consistent formula of dualization along both directions. 
This is different from the previous work \cite{Kimura:2013fda} in which the imaginary parts of $\Theta$ and $\Psi$ are compactified and dualized.
The case of a single dualization of such irreducible superfields has been established by \cite{Rocek:1991ps, Hori:2000kt}.  
However, 
a double dualization of a twisted chiral superfield has not been established.
Fortunately, we have already known that duality transformations in terms of reducible superfields are useful even when global isometries on the sigma model target space are unclear \cite{Gates:1984nk, Grisaru:1997ep}.
We would like to apply this formulation in the current situation.
Indeed, the present author has already tried to use this formulation in the exotic brane \cite{Kimura:2015qze}, and partially succeeded. 
In this paper, we would like to develop this formulation and complete it.
Hence, 
the clear statement of the issues we study is as follows:
``How do we construct the double dualization of a twisted chiral?'' 
``How is this procedure applied to the GLSM for five-branes of codimension two?'' and 
``How can we capture the string worldsheet instanton corrections to the
five-branes?''
We will show that the notion of the ``semi-doubled GLSM'' proposed in this paper plays a crucial role to answer these questions.

The organization of this paper is as follows.
In section \ref{S:twistedlinear},
we introduce a complex twisted linear superfield.
This plays a central role in dualizing both real and imaginary parts of the twisted chiral superfield $\Theta$.
In this discussion we demonstrate computations in detail for readers.
In section \ref{S:SDGLSM},
we first exhibit the GLSM proposed by Tong
for H-monopoles \cite{Tong:2002rq}.
This is a gauge theory in which one of the four transverse directions of the five-branes has an isometry by gauge symmetry. 
Next, we extend it by introducing another gauge symmetry which governs another isometry in the second direction.
We also introduce the complex twisted linear superfield.
Instead of the $\N=(4,4)$ superfields, we construct the system in the $\N=(2,2)$ superfield formalism. 
Here we focus on $SU(2)$ R-symmetry which constrains the system.
In the end of this section, we prepare the Lagrangian described by the component fields. 
In section \ref{S:SDNLSM},
we investigate the IR effective theory of the semi-doubled GLSM obtained in section \ref{S:SDGLSM}.
In this section, we will reconstruct various NLSMs whose target spaces are five-branes of codimension two, and will conclude that the dualization procedure in section \ref{S:twistedlinear} is applicable in the IR sigma models. 
In section \ref{S:standard},
we first perform the dualization procedure and obtain various standard GLSMs. 
We investigate their IR limit and find the NLSMs which have been derived in section \ref{S:SDNLSM}. 
Then we conclude that the dualization procedure in terms of the complex twisted linear superfield are applicable at the UV level, as well as at the IR level. 
In section \ref{S:instantons},
we study non-perturbative quantum corrections in gauge theories at the UV level.
After we discuss the Abrikosov-Nielsen-Olesen (ANO) vortex corrections, we find that they are mapped to the worldsheet instanton corrections to the background configurations of the defect five-branes. 
This analysis confirms the investigation in the framework of DFT in \cite{Kimura:2018hph}.
Section \ref{S:conclusion} is devoted to the conclusion and discussions.
In appendix \ref{A:SF}, we introduce two-dimensional $\N=(2,2)$ superfields and their component fields.
In appendix \ref{A:Omega}, we discuss the feature of one-forms and their polarizations which play a significant role in the background configurations of five-branes.

\section{Double dualization of twisted chiral} 
\label{S:twistedlinear}

In this section, we begin by a Lagrangian in which a twisted chiral superfield $\Theta$ is topologically coupled to gauge field strengths discussed in section \ref{S:introduction}:
\begin{align}
\Scr{L}_{\Theta}
\ &= \ 
- \int \d^4 \theta\, \frac{1}{g^2} |\Theta|^2
- \Big\{
\sqrt{2} \int \d^2 \wt{\theta} \, \Theta \big( \Sigma + \I \wh{\Sigma} \big)
+ \text{(h.c.)}
\Big\}
\nn \\
\ &= \ 
\int \d^4 \theta \, \Big\{
- \frac{1}{g^2} |\Theta|^2
- 2 \big( \Theta + \ol{\Theta} \big) V 
- 2 \I \big( \Theta - \ol{\Theta} \big) \wh{V}
\Big\}
\nn \\
\ & \ \ \ \ 
+ \sqrt{2} \, \eps^{mn} \del_m \big( \vartheta A_n \big)
+ \sqrt{2} \, \eps^{mn} \del_m \big( r^3 \wh{A}_n \big)
\, . \label{L-Theta}
\end{align}
Notice that the total derivative terms appear when we rewrite twisted superpotential terms to D-terms.
The component fields $\vartheta$ and $r^3$ come from the twisted chiral superfield $\Theta$, while $A_m$ and $\wh{A}_m$ are involved in the Abelian vector superfields $V$ and $\wh{V}$, respectively. 
$\eps^{mn}$ is the Levi-Civita antisymmetric symbol in two-dimensional spacetime, whose normalization is $\eps^{01} = +1 = - \eps^{10}$ and $\eps_{01} = -1 = - \eps_{10}$.
For later convenience, we explicitly express the expansion (for details, see appendix \ref{A:SF}):
\bsubeq \label{ex-Theta-Sigma}
\begin{align}
\Theta
\ &= \ 
\frac{1}{\sqrt{2}} (r^3 + \I \vartheta)
+ \I \sqrt{2} \, \theta^+ \ol{\wt{\chi}}{}_+
- \I \sqrt{2} \, \ol{\theta}{}^- \wt{\chi}_-
+ 2 \I \, \theta^+ \ol{\theta}{}^- \wt{\sf G}
+ \ldots
\, , \\
\Sigma
\ &= \ 
\sigma
+ \I \sqrt{2} \, \theta^+ \ol{\lambda}{}_+
- \I \sqrt{2} \, \ol{\theta}{}^- \lambda_-
- \sqrt{2} \, \theta^+ \ol{\theta}{}^- ({\sf D}_V - \I F_{01})
+ \ldots 
\, , \\
\wh{\Sigma}
\ &= \ 
\wh{\sigma}
+ \I \sqrt{2} \, \theta^+ \ol{\wh{\lambda}}{}_+
- \I \sqrt{2} \, \ol{\theta}{}^- \wh{\lambda}_-
- \sqrt{2} \, \theta^+ \ol{\theta}{}^- (\wh{\sf D}_V - \I \wh{F}_{01})
+ \ldots 
\, , 
\end{align}
\esubeq
where $F_{mn} = \del_m A_n - \del_n A_m$ and $\wh{F}_{mn} = \del_m \wh{A}_n - \del_n \wh{A}_m$.
We have already applied the Wess-Zumino gauge.
The term $\ldots$ in each expansion represents derivative terms described in appendix \ref{A:SF}.

\subsection{Dual Lagrangian}

Since we would like to study various theories related via dualities, we study a dual description of (\ref{L-Theta}) by introducing a new Lagrangian
\begin{align}
\Scr{L}_{R\wt{L}}
\ &= \ 
\int \d^4 \theta \, \Big\{
- \frac{1}{g^2} |R|^2
- 2 \big( R + \ol{R} \big) V 
- 2 \I \big( R - \ol{R} \big) \wh{V}
- R \wt{L} 
- \ol{R} \ol{\wt{L}}
\Big\}
\nn \\
\ & \ \ \ \ 
+ \sqrt{2} \, \eps^{mn} \del_m \big( \vartheta A_n \big)
+ \sqrt{2} \, \eps^{mn} \del_m \big( r^3 \wh{A}_n \big)
\, , \label{L-RLV}
\end{align}
where $R$ is an unconstrained complex superfield, 
and $\wt{L}$ is a complex twisted linear superfield whose definition is $0 = \ol{D}{}_+ D_- \wt{L}$.
The Lagrangian (\ref{L-RLV}) can be interpreted as a generating functional which provides the original Lagrangian (\ref{L-Theta}) and a new one in the following way: 
\begin{itemize}
\item 
Integrating out $\wt{L}$ in (\ref{L-RLV}), we find a constraint on $R$.
This indicates that the $R$ is reduced to a twisted chiral:
\begin{align}
0 \ &= \ 
\ol{D}_+ R
\ = \ D_- R
\, , \ls
\therefore \ \ \ 
R \ = \ 
\text{twisted chiral $\Theta$}
\, . \label{sol-R_1}
\end{align}
Substituting (\ref{sol-R_1}) into (\ref{L-RLV}), we easily obtain the original Lagrangian (\ref{L-Theta}).

\item 
Integrating out $R$, we obtain another solution of $R$:
\begin{align}
0 \ &= \ 
- \frac{1}{g^2} \ol{R}
- 2 (V + \I \wh{V}) - \wt{L}
\, , \ls
\therefore \ \ \ 
- \frac{1}{g^2} \ol{R}
\ = \ 
\wt{L} + 2 (V + \I \wh{V})
\, . \label{sol-R_2}
\end{align}
Substituting this into (\ref{L-RLV}), we obtain a new Lagrangian 
\begin{align}
\Scr{L}_{\wt{L}V}
\ &= \ 
g^2 \int \d^4 \theta \, \Big| \wt{L} + 2 (V + \I \wh{V}) \Big|^2
+ \sqrt{2} \, \eps^{mn} \del_m \big( \vartheta A_n \big)
+ \sqrt{2} \, \eps^{mn} \del_m \big( r^3 \wh{A}_n \big)
\, . \label{L-LV}
\end{align}

\end{itemize}
The above analysis is well known when duality transformations between (ir)reducible superfields are discussed \cite{Rocek:1991ps, Hori:2000kt} and \cite{Gates:1984nk, Grisaru:1997ep, Kimura:2015qze}.

We should emphasize that the Lagrangian (\ref{L-LV}) has not yet fixed as a dual Lagrangian of the original one (\ref{L-Theta}), because (\ref{L-LV}) is given by the reducible superfield $\wt{L}$.
Indeed, (\ref{L-LV}) carries many redundant degrees of freedom.
Choosing appropriate fields, we can obtain a correct dual Lagrangian, 
or we can realize the original Lagrangian itself.
In order to understand this statement, we carefully investigate the new Lagrangian (\ref{L-LV}).
It might be better to describe a complex twisted linear superfield $\wt{L}$ in terms of other irreducible ones, 
because $\wt{L}$ is equivalent to the sum of a chiral $X$, an anti-chiral $\ol{W}$ and a twisted chiral $Y$, i.e., $\wt{L} = X + \ol{W} + Y$ (see, appendix \ref{A:SF}).
Then (\ref{L-LV}) is rewritten as
\begin{align}
\Scr{L}_{\wt{L}V}
\ &= \ 
g^2 \int \d^4 \theta \, \Big| (X + \ol{W} + Y) + 2 (V + \I \wh{V}) \Big|^2
+ \sqrt{2} \, \eps^{mn} \del_m \big( \vartheta A_n \big)
+ \sqrt{2} \, \eps^{mn} \del_m \big( r^3 \wh{A}_n \big)
\, . \label{L-XYWV}
\end{align}
In order to study this Lagrangian, we expand the superfields in the following way:
\bsubeq \label{ex-XWY}
\begin{align}
X \ &= \
\frac{1}{\sqrt{2}} (\phi_{X,\text{R}} + \I \phi_{X,\text{I}})
+ \I \sqrt{2} \, \theta^+ \psi_{X+}
+ \I \sqrt{2} \, \theta^- \psi_{X-}
+ 2 \I \, \theta^+ \theta^- {\sf F}_X
+ \ldots
\, , \\
\ol{W} \ &= \ 
\frac{1}{\sqrt{2}} (\phi_{W,\text{R}} - \I \phi_{W,\text{I}})
+ \I \sqrt{2} \, \ol{\theta}{}^+ \ol{\psi}{}_{W+}
+ \I \sqrt{2} \, \ol{\theta}{}^- \ol{\psi}{}_{W-}
+ 2 \I \, \ol{\theta}{}^+ \ol{\theta}{}^- \ol{\sf F}{}_W
+ \ldots
\, , \\
Y \ &= \ 
\frac{1}{\sqrt{2}} (\sigma_{Y,\text{R}} + \I \sigma_{Y,\text{I}}) 
+ \I \sqrt{2} \, \theta^+ \ol{\wt{\chi}}{}_{Y+}
- \I \sqrt{2} \, \ol{\theta}{}^- \wt{\chi}_{Y-}
+ 2 \I \, \theta^+ \ol{\theta}{}^- \wt{\sf G}_Y
+ \ldots
\, .
\end{align}
\esubeq
Due to the linear combinations $X + \ol{W}$ and $V + \I \wh{V}$, 
the derivatives $\del_m \phi_{X,\text{I}}$ and $\del_m \phi_{W,\text{I}}$ are promoted to the covariant derivatives involving the gauge potential $A_m$ in such a way that
\begin{align}
D_m \phi_{X,\text{I}}
\ &= \ 
\del_m \phi_{X,\text{I}} - \sqrt{2} A_m
\, , \ls
D_m \phi_{W,\text{I}}
\ = \
\del_m \phi_{W,\text{I}} - \sqrt{2} A_m
\, .
\end{align}
In the same way, the derivatives $\del_m \phi_{X,\text{R}}$ and $\del_m \phi_{W,\text{R}}$ are also promoted to the covariant derivatives:
\begin{align}
D_m \phi_{X,\text{R}} 
\ &= \ 
\del_m \phi_{X,\text{R}} + \sqrt{2} \wh{A}_m
\, , \ls
D_m \phi_{W,\text{R}}
\ = \ 
\del_m \phi_{W,\text{R}} - \sqrt{2} \wh{A}_m
\, .
\end{align}
Furthermore, these covariant derivatives appear as linear combinations 
$D_m \phi_{W,\text{I}} + D_m \phi_{X,\text{I}}$ and 
$D_m \phi_{W,\text{R}} - D_m \phi_{X,\text{R}}$.
We again perform the field redefinition:
\bsubeq
\begin{alignat}{2}
2 \varphi_{\text{I}\pm}
\ &:= \ 
\phi_{W,\text{I}} \pm \phi_{X,\text{I}}
\, , &\ls
&\left\{
\renewcommand{\arraystretch}{1.4}
\begin{array}{rl}
D_m \phi_{W,\text{I}} + D_m \phi_{X,\text{I}}
&= \ 
2 D_m \varphi_{\text{I}+}
\, , \\
D_m \phi_{W,\text{I}} - D_m \phi_{X,\text{I}}
&= \ 
2 \del_m \varphi_{\text{I}-} 
\, , 
\end{array}
\right.
\\
2 \varphi_{\text{R}\pm}
\ &:= \ 
\phi_{W,\text{R}} \pm \phi_{X,\text{R}}
\, , &\ls
&\left\{
\renewcommand{\arraystretch}{1.4}
\begin{array}{rl}
D_m \phi_{W,\text{R}} - D_m \phi_{X,\text{R}}
&= \
2 D_m \varphi_{\text{R}-}
\, , \\
D_m \phi_{W,\text{R}} + D_m \phi_{X,\text{R}}
&= \ 
2 \del_m \varphi_{\text{R}+} 
\, .
\end{array}
\right.
\end{alignat}
\esubeq
By using the above expressions, we write down the Lagrangian (\ref{L-XYWV}) in terms of the component fields:
\begin{align}
\Scr{L}_{\wt{L}V}
\ &= \ 
- g^2 \Big\{ 
(\del_m \varphi_{\text{R}+})^2
+ (\del_m \varphi_{\text{I}-})^2  
+ (D_m \varphi_{\text{R}-})^2
+ (D_m \varphi_{\text{I}+})^2
\Big\}
\nn \\
\ & \ \ \ \ 
+ \frac{g^2}{2} \Big\{ (\del_m \sigma_{Y,\text{R}})^2 + (\del_m \sigma_{Y,\text{I}})^2 \Big\} 
+ g^2 \eps^{mn} (\del_m \sigma_{Y,\text{R}}) (D_n \varphi_{\text{R}-})
- g^2 \eps^{mn} (\del_m \sigma_{Y,\text{I}}) (D_n \varphi_{\text{I}+})
\nn \\
\ & \ \ \ \ 
- \sqrt{2} \, g^2 {\sf D}_{V} \big( 2 \varphi_{\text{R}+} + \sigma_{Y,\text{R}} \big)
- \sqrt{2} \, g^2 \wh{\sf D}_{V} \big( 2 \varphi_{\text{I}-} + \sigma_{Y,\text{I}} \big)
\nn \\
\ & \ \ \ \ 
+ g^2 |{\sf F}_X|^2
+ g^2 |{\sf F}_W|^2
- g^2 \big| \ol{\wt{\sf G}}{}_Y - \I \sqrt{2} \, (\sigma - \I \wh{\sigma}) \big|^2
- 2 g^2 |\sigma + \I \wh{\sigma}|^2
\nn \\
\ & \ \ \ \
+ \I g^2 \, \ol{\psi}{}_{X-} \del_+ \psi_{X-} 
+ \I g^2 \, \ol{\psi}{}_{W+} \del_- \psi_{W+} 
\nn \\
\ & \ \ \ \ 
+ \frac{\I g^2}{2} (\psi_{X+} + \ol{\wt{\chi}}{}_{Y+}) \del_- (\ol{\psi}{}_{X+} - \wt{\chi}_{Y+})
+ \frac{\I g^2}{2} (\ol{\psi}{}_{X+} + \wt{\chi}_{Y+}) \del_- (\psi_{X+} - \ol{\wt{\chi}}{}_{Y+}) 
\nn \\
\ & \ \ \ \ 
+ \frac{\I g^2}{2} (\ol{\psi}{}_{W-} - \wt{\chi}_{Y-}) \del_+ (\psi_{W-} + \ol{\wt{\chi}}{}_{Y-})
+ \frac{\I g^2}{2} (\psi_{W-} - \ol{\wt{\chi}}{}_{Y-}) \del_+ (\ol{\psi}{}_{W-} + \wt{\chi}_{Y-})
\nn \\
\ & \ \ \ \
+ \sqrt{2} \, g^2 \, \psi_{X-} (\lambda_{+} - \I \wh{\lambda}_{+}) 
- \sqrt{2} \, g^2 \, \ol{\psi}{}_{X-} (\ol{\lambda}{}_{+} + \I \ol{\wh{\lambda}}{}_{+}) 
\nn \\
\ & \ \ \ \ 
- \sqrt{2} \, g^2 \, \psi_{W+} (\lambda_{-} + \I \wh{\lambda}_{-})  
+ \sqrt{2} \, g^2 \, \ol{\psi}{}_{W+} (\ol{\lambda}{}_{-} - \I \ol{\wh{\lambda}}{}_{-}) 
\nn \\
\ & \ \ \ \ 
- \sqrt{2} \, g^2 (\psi_{X+} + \ol{\wt{\chi}}{}_{Y+}) (\lambda_{-} - \I \wh{\lambda}_{-}) 
+ \sqrt{2} \, g^2 (\ol{\psi}{}_{X+} + \wt{\chi}_{Y+}) (\ol{\lambda}{}_{-} + \I \ol{\wh{\lambda}}{}_{-}) 
\nn \\
\ & \ \ \ \ 
- \sqrt{2} \, g^2 (\ol{\psi}{}_{W-} - \wt{\chi}_{Y-}) (\ol{\lambda}{}_{+} - \I \ol{\wh{\lambda}}{}_{+}) 
+ \sqrt{2} \, g^2 (\psi_{W-} - \ol{\wt{\chi}}{}_{Y-}) (\lambda_{+} + \I \wh{\lambda}_{+}) 
\nn \\
\ & \ \ \ \ 
+ \sqrt{2} \, \eps^{mn} \del_m \big( \vartheta A_n \big)
+ \sqrt{2} \, \eps^{mn} \del_m \big( r^3 \wh{A}_n \big)
\, . \label{L-XYWV_2}
\end{align}

\subsection{Dualization procedure}

As mentioned before, 
the Lagrangian (\ref{L-XYWV_2}) carries many redundant degrees of freedom compared with the original one (\ref{L-Theta}).
In order to reduce them, we consider the duality relation among component fields.
Previously, we have found two expressions of the superfield $R$ 
as in (\ref{sol-R_1}) and (\ref{sol-R_2}).
Connecting these two, we obtain the duality relation 
\begin{align}
- \frac{1}{g^2} \ol{\Theta}
\ &\doteq \
\wt{L} + 2 (V + \I \wh{V})
\, . \label{dual-Theta} 
\end{align}
Here the symbol $\doteq$ means that it specifies the duality transformation rule.
Hence, it turns out that the duality relations among component fields can be understood.
The relations among bosonic fields are
\bsubeq \label{dual-Theta_4b}
\begin{alignat}{2}
\frac{1}{g^2} r^3 
\ &\doteq \ 
- 2 \varphi_{\text{R}+} 
- \sigma_{Y,\text{R}}
\, , &\ls
\frac{1}{g^2} \vartheta
\ &\doteq \ 
- 2 \varphi_{\text{I}-}
+ \sigma_{Y,\text{I}}
\, , \label{4b_1} \\
\frac{1}{g^2} \del_m r^3 
\ &\doteq \ 
+ 2 \eps_{mn} D^n \varphi_{\text{R}-}
+ \del_m \sigma_{Y,\text{R}}
\, , &\ls
\frac{1}{g^2} \del_m \vartheta
\ &\doteq \
+ 2 \eps_{mn} D^n \varphi_{\text{I}+}
- \del_m \sigma_{Y,\text{I}}
\, , \label{4b_2} \\
0 
\ &\doteq \ 
{\sf F}_X
\, , &\ls
0
\ &\doteq \ 
\ol{\sf F}{}_W
\, , \label{4b_3} \\
0
\ &\doteq \ 
\wt{\sf G}_Y 
+ \I \sqrt{2} \, (\ol{\sigma} + \I \ol{\wh{\sigma}})
\, , &\ls
\frac{1}{g^2} \ol{\wt{\sf G}}
\ &\doteq \ 
\I \sqrt{2} \, (\sigma + \I \wh{\sigma})
\, . \label{4b_4}
\end{alignat}
\esubeq
The relations among fermionic fields are
\bsubeq \label{dual-Theta_4f}
\begin{alignat}{2}
\ol{\wt{\chi}}{}_{-}
\ &\doteq \ 
+ g^2 \psi_{X-}
\, , &\ls
\wt{\chi}_+
\ &\doteq \ 
- g^2 \ol{\psi}{}_{W+} 
\, , \label{4f_1} \\
\del_+ \ol{\wt{\chi}}{}_-
\ &\doteq \ 
- g^2 \Big\{
\del_+ \psi_{X-} 
+ 2 \I \sqrt{2} \, (\ol{\lambda}{}_{+,1} + \I \ol{\lambda}{}_{+,2})
\Big\}
\, , &\ls
\del_- \wt{\chi}_+
\ &\doteq \ 
+ g^2 \Big\{
\del_- \ol{\psi}{}_{W+} 
+ 2 \I \sqrt{2} \, (\lambda_{-,1} + \I \lambda_{-,2})
\Big\}
\, , \label{4f_2} \\
0
\ &\doteq \ 
\psi_{X+} + \ol{\wt{\chi}}{}_{Y+}
\, , &\ls
0
\ &\doteq \ 
\ol{\psi}{}_{W-} - \wt{\chi}{}_{Y-}
\, , \label{4f_3} \\
0
\ &\doteq \ 
\del_- (\psi_{X+} - \ol{\wt{\chi}}{}_{Y+}) 
- 2 \I \sqrt{2} \, (\ol{\lambda}{}_{-,1} + \I \ol{\lambda}{}_{-,2})
\, , &\ls
0
\ &\doteq \ 
\del_+ (\psi_{X+} + \wt{\chi}{}_{Y-}) 
- 2 \I \sqrt{2} \, (\lambda_{+,1} + \I \lambda_{+,2})
\, . \label{4f_4}
\end{alignat}
\esubeq
We should have comments on the duality relation between superfields (\ref{dual-Theta}) and those among component fields (\ref{dual-Theta_4b}) and (\ref{dual-Theta_4f}).
The relations among non-derivative terms describe equivalence,
while equivalence is not applied to derivative terms.
This is because the signature of derivative $\del_-$ in the left-hand
side of (\ref{dual-Theta}) is twisted compare with the right-hand side.
Indeed, the Levi-Civita antisymmetric symbol $\eps_{mn}$ appears in the bosonic relations (\ref{4b_1}) and (\ref{4b_2}). 
This originates from the twisting of the signature in front of the derivatives. 
The symbol plays a essential role in dualization in the component field framework.

We discuss the relations among various fields more carefully.
Since the derivative of the scalar field $\varphi_{\text{R}-}$ has the antisymmetric symbol $\eps_{mn}$, we understand that $\varphi_{\text{R}-}$ is dual to the original field $r^3$. 
In the same way, the field $\varphi_{\text{I}+}$ is dual to $\vartheta$.
We also notice that $\varphi_{\text{R}+}$ and $\sigma_{Y,\text{R}}$ might equally contribute to the original field $r^3$.
However, $\sigma_{Y,\text{R}}$ has a negative norm in (\ref{L-XYWV_2}). 
Then we regard this as an unphysical field which should be integrated out from the Lagrangian.
The role of $\sigma_{Y,\text{I}}$ is the same as $\sigma_{Y,\text{R}}$.
After the integration, the duality relation between the original and dual fields would be complemented.
The fermionic fields should be treated in the same way as their bosonic supersymmetric partners.
We summarize the relations among bosonic fields in Table \ref{Tb:duality}.
\begin{center}
\slb{1}{\setlength\dashlinedash{1.4pt}
\setlength\dashlinegap{2.3pt}
\renewcommand{\arraystretch}{1.35}
\begin{tabular}{c||p{27mm}:p{27mm}:p{27mm}p{-2mm}} \hline
\multirow{1}{*}{$\ol{\Theta}$} & \multicolumn{3}{|c}{\multirow{1}{*}{$\wt{L} = X + \ol{W} + Y$}} 
\\ \hline
\text{original} 
& \centering \text{equivalent} 
& \centering \text{dual} 
& \centering \text{unphysical} 
&
\\ \hline \hline
\multirow{1}{*}{$r^3$} 
& \centering \multirow{1}{*}{$\varphi_{\text{R}+}$}
& \centering \multirow{1}{*}{$\varphi_{\text{R}-}$}
& \centering \multirow{1}{*}{$\sigma_{Y,\text{R}}$}
&
\\ \hdashline
\multirow{1}{*}{$\vartheta$} 
& \centering \multirow{1}{*}{$\varphi_{\text{I}-}$}
& \centering \multirow{1}{*}{$\varphi_{\text{I}+}$}
& \centering \multirow{1}{*}{$\sigma_{Y,\text{I}}$}
&
\\ \hline
\end{tabular}
}
\tbcaption{\footnotesize The duality relations between bosonic fields in $\Theta$ and $\wt{L}$ in (\ref{dual-Theta_4b}). We ignore auxiliary fields which are not so significant.}
\label{Tb:duality}
\end{center}

Now we apply the dualization procedure to the Lagrangian (\ref{L-XYWV_2}).
Unfortunately, in the previous work \cite{Kimura:2015qze}, the method has not been established.
But in the current analysis, we propose the following procedure:
\begin{enumerate}
\item[1.] By using (\ref{4b_1}), remove $\varphi_{\text{R}+}$ and $\varphi_{\text{I}-}$ from the Lagrangian (\ref{L-XYWV_2}).

\item[2.] Integrate out $\sigma_{Y,\text{R}}$ and $\sigma_{Y,\text{I}}$.

\item[3-(i).] 
Integrate out $r^3$ (or $\varphi_{\text{R}-}$) if the Lagrangian of $\varphi_{\text{R}-}$ (or $r^3$) would like to be obtained. 

\item[3-(ii).] 
Integrate out $\vartheta$ (or $\varphi_{\text{I}+}$) if the Lagrangian of $\varphi_{\text{I}+}$ (or $\vartheta$) would like to be obtained. 
\end{enumerate}
This is simple and quite natural.
Actually the relations between derivative terms (\ref{4b_2}) themselves are not explicitly utilized, while similar forms will be obtained.
In order to confirm that this procedure really works, let us demonstrate this dualization procedure.

\subsection{Demonstration}

To simplify the discussion, we decompose the Lagrangian (\ref{L-XYWV_2}) in the following way:
\bsubeq
\begin{align}
\Scr{L}_{\wt{L}V}
\ &:= \ 
\Scr{L}_{\text{R}}
+ \Scr{L}_{\text{I}}
+ \Scr{L}_{\text{O}}
\, , \\
\Scr{L}_{\text{R}}
\ &:= \ 
- g^2 \Big\{ 
(\del_m \varphi_{\text{R}+})^2
+ (D_m \varphi_{\text{R}-})^2
\Big\}
+ \frac{g^2}{2} (\del_m \sigma_{Y,\text{R}})^2 
+ g^2 \eps^{mn} (\del_m \sigma_{Y,\text{R}}) (D_n \varphi_{\text{R}-})
\nn \\
\ & \ \ \ \ 
- \sqrt{2} \, g^2 {\sf D}_{V} \big( 2 \varphi_{\text{R}+} + \sigma_{Y,\text{R}} \big)
+ \sqrt{2} \, \eps^{mn} \del_m \big( r^3 \wh{A}_n \big)
\, , \label{L-XYWV_R} \\
\Scr{L}_{\text{I}}
\ &:= \ 
- g^2 \Big\{ 
(\del_m \varphi_{\text{I}-})^2  
+ (D_m \varphi_{\text{I}+})^2
\Big\}
+ \frac{g^2}{2} (\del_m \sigma_{Y,\text{I}})^2 
- g^2 \eps^{mn} (\del_m \sigma_{Y,\text{I}}) (D_n \varphi_{\text{I}+})
\nn \\
\ & \ \ \ \ 
- \sqrt{2} \, g^2 \wh{\sf D}_{V} \big( 2 \varphi_{\text{I}-} + \sigma_{Y,\text{I}} \big)
+ \sqrt{2} \, \eps^{mn} \del_m \big( \vartheta A_n \big)
\, , \label{L-XYWV_I} \\
\Scr{L}_{\text{O}}
\ &:= \ 
\text{(other terms)}
\, . 
\end{align}
\esubeq
Let us demonstrate the dualization procedure.

\begin{enumerate}
\item[1.] {Remove $\varphi_{\text{R}+}$ and $\varphi_{\text{I}-}$:}

By using (\ref{4b_1}), we first remove the equivalent fields $\varphi_{\text{R}+}$ and $\varphi_{\text{I}-}$. The Lagrangians (\ref{L-XYWV_R}) and (\ref{L-XYWV_I}) are rewritten as
\bsubeq
\begin{align}
\Scr{L}_{\text{R}}
\ &= \ 
- \frac{1}{4 g^2} (\del_m r^3)^2
- g^2 (D_m \varphi_{\text{R}-})^2
+ \sqrt{2} \, {\sf D}_{V} r^3
+ \sqrt{2} \, \eps^{mn} \del_m \big( r^3 \wh{A}_n \big)
\nn \\
\ & \ \ \ 
+ \frac{g^2}{4} (\del_m \sigma_{Y,\text{R}})^2
- \frac{1}{2} (\del_m r^3) (\del^m \sigma_{Y,\text{R}})
+ g^2 \eps^{mn} (\del_m \sigma_{Y,\text{R}}) (D_n \varphi_{\text{R}-})
\, , \label{LR_2} \\
\Scr{L}_{\text{I}}
\ &= \ 
- \frac{1}{4 g^2} (\del_m \vartheta)^2  
- g^2 (D_m \varphi_{\text{I}+})^2
- \sqrt{2} \, \wh{\sf D}_{V} \, \vartheta
+ \sqrt{2} \, \eps^{mn} \del_m \big( \vartheta A_n \big)
\nn \\
\ & \ \ \ \ 
+ \frac{g^2}{4} (\del_m \sigma_{Y,\text{I}})^2
+ \frac{1}{2} (\del_m \vartheta) (\del^m \sigma_{Y,\text{I}})
- g^2 \eps^{mn} (\del_m \sigma_{Y,\text{I}}) (D_n \varphi_{\text{I}+})
\, . \label{LI_2}
\end{align}
\esubeq

\item[2.] {Integrate out $\sigma_{Y,\text{R}}$ and $\sigma_{Y,\text{I}}$:}

Next, we solve the equation of motion for $\sigma_{Y,\text{R}}$ in (\ref{LR_2}):
\bsubeq
\begin{align}
0 \ &= \ 
\del^m \Big\{
- \frac{g^2}{2} (\del_m \sigma_{Y,\text{R}})
+ \half \del_m r^3
- g^2 \eps_{mn} (D^n \varphi_{\text{R}-})
\Big\}
\, , \\
\therefore \ \ \ 
\del_m \sigma_{Y,\text{R}}
\ &= \ 
\frac{1}{g^2} \del_m r^3
- 2 \eps_{mn} (D^n \varphi_{\text{R}-})
+ \del^n a_{mn} 
\, . 
\end{align}
\esubeq
Here $a_{mn}$ is an arbitrary antisymmetric tensor.
Substituting this solution into (\ref{LR_2}), we obtain
\begin{align}
\Scr{L}_{\text{R}}
\ &= \ 
- \frac{1}{2 g^2} (\del_m r^3)^2
+ \eps^{mn} (\del_m r^3) (D_n \varphi_{\text{R}-})
+ \sqrt{2} \, {\sf D}_V \, r^3
+ \sqrt{2} \, \eps^{mn} \del_m \big( r^3 \wh{A}_n \big)
\, , \label{LR_3}
\end{align}
where we simply set $a_{mn}$ to zero.

Analogously, we integrate out $\sigma_{Y,\text{I}}$ from (\ref{LI_2}).
The result is
\begin{align}
\Scr{L}_{\text{I}}
\ &= \ 
- \frac{1}{2 g^2} (\del_m \vartheta)^2
+ \eps^{mn} (\del_m \vartheta) (D_n \varphi_{\text{I}+})
- \sqrt{2} \, \wh{\sf D}_V \, \vartheta
+ \sqrt{2} \, \eps^{mn} \del_m \big( \vartheta A_n \big)
\, . \label{LI_3}
\end{align}

The expressions of (\ref{LR_3}) and (\ref{LI_3}) are quite simple to explore the duality relation. 
In the final step, we derive the original Lagrangians and the dual ones.

\item[3-(i).] {Dualize original $r^3$ or dual $\varphi_{\text{R}-}$:}

We focus on (\ref{LR_3}) and first integrate out the field $\varphi_{\text{R}-}$.
The equation of motion is 
\begin{align}
0 \ &= \ 
\del^n \Big\{ - \eps_{mn} (\del^m r^3) \Big\}
\, , \label{eq_1-1}
\end{align}
whose solution is trivial. Then the Lagrangian under this equation is reduced to
\begin{align}
\Scr{L}_{\text{R}} \big|_{\text{(\ref{eq_1-1})}}
\ &= \ 
- \frac{1}{2 g^2} (\del_m r^3)^2
+ \sqrt{2} \, {\sf D}_V \, r^3
+ \sqrt{2} \, r^3 \wh{F}_{01}
\, . \label{LR_4-1}
\end{align}
This is nothing but the original Lagrangian for $r^3$ derived from (\ref{L-Theta}).

On the other hand, we study the equation of motion for $r^3$ in (\ref{LR_3}):
\bsubeq \label{dual-r3_toy}
\begin{align}
0 \ &= \ 
\sqrt{2} \, {\sf D}_V
- \del^m \Big\{ - \frac{1}{g^2} \del_m r^3 + \eps_{mn} (D^n \varphi_{\text{R}-}) \Big\}
\, .
\end{align}
Its formal solution is
\begin{align}
\frac{1}{g^2} (\del^m r^3)
\ &= \ 
\eps^{mn} (D_n \varphi_{\text{R}-})
- \sqrt{2} \, \mathscr{A}^m 
- \del_n b^{mn}
\, , \ls
\mathscr{A}^m
\ := \ 
\int \d x^m \, {\sf D}_V
\, . 
\end{align}
\esubeq
Here $b^{mn}$ is an arbitrary antisymmetric tensor.
This is genuinely the duality relation between $r^3$ and $\varphi_{\text{R}-}$, rather than that in (\ref{4b_2}).
Substituting this into the Lagrangian (\ref{LR_3}), we obtain
\begin{align}
\Scr{L}_{\text{R}} \big|_{\text{(\ref{dual-r3_toy})}}
\ &= \
- \frac{g^2}{2} (D_m \varphi_{\text{R}-})^2
+ \sqrt{2} \, {\sf D}_V \, \mr{r}^3
+ \sqrt{2} \, \eps^{mn} \del_m \big( \mr{r}^3 \wh{A}_n \big)
- g^2 \mathscr{A}_m \mathscr{A}^m 
\, . \label{LR_4-2}
\end{align}
Here the symbol $\mr{r}^3$ implies that the field $r^3$ is governed by the equation of motion (\ref{dual-r3_toy}).
Without loss of generality, we can set the tensor $b^{mn}$ to zero.
Compared with (\ref{LR_4-1}), this model is genuinely dual, 
because the coupling constant $g^2$ appears inversely.
This phenomenon can be seen in the developed dualization procedure \cite{Rocek:1991ps, Hori:2000kt} and a kind of its applications \cite{Tong:2002rq}. 
In this dualization procedure, $\vartheta$ is not related to this Lagrangian at all. This indicates that we can perform the duality transformation of $r^3$ and $\vartheta$ separately.

\item[3-(ii).] {Dualize original $\vartheta$ or dual $\varphi_{\text{I}+}$:}

Independent of $r^3$ and $\varphi_{\text{R}-}$, we discuss the duality relation between $\vartheta$ and $\varphi_{\text{I}+}$.
The technique itself is completely parallel to the previous discussion.
First, we construct the Lagrangian for $\vartheta$ by integrating out $\varphi_{\text{I}+}$ from (\ref{LI_3}).
The equation of motion for $\varphi_{\text{I}+}$ is trivial:
\begin{align}
0 \ &= \ 
\del^n \Big\{ \eps_{mn} (\del^m \vartheta) \Big\}
\, . \label{eq_2-1}
\end{align}
Substituting this into (\ref{LI_3}), we obtain the original Lagrangian for $\vartheta$ derived from (\ref{L-Theta}):
\begin{align}
\Scr{L}_{\text{I}} \big|_{\text{(\ref{eq_2-1})}}
\ &= \ 
- \frac{1}{2 g^2} (\del_m \vartheta)^2
- \sqrt{2} \, \wh{\sf D}_V \, \vartheta
+ \sqrt{2} \, \vartheta F_{01}
\, . \label{LI_4-1}
\end{align}

On the other hand, we study the equation of motion for $\vartheta$ in (\ref{LI_3})
\bsubeq \label{dual-theta_toy}
\begin{align}
0 \ &= \ 
- \sqrt{2} \, \wh{D}_V
- \del^m \Big\{
- \frac{1}{g^2} (\del_m \vartheta)
+ \eps_{mn} (D^n \varphi_{\text{I}+})
\Big\}
\, , 
\end{align}
whose formal solution is
\begin{align}
\frac{1}{g^2} (\del^m \vartheta)
\ &= \ 
\eps^{mn} (D_n \varphi_{\text{I}+})
+ \sqrt{2} \, \wh{\mathscr{A}}{}^m
- \del_n c^{mn}
\, , \ls
\wh{\mathscr{A}}{}^m
\ := \ 
\int \d x^m \, \wh{D}_V
\, , 
\end{align}
\esubeq
where $c^{mn}$ is an arbitrary antisymmetric tensor.
This is nothing but the duality relation between $\vartheta$ and $\varphi_{\text{I}+}$. 
This is different from (\ref{4b_2}).
Substituting this into (\ref{LI_3}), 
we obtain the dual Lagrangian
\begin{align}
\Scr{L}_{\text{I}} \big|_{\text{(\ref{dual-theta_toy})}}
\ &= \ 
- \frac{g^2}{2} (D_m \varphi_{\text{I}+})^2
- \sqrt{2} \, \wh{D}_V \, \mr{\vartheta}
+ \sqrt{2} \, \eps^{mn} \del_m \big( \mr{\vartheta} A_n \big)
- g^2 \wh{\mathscr{A}}_m \wh{\mathscr{A}}{}^m
\, . \label{LI_4-2}
\end{align}
Here $\mr{\vartheta}$ is the original field $\vartheta$ governed by the equation of motion (\ref{dual-theta_toy}).
We have also set the function $c^{mn}$ to zero, for simplicity.
Applying the same analysis in the previous case, we conclude that the Lagrangian (\ref{LI_4-2}) is dual to the original one (\ref{LI_4-1}).

\end{enumerate}

Now we understood that the superfield formulation (\ref{L-XYWV}) and the duality relation (\ref{dual-Theta}), or the component field formulation (\ref{L-XYWV_2}), (\ref{dual-Theta_4b}) and (\ref{dual-Theta_4f}) are quite powerful to analyze the duality relation explicitly.
Indeed, we can simultaneously perform dualization both the real and imaginary part of the original twisted chiral $\Theta$.
We refer to this as the ``double dualization''. 
The reduced forms (\ref{LR_3}) and (\ref{LI_3}) contain both the original and dual fields, though only the original fields possess the kinetic terms.
As proposed in \cite{Kimura:2015qze}, we call these forms as the ``semi-doubled'' Lagrangians.

\section{Semi-doubled GLSM}
\label{S:SDGLSM}

In this section, we propose a semi-doubled GLSM for five-branes of codimension two. 
This is an extension of the $\N=(4,4)$ GLSM for H-monopoles provided by Tong \cite{Tong:2002rq} and developed by Harvey and Jensen \cite{Harvey:2005ab}, Okuyama \cite{Okuyama:2005gx}.
We will read off, from the prototypical example of GLSM proposed by Tong, 
how isometry of the background configuration is generated, and how T-duality is realized by using two-dimensional gauge theory.
We have three reasons why we study and develop this gauge theory:
\begin{itemize}
\item[(i)] The GLSM by Tong represents the background geometry of multiple H-monopoles in the IR limit. 
An H-monopole is nothing but an NS5-brane of codimension {\it three}, i.e., one of the four transverse directions of the NS5-brane is compactified and smeared.
The remaining three directions are governed by $SU(2)$ R-symmetry.
\item[(ii)] This model has a duality transformation at the UV level which captures T-duality transformation at the IR level.
Then the KK-monopole, whose geometry is described as the Taub-NUT space, is successfully obtained.
\item[(iii)] The ANO vortex corrections, a kind of non-perturbative quantum corrections to gauge theory at the UV level, describe the worldsheet instanton corrections in the NLSM at the IR level.
Indeed, this corrections can be interpreted as the string KK-modes or winding mode corrections to the background geometry.
\end{itemize}

We first briefly study the field contents and $SU(2)$ R-symmetry.
Next, we develop this model to our semi-doubled GLSM for five-branes of codimension {\it two}.
For later convenience, we often refer to such branes as ``defect'' five-branes \cite{Bergshoeff:2011se}.
There we study how the field contents are extended and how the $SU(2)$ R-symmetry is modified.
In this section we focus only on the structure of the semi-doubled GLSM.
The discussion of the IR effective theory of the semi-doubled GLSM will be shown in section \ref{S:SDNLSM}.
The duality transformations at the IR level will be discussed also in section \ref{S:SDNLSM}, and those at the UV level will be in section \ref{S:standard}.
The ANO vortex corrections will be discussed in section \ref{S:instantons}.
Throughout these sections, we will find an alternative construction of the gauge theory for exotic $5^2_2$-brane and its corrections discussed in \cite{Kimura:2013fda, Kimura:2013zva, Kimura:2013khz}.
Furthermore, we will understand that our new model in this paper is much more natural to consider the background geometries of defect five-branes,
and simple to study further analyses such as path integral evaluations, analogous to \cite{Harvey:2014nha}.

\subsection{GLSM for H-monopole: NS5-brane of codimension three}

In \cite{Tong:2002rq}, the author successfully described the GLSM for H-monopoles and its T-dual system, i.e., a multi-centered Taub-NUT space,
by using $SU(2)$ R-symmetry in $\N=(4,4)$ supersymmetry.
The detail of the construction can be seen in appendix of \cite{Tong:2002rq}.
Here we exhibit the Lagrangian which contains three constituents:
\bsubeq \label{Tong-L}
\begin{align}
\Scr{L}
\ &= \ 
\Scr{L}^{\text{gauge}}
+ \Scr{L}^{\text{CHM}}
+ \Scr{L}^{\text{NHM}}
\, , \\
\Scr{L}^{\text{gauge}}
\ &= \ 
\int \d^4 \theta \, \sum_{a = 1}^k \frac{1}{e_a^2} \Big\{ - |\Sigma_a|^2 + |\Phi_a|^2 \Big\}
\, , \label{TongL-gauge} \\
\Scr{L}^{\text{CHM}}
\ &= \ 
\int \d^4 \theta \, \sum_a \Big\{ |Q_a|^2 \, \e^{+ 2 V_a} + |\wt{Q}_a|^2 \, \e^{- 2 V_a} \Big\}
- \Big\{ \sqrt{2} \int \d^2 \theta \, \sum_a \wt{Q}_a \Phi_a Q_a + \text{(h.c.)} \Big\}
\, , \label{TongL-CHM} \\
\Scr{L}^{\text{NHM}}
\ &= \ 
\frac{1}{g^2} \int \d^4 \theta \, \Big\{ - |\Theta|^2 + |\Psi|^2 \Big\}
\nn \\
\ & \ \ \ \ 
+ \sum_a \Big\{ 
\sqrt{2} \int \d^2 \wt{\theta} \, (t_a - \Theta) \Sigma_a 
+ \sqrt{2} \int \d^2 \theta \, (s_a - \Psi) \Phi_a 
+ \text{(h.c.)}
\Big\}
\, . \label{TongL-NHM}
\end{align}
\esubeq
We note that all of the field contents are given in terms of $\N=(2,2)$ superfields with $SU(2)$ R-symmetry. 
First, $(V_a, \Phi_a)$ are $\N=(4,4)$ $U(1)^k$ gauge multiplets with gauge coupling constants $e_a$ of mass dimension one.
In particular, we often use $\Sigma_a = \frac{1}{\sqrt{2}} \ol{D}{}_+ D_- V_a$.
Their component fields are represented as $\Sigma_a (\sigma_a, \ol{\lambda}_{+,a}, \lambda_{-,a}, A_{m,a}, {\sf D}_{V,a})$ and $\Phi_a (\phi_a, \wt{\lambda}_{\pm,a}, {\sf F}_{\Phi,a})$.
Roughly speaking, each $U(1)$ gauge symmetry describes a (gauged) isometry along one of the four transverse directions (i.e., $\vartheta$ in $\Theta$) of each five-brane.
Second, $(Q_a, \wt{Q}_a)$ are $k$ charged hypermultiplets coupled to $(V_a, \Phi_a)$.
The component fields are given as $Q_a(q_a, \psi_{\pm,q,a}, {\sf F}_{q,a})$ and $\wt{Q}_a(\wt{q}_a, \wt{\psi}_{\pm,q,a}, \wt{\sf F}_{q,a})$.
On a supersymmetric vacuum, they make the target space geometry be curved.
Third, $(\Psi, \Theta)$ is a neutral hypermultiplet which describes the four transverse directions of the five-branes. 
Their field contents are $\Psi(\tfrac{r^1 + \I r^2}{\sqrt{2}}, \chi_{\pm}, {\sf F})$ and $\Theta(\tfrac{r^3 + \I \vartheta}{\sqrt{2}}, \ol{\wt{\chi}}{}_+, \wt{\chi}_-, {\sf G})$.
This multiplet carries the sigma model coupling constant $g$ which controls the asymptotic radius of the isometry direction $\vartheta$, 
and $k$ Fayet-Iliopoulos (FI) parameters $s_a = \frac{1}{\sqrt{2}} (s^1_a + \I s^2_a)$ and $t_a = \frac{1}{\sqrt{2}} (t^3_a + \I t^4_a)$ which represent the positions of $k$ five-branes in the transverse space.

Here we mention $SU(2)$ R-symmetry which assigns $\N=(4,4)$ supersymmetry.
Strictly speaking, there exists three $SU(2)$ R-symmetries in the system,
i.e., $SO(4) \times SU(2)_R \simeq SU(2)_1 \times SU(2)_2 \times SU(2)_R$ R-symmetries
(for details, see \cite{Tong:2002rq, Harvey:2005ab, Harvey:2014nha}).
The component fields are assigned as appropriate representations under these symmetries in such a way that 
\begin{align}
\begin{array}{r@{\hspace{-1mm}}rl}
\multirow{2}{*}{$\text{gauge multiplet $(V_a, \Phi_a)$} \ \ \left\{ 
\begin{array}{c}
\vphantom{(\sigma_a, \phi_a)}
\cr
\vphantom{A_{m,a}}
\cr
\vphantom{(\lambda_{\pm,a}, \wt{\lambda}_{\pm,a})}
\end{array}
\right.$}
&
(\sigma_a, \phi_a) 
\ &: \ \ 
({\bf 2}, {\bf 2}, {\bf 1})
\\
&
A_{m,a}
\ &: \ \ 
({\bf 1}, {\bf 1}, {\bf 1})
\\
&
(\lambda_{\pm,a}, \wt{\lambda}_{\pm,a}) 
\ &: \ \ 
({\bf 2}, {\bf 1}, {\bf 2})_- \oplus ({\bf 1}, {\bf 2}, {\bf 2})_+
\\
\\
\multirow{2}{*}{$\text{charged hypermultiplet $(Q_a, \wt{Q}_a)$} \ \ \left\{ 
\begin{array}{c}
\vphantom{(q_a , \wt{q}_a)}
\cr
\vphantom{(\psi_{\pm,q,a}, \wt{\psi}_{\pm,q,a})}
\end{array}
\right.$}
&
(q_a , \wt{q}_a) 
\ &: \ \ 
({\bf 1}, {\bf 1}, {\bf 2})
\\
&
(\psi_{\pm,q,a}, \wt{\psi}_{\pm,q,a}) 
\ &: \ \ 
({\bf 1}, {\bf 2}, {\bf 1})_- \oplus ({\bf 2}, {\bf 1}, {\bf 1})_+
\\
\\
\multirow{3}{*}{$\text{neutral hypermultiplet $(\Psi, \Theta)$} \ \ \left\{ 
\begin{array}{c}
\vphantom{(r^1, r^2, r^3)}
\cr
\vphantom{\vartheta}
\cr
\vphantom{(\chi_{\pm}, \wt{\chi}_{\pm})}
\end{array}
\right.$}
&
(r^1, r^2, r^3)
\ &: \ \ 
({\bf 1}, {\bf 1}, {\bf 3})
\\
&
\vartheta \
\ &: \ \ 
({\bf 1}, {\bf 1}, {\bf 1})
\\
&
(\chi_{\pm}, \wt{\chi}_{\pm})
\ &: \ \  
({\bf 1}, {\bf 2}, {\bf 2})_- \oplus ({\bf 2}, {\bf 1}, {\bf 2})_+
\\
\\
\multirow{2}{*}{$\text{Fayet-Iliopoulos parameters $(s_a, t_a)$} \ \ \left\{ 
\begin{array}{c}
\vphantom{(s^1_a, s^2_a, t^3_a)}
\cr
\vphantom{t^4_a}
\end{array}
\right.$}
&
(s^1_a, s^2_a, t^3_a)
\ &: \ \ 
({\bf 1}, {\bf 1}, {\bf 3})
\\
&
t^4_a \
\ &: \ \ 
({\bf 1}, {\bf 1}, {\bf 1})
\end{array}
\label{3-SU2}
\end{align}
Here, the subscripts $\pm$ imply the chirality of the fermionic fields.
Unfortunately, these $SU(2)$ R-symmetries are not so explicitly expressed in the above formulation because we describe the system in terms of the $\N=(2,2)$ superfields, rather than $\N=(4,4)$ superfields.
In the GLSM system by Tong, and in our model which will be introduced in next subsection, the third $SU(2)_R$ symmetry is the most important in the following viewpoint:
We choose $\vartheta$ as a special field coupled to the gauge field strength $F_{01,a}$ as mentioned in section \ref{S:introduction}.
The remaining three fields $(r^1,r^2,r^3)$ are organized into an $SU(2)$ triplet, 
which defines the $SU(2)_R$ symmetry.
If we choose another field as the special direction coupled to the gauge field,
the other three fields are subject to the $SU(2)_R$ symmetry.

\subsection{Semi-doubled GLSM for five-branes of codimension two}

Now, we are ready to propose a new model which will genuinely describes the configurations of five-branes of codimension {\it two} (or defect five-branes, for short), which means two of the four transverse directions of the five-branes are compactified and smeared.
In the previous discussion we mentioned that the GLSM by Tong describes the codimension three configuration. The one smeared direction is given by an imaginary part $\vartheta$ of the twisted chiral $\Theta$, which is topologically coupled to $U(1)^k$ gauge field strengths $\sum_a F_{01,a}$.

Here we deform the configuration along the real part $r^3$ of $\Theta$, which is coupled to another set of $U(1)$ gauge field strengths $\sum_{a'} \wh{F}_{01,a'}$, as shown in (\ref{2-top}). They are involved in $U(1)^{\ell}$ gauge multiplets $(\wh{\Sigma}_{a'}, \wh{\Phi}_{a'})$.
Simultaneously, we also introduce another set of charged hypermultiplets $(P_{a'}, \wt{P}_{a'})$ and FI parameters 
$\wh{s}_{a'} = \frac{1}{\sqrt{2}} (\wh{s}^1_{a'} + \I \wh{s}^2_{a'})$,
$\wh{t}_{a'} = \frac{1}{\sqrt{2}} (\wh{t}^3_{a'} + \I \wh{t}^4_{a'})$
in the following form\footnote{The component fields of the new gauge multiplets are assigned as
$\wh{\Sigma}_{a'}(\wh{\sigma}_{a'}, \ol{\wh{\lambda}}{}_{+,a'}, \wh{\lambda}_{-,a'}, \wh{A}_{m,a'}, \wh{\sf D}_{V,a'})$ and
$\wh{\Phi}_{a'}(\wh{\phi}_{a'}, \wh{\wt{\lambda}}_{\pm,a'}, \wh{\sf F}_{\Phi,a'})$.
In the same way, the component fields of the new charged hypermultiplets are
assigned as $P_{a'}(p_{a'}, \psi_{\pm,p,a'}, {\sf F}_{p,a'})$ and 
$\wt{P}_{a'}(\wt{p}_{a'}, \wt{\psi}_{\pm,p,a'}, \wt{\sf F}_{p,a'})$.}:
\bsubeq \label{SDL}
\begin{align}
\Scr{L}
\ &= \ 
\Scr{L}^{\text{gauge}}
+ \Scr{L}^{\text{CHM}}
+ \Scr{L}^{\text{NHM}}
\, , \\
\Scr{L}^{\text{gauge}}
\ &= \ 
\int \d^4 \theta \, \sum_{a = 1}^k \frac{1}{e_a^2} \Big\{ - |\Sigma_a|^2 + |\Phi_a|^2 \Big\}
+ \int \d^4 \theta \, \sum_{a' = 1}^{\l} \frac{1}{\wh{e}_{a'}^2} \Big\{ - |\wh{\Sigma}_{a'}|^2 + |\wh{\Phi}_{a'}|^2 \Big\}
\, , \label{SDL-gauge} \\
\Scr{L}^{\text{CHM}}
\ &= \ 
\int \d^4 \theta \, \sum_a \Big\{ |Q_a|^2 \, \e^{+ 2 V_a} + |\wt{Q}_a|^2 \, \e^{- 2 V_a} \Big\}
- \Big\{ \sqrt{2} \int \d^2 \theta \, \sum_a \wt{Q}_a \Phi_a Q_a + \text{(h.c.)} \Big\}
\nn \\
\ & \ \ \ \ 
+ \int \d^4 \theta \, \sum_{a'} \Big\{ |P_{a'}|^2 \, \e^{+ 2 \wh{V}_{a'}} + |\wt{P}_{a'}|^2 \, \e^{- 2 \wh{V}_{a'}} \Big\}
- \Big\{ \sqrt{2} \int \d^2 \theta \, \sum_{a'} \wt{P}_{a'} \wh{\Phi}_{a'} P_{a'} + \text{(h.c.)} \Big\}
\, , \label{SDL-CHM} \\
\Scr{L}^{\text{NHM}}
\ &= \ 
\frac{1}{g^2} \int \d^4 \theta \, \Big\{ - |\Theta|^2 + |\Psi|^2 \Big\}
\nn \\
\ & \ \ \ \ 
+ \sum_a \Big\{ 
\sqrt{2} \int \d^2 \wt{\theta} \, (t_a - \Theta) \Sigma_a 
+ \sqrt{2} \int \d^2 \theta \, (s_a - \Psi) \Phi_a 
+ \text{(h.c.)}
\Big\}
\nn \\
\ & \ \ \ \ 
+ \sum_{a'} \Big\{
\I \sqrt{2} \int \d^2 \wt{\theta} \, (\wh{t}_{a'} - \Theta) \wh{\Sigma}_{a'} 
+ \I \sqrt{2} \int \d^2 \theta \, (\wh{s}_{a'} - \Psi) \wh{\Phi}_{a'} 
+ \text{(h.c.)}
\Big\}
\, . \label{SDL-NHM}
\end{align}
In order to obtain various five-brane configurations under duality transformations, we promote the sector $\Scr{L}^{\text{NHM}}$ (\ref{SDL-NHM}) containing twisted chiral superfield $\Theta$ to that of the complex twisted linear superfield $\wt{L}$, as discussed in section \ref{S:twistedlinear}:
\begin{align}
\Scr{L}^{\text{NHM}}
\ &= \
\int \d^4 \theta \, \Big\{ 
g^2 \Big| \wt{L} + 2 (\sum_a V_a + \I \sum_{a'} \wh{V}_{a'}) \Big|^2
+ \frac{1}{g^2} |\Psi|^2 
\Big\}
\nn \\
\ & \ \ \ \ 
+ \sum_a \sqrt{2} \, \eps^{mn} \del_m (\vartheta A_{n,a})
+ \sum_a \sqrt{2} \, \eps^{mn} \del_m (r^3 \wh{A}_{n,a'})
\nn \\
\ & \ \ \ \ 
+ \sum_a \Big\{ 
\sqrt{2} \int \d^2 \wt{\theta} \, t_a \, \Sigma_a 
+ \sqrt{2} \int \d^2 \theta \, (s_a - \Psi) \Phi_a 
+ \text{(h.c.)}
\Big\}
\nn \\
\ & \ \ \ \ 
+ \sum_{a'} \Big\{
\I \sqrt{2} \int \d^2 \wt{\theta} \, \wh{t}_{a'} \, \wh{\Sigma}_{a'} 
+ \I \sqrt{2} \int \d^2 \theta \, (\wh{s}_{a'} - \Psi) \wh{\Phi}_{a'} 
+ \text{(h.c.)}
\Big\} 
\, . \label{SDL-NHM_2}
\end{align}
\esubeq

Since we describe the two smeared directions as on equal footing, 
this system is symmetric under the exchange 
$(\Sigma_a, \Phi_a; Q_a, \wt{Q}_a; s_a, t_a)$ 
for
$(\wh{\Sigma}_{a'}, \wh{\Phi}_{a'}; P_{a'}, \wt{P}_{a'}; \wh{s}_{a'}, \wh{t}_{a'})$.
Due to the coexistence of these two, 
this system ``approximately'' possesses not only the previous $SO(4) \times SU(2)_R$ symmetries in (\ref{3-SU2}), but also another type of $SU(2)$ R-symmetries $SO(4) \times \wh{SU(2)}_R \simeq SU(2)_1 \times SU(2)_2 \times \wh{SU(2)}_R$ assigned in (\ref{wh3-SU2}), where the $SO(4) \simeq SU(2)_1 \times SU(2)_2$ part is common:
\begin{align}
\begin{array}{r@{\hspace{-1mm}}rl}
\multirow{2}{*}{$\text{gauge multiplet $(\wh{V}_{a'}, \wh{\Phi}_{a'})$} \ \ \left\{ 
\begin{array}{c}
\vphantom{(\wh{\sigma}_{a'}, \wh{\phi}_{a'})}
\cr
\vphantom{\wh{A}_{m,a'}}
\cr
\vphantom{(\wh{\lambda}_{\pm,a'}, \wt{\wh{\lambda}}_{\pm,a'})}
\end{array}
\right.$}
&
(\wh{\sigma}_{a'}, \wh{\phi}_{a'}) 
\ &: \ \ 
({\bf 2}, {\bf 2}, {\bf 1})
\\
&
\wh{A}_{m,a'}
\ &: \ \ 
({\bf 1}, {\bf 1}, {\bf 1})
\\
&
(\wh{\lambda}_{\pm,a'}, \wt{\wh{\lambda}}_{\pm,a'}) 
\ &: \ \ 
({\bf 2}, {\bf 1}, {\bf 2})_- \oplus ({\bf 1}, {\bf 2}, {\bf 2})_+
\\
\\
\multirow{2}{*}{$\text{charged hypermultiplet $(P_{a'}, \wt{P}_{a'})$} \ \ \left\{ 
\begin{array}{c}
\vphantom{(p_{a'} , \wt{p}_{a'})}
\cr
\vphantom{(\psi_{\pm,p,a'}, \wt{\psi}_{\pm,p,a'})}
\end{array}
\right.$}
&
(p_{a'} , \wt{p}_{a'}) 
\ &: \ \ 
({\bf 1}, {\bf 1}, {\bf 2})
\\
&
(\psi_{\pm,p,a'}, \wt{\psi}_{\pm,p,a'}) 
\ &: \ \ 
({\bf 1}, {\bf 2}, {\bf 1})_- \oplus ({\bf 2}, {\bf 1}, {\bf 1})_+
\\
\\
\multirow{3}{*}{$\text{neutral hypermultiplet $(\Psi, \Theta)$} \ \ \left\{ 
\begin{array}{c}
\vphantom{(r^1, r^2, \vartheta)}
\cr
\vphantom{\vartheta}
\cr
\vphantom{(\chi_{\pm}, \wt{\chi}_{\pm})}
\end{array}
\right.$}
&
(r^1, r^2, \vartheta)
\ &: \ \ 
({\bf 1}, {\bf 1}, {\bf 3})
\\
&
r^3 \
\ &: \ \ 
({\bf 1}, {\bf 1}, {\bf 1})
\\
&
(\chi_{\pm}, \wt{\chi}_{\pm})
\ &: \ \  
({\bf 1}, {\bf 2}, {\bf 2})_- \oplus ({\bf 2}, {\bf 1}, {\bf 2})_+
\\
\\
\multirow{2}{*}{$\text{Fayet-Iliopoulos parameters $(\wh{s}_{a'}, \wh{t}_{a'})$} \ \ \left\{ 
\begin{array}{c}
\vphantom{(\wh{s}^1_{a'}, \wh{s}^2_{a'}, t^4_{a'})}
\cr
\vphantom{\wh{t}^3_{a'}}
\end{array}
\right.$}
&
(\wh{s}^1_{a'}, \wh{s}^2_{a'}, \wh{t}^4_{a'})
\ &: \ \ 
({\bf 1}, {\bf 1}, {\bf 3})
\\
&
\wh{t}^3_{a'} \
\ &: \ \ 
({\bf 1}, {\bf 1}, {\bf 1})
\end{array}
\label{wh3-SU2}
\end{align}
Strictly speaking, (\ref{3-SU2}) and (\ref{wh3-SU2}) are not satisfied simultaneously.
This means that the $SU(2)$ rotational symmetry among $(r^1,r^2,r^3)$ conflicts with the $\wh{SU(2)}$ rotational symmetry among $(r^1,r^2,\vartheta)$.
Then the original $SU(2)_R$ and the additional $\wh{SU(2)}_R$ symmetries are broken. 
$\N=(4,4)$ supersymmetry is also broken to $\N=(2,2)$.
However, if the $U(1)^{\ell}$ gauge coupling constants $\wh{e}_{a'}$ are zero, the gauge multiplets $(\wh{\Sigma}_{a'}, \wh{\Phi}_{a'})$ disappear and the charged hypermultiplets $(P_{a'}, \wt{P}_{a'})$ are decoupled from the system.
In this reduction the system goes back to the system by Tong and the $SU(2)_R$ symmetry is restored.
Analogously, in the vanishing $e_a$ case, $(\Sigma_a, \Phi_a)$ and $(Q_a, \wt{Q}_a)$ are decoupled from the system, and the $\wh{SU(2)}_R$ symmetry is restored.
Because of this phenomenon, we interpret that the system (\ref{SDL}) ``approximately'' possesses both (\ref{3-SU2}) and (\ref{wh3-SU2}).

In later discussions, we will explore supersymmetric vacua and construct a low energy effective theory, and we will obtain two one-form fields living in the target space, named $\Omega_a = \Omega_{i,a} \d r^i$ and $\wh{\Omega}_{a'} = \wh{\Omega}_{j',a'} \d r^{j'}$.
The former originates from the integration of the charged hypermultiplets $(Q_a, \wt{Q}_a)$, the latter from $(P_{a'}, \wt{P}_{a'})$.
In the original Lagrangian (\ref{SDL}), each of $\Omega_a$ and $\wh{\Omega}_{a'}$ will be polarized along a certain direction (see appendix \ref{A:Omega}), which is not appropriate to analyze the background configuration. 
Then, for later discussions, we perform the following $SU(2)_R \times \wh{SU(2)}_R$ rotation to change the system (\ref{SDL}) to a suitable configuration:
\bsubeq \label{SU2-rotation}
\begin{alignat}{2}
SU(2)_R:& \ \ \left(
{\renewcommand{\arraystretch}{1}
\begin{array}{r}
r^1 - s^1_a
\\
r^2 - s^2_a
\\
r^3 - t^3_a
\end{array}
}
\right)
\ &\to \ 
\left(
{\renewcommand{\arraystretch}{1}
\begin{array}{r}
r^1 - s^1_a
\\
- r^3 + t^3_a
\\
r^2 - s^2_a
\end{array}
}
\right)
\ = \
\left(
{\renewcommand{\arraystretch}{1}
\begin{array}{ccc}
1 & & \\
& & -1 \\
& 1 & 
\end{array}
}
\right)
\left(
{\renewcommand{\arraystretch}{1}
\begin{array}{r}
r^1 - s^1_a
\\
r^2 - s^2_a
\\
r^3 - t^3_a
\end{array}
}
\right)
\, , \label{SU2-rot} \\
\wh{SU(2)}_R:& \ \ 
\left(
{\renewcommand{\arraystretch}{1}
\begin{array}{r}
r^1 - \wh{s}^1_{a'}
\\
r^2 - \wh{s}^2_{a'}
\\
\vartheta - \wh{t}^4_{a'}
\end{array}
}
\right)
\ &\to \ 
\left(
{\renewcommand{\arraystretch}{1}
\begin{array}{r}
r^1 - \wh{s}^1_{a'}
\\
\vartheta - \wh{t}^4_{a'}
\\
- r^2 + \wh{s}^2_{a'}
\end{array}
}
\right)
\ = \
\left(
{\renewcommand{\arraystretch}{1}
\begin{array}{ccc}
1 & & \\
  & & 1 \\
& -1 & 
\end{array}
}
\right)
\left(
{\renewcommand{\arraystretch}{1}
\begin{array}{r}
r^1 - \wh{s}^1_{a'}
\\
r^2 - \wh{s}^2_{a'}
\\
\vartheta - \wh{t}^4_{a'}
\end{array}
}
\right)
\, . \label{whSU2-rot}
\end{alignat}
\esubeq
Under this transformation, all the antisymmetric terms derived from (\ref{SDL-NHM_2}) are invariant. 
Each kinetic term $(\del_m r^i)^2$ is rotated, but the whole is invariant.
We then obtain the following form:
\begin{align}
\Scr{L}
\ &= \ 
\sum_{a=1}^k \frac{1}{2 e_a^2} (F_{01,a})^2 
+ \sum_{a'=1}^{\l} \frac{1}{2 \wh{e}_{a'}^2} (\wh{F}_{01,a'})^2 
\nn \\
\ & \ \ \ \ 
- \sum_a \frac{1}{e_a^2} \Big\{ |\del_m \phi_a|^2 + |\del_m \sigma_a|^2 \Big\}
- \sum_{a'} \frac{1}{\wh{e}_{a'}^2} \Big\{ |\del_m \wh{\phi}_{a'}|^2 + |\del_m \wh{\sigma}_{a'}|^2 \Big\}
\nn \\
\ & \ \ \ \ 
- \sum_a 2 \big( |q_a|^2 + |\wt{q}_a|^2 \big) \big( |\phi_a|^2 + |\sigma_a|^2 \big) 
- \sum_{a'} 2 \big( |p_{a'}|^2 + |\wt{p}_{a'}|^2 \big) \big( |\wh{\phi}_{a'}|^2 + |\wh{\sigma}_{a'}|^2 \big)
\nn \\
\ & \ \ \ \ 
- 2 g^2 \Big| \sum_a \phi_a + \I \sum_{a'} \wh{\phi}_{a'} \Big|^2
- 2 g^2 \Big| \sum_a \sigma_a + \I \sum_{a'} \wh{\sigma}_{a'} \Big|^2
\nn \\
\ & \ \ \ \ 
- \sum_a \Big\{ |D_m q_a|^2 + |D_m \wt{q}_a|^2 \Big\}
- \sum_{a'} \Big\{ |D_m p_{a'}|^2 + |D_m \wt{p}_{a'}|^2 \Big\}
\nn \\
\ & \ \ \ \ 
- \frac{1}{2 g^2} \Big\{
(\del_m r^1)^2
+ (\del_m r^2)^2
+ (\del_m r^3)^2 
+ (\del_m \vartheta)^2 
\Big\}
+ \eps^{mn} (\del_m r^3) (D_n \varphi_{\text{R}-})
+ \eps^{mn} (\del_m \vartheta) (D_n \varphi_{\text{I}+})
\nn \\
\ & \ \ \ \ 
+ \sum_a \sqrt{2} \, \eps^{mn} \del_m \big( (\vartheta - t^4_a) A_{n,a} \big)
+ \sum_{a'} \sqrt{2} \, \eps^{mn} \del_m \big( (r^3 - \wh{t}^3_{a'}) \wh{A}_{n,a'} \big) 
\nn \\
\ & \ \ \ \ 
- \sum_a \frac{e_a^2}{2} \Big\{ (|q_a|^2 - |\wt{q}_a|^2) - \sqrt{2} \, (r^2 - s^2_a) \Big\}^2
- \sum_a e_a^2 \Big| \sqrt{2} \, q_a \wt{q}_a + \big( (r^1 - s^1_a) - \I (r^3 - t^3_a) \big) \Big|^2
\nn \\
\ & \ \ \ \ 
- \sum_{a'} \frac{\wh{e}_{a'}^2}{2} \Big\{ (|p_{a'}|^2 - |\wt{p}_{a'}|^2) - \sqrt{2} \, (r^2 - \wh{s}^2_{a'}) \Big\}^2
- \sum_{a'} \wh{e}_{a'}^2 \Big| \sqrt{2} \, p_{a'} \wt{p}_{a'} + \I \big( (r^1 - \wh{s}^1_{a'}) + \I (\vartheta - \wh{t}^4_{a'}) \big) \Big|^2
\nn \\
\ & \ \ \ \ 
+ \text{(fermionic terms)}
\, . \label{SD-GLSM}
\end{align}
We emphasize that the scalar potentials are different from those in (\ref{SDL}), because we changed the system by using the $SU(2)_R \times \wh{SU(2)}_R$ rotation.
However, this is merely a technical issue. 
As mentioned in next section, this does not change physics.
This rotation is convenient to analyze the effective theory.
From now on, we study the new system (\ref{SD-GLSM}) which approximately possesses (\ref{3-SU2}) and (\ref{wh3-SU2}).

Before going to next sections, we have comments on the description of the system (\ref{SD-GLSM}).
In the twisted linear sector (\ref{SDL-NHM_2}), 
we first replaced $\wt{L}$ with $X + \ol{W} + Y$ as discussed in (\ref{L-XYWV}), and integrated out the unphysical fields $\sigma_{Y,\text{R}}$ and $\sigma_{Y,\text{I}}$.
The charged scalar fields $(q_a, \wt{q}_a)$ and $(p_{a'}, \wt{p}_{a'})$, and the dual scalar fields $(\varphi_{\text{R}-}, \varphi_{\text{I}+})$ are governed by the following gauge covariant derivatives: 
\bsubeq
\begin{alignat}{2}
D_m q_a 
\ &= \ 
\del_m q_a - \I A_{m,a} \, q_a
\, , &\ls
D_m \wt{q}_a
\ &= \ 
\del_m \wt{q}_a + \I A_{m,a} \, \wt{q}_a
\, , \\
D_m p_{a'} 
\ &= \ 
\del_m p_{a'} - \I \wh{A}_{m,a'}
\, , &\ls
D_m \wt{p}_{a'}
\ &= \ 
\del_m \wt{p}_{a'} + \I \wh{A}_{m,a'}
\, , \\
D_m \varphi_{\text{R}-}
\ &= \ 
\del_m \varphi_{\text{R}-} - \sqrt{2} \sum_{a'} \wh{A}_{m,a'}
\, , &\ls
D_m \varphi_{\text{I}+}
\ &= \ 
\del_m \varphi_{\text{I}+} - \sqrt{2} \sum_a A_{m,a}
\, . 
\end{alignat}
\esubeq

We mention our direction in later sections.
In section \ref{S:SDNLSM}, we will first take the IR limit of (\ref{SD-GLSM}) and analyze the effective theory, called the semi-doubled NLSM.
Second, we will integrate out some scalar fields in $\wt{L} = X + \ol{W} + Y$ to obtain three standard NLSMs. 
Each target space represents the background of a defect five-brane in \cite{Tong:2002rq, Harvey:2005ab, Okuyama:2005gx, Kimura:2013fda}.
On the other hand, in section \ref{S:standard}, we first integrate out the scalar fields to reduce the semi-doubled GLSM (\ref{SD-GLSM}) to three standard GLSMs given in \cite{Tong:2002rq, Harvey:2005ab, Okuyama:2005gx, Kimura:2013fda}.
Next, we perform their IR limit and obtain the correct NLSMs.
Summarizing these two sections, we will find that the IR limit and the reduction are commutative to analyze the configurations of defect five-branes.
Based on this, section \ref{S:instantons} is devoted to investigation of non-perturbative corrections of (\ref{SD-GLSM}).
The corrections will correspond to stringy corrections to the above configurations. Some of them have already discussed in \cite{Tong:2002rq, Harvey:2005ab, Okuyama:2005gx, Kimura:2013zva}, and the other is discussed in \cite{Kimura:2018hph}.

\section{IR effective theory of semi-doubled GLSM}
\label{S:SDNLSM}

In this section, we investigate the supersymmetric vacuum of the semi-doubled GLSM (\ref{SD-GLSM}), and construct its IR effective theory.
In particular, we first obtain the IR theory as a semi-doubled form in which the original scalar fields and their dual fields in $\wt{L} = X + \ol{W} + Y$ are involved.
In the analysis, we perform the smearing procedure that generates two isometry directions of the target space.
After that procedure, we integrate out the original or dual fields to reduce the semi-doubled theory to a standard NLSM whose target space describes the correct background configuration of a defect five-brane.

\subsection{Supersymmetry vacuum}

First of all, we consider supersymmetric vacuum of the semi-doubled GLSM (\ref{SD-GLSM}).
The vanishing scalar potentials give a set of algebraic equations:
\bsubeq \label{Higgs-vacuum-SDL}
\begin{alignat}{2}
0 \ &= \ 
2 (|\sigma_a|^2 + |\phi_a|^2) (|q_a|^2 + |\wt{q}_a|^2)
\, , &\ls
0 \ &= \ 
2 (|\wh{\sigma}_{a'}|^2 + |\wh{\phi}_{a'}|^2) (|p_{a'}|^2 + |\wt{p}_{a'}|^2)
\, , \\
0 \ &= \ 
\sum_a \sigma_a + \I \sum_{a'} \wh{\sigma}_{a'}
\, , &\ls
0 \ &= \ 
\sum_a \phi_a + \I \sum_{a'} \wh{\phi}_{a'}
\, , \\
0 \ &= \ 
(|q_a|^2 - |\wt{q}_a|^2) - \sqrt{2} (r^2 - s^2_a)
\, , &\ls
0 \ &= \ 
\sqrt{2} \, q_a \wt{q}_a + \big( (r^1 - s^1_a) - \I (r^3 - t^3_a) \big)
\, , \label{qq-SDL} \\
0 \ &= \ 
(|p_{a'}|^2 - |\wt{p}_{a'}|^2) - \sqrt{2} (r^2 - \wh{s}^2_{a'})
\, , &\ls
0 \ &= \ 
\sqrt{2} \, p_{a'} \wt{p}_{a'} - \big( (\vartheta - \wh{t}^4_{a'}) - \I (r^1 - \wh{s}^1_{a'}) \big)
\, . \label{pp-SDL}
\end{alignat}
\esubeq
We can immediately find that this gives the Higgs phase assigned by the vanishing scalar fields of the gauge multiplets: $\sigma_a = 0 = \phi_a$, $\wh{\sigma}_{a'} = 0 = \wh{\phi}_{a'}$.
Furthermore, as demonstrated in \cite{Tong:2002rq}, the vacuum configuration of the charged scalar fields are given as
\bsubeq \label{qqRppR-SDL}
\begin{align}
(|q_a|^2 + |\wt{q}_a|^2)^2
\ &= \
(|q_a|^2 - |\wt{q}_a|^2)^2 + 4 |q_a \wt{q}_a|^2
\ = \ 
2 \Big\{
(r^1 - s^1_a)^2 + (r^2 - s^2_a)^2 + (r^3 - t^3_a)^2
\Big\}
\nn \\
\ &=: \
2 (R_a)^2
\, , \label{qqR-SDL} \\
(|p_{a'}|^2 + |\wt{p}_{a'}|^2)^2
\ &= \
(|p_{a'}|^2 - |\wt{p}_{a'}|^2)^2 + 4 |p_{a'} \wt{p}_{a'}|^2
\ = \ 
2 \Big\{
(r^1 - \wh{s}^1_{a'})^2 + (r^2 - \wh{s}^2_{a'})^2 + (\vartheta - \wh{t}^4_{a'})^2
\Big\}
\nn \\
\ &=: \ 
2 (\wh{R}_{a'})^2
\, . \label{ppR-SDL}
\end{align}
\esubeq
By using these two values $R_a$ and $\wh{R}_{a'}$, we can solve the charge scalar fields in such a way that
\bsubeq \label{qqpp-SU2}
\begin{alignat}{2}
q_a 
\ &= \ 
\frac{\I}{2^{1/4}} \, \e^{+ \I \alpha_a} \sqrt{R_a + (r^2 - s^2_a)}
\, , &\ls
\wt{q}_a
\ &= \ 
\frac{\I}{2^{1/4}} \, \e^{- \I \alpha_a} \frac{(r^1 - s^1_a) - \I (r^3 - t^3_a)}{\sqrt{R_a + (r^2 - s^2_a)}}
\, , \\
p_{a'} 
\ &= \ 
\frac{\I}{2^{1/4}} \, \e^{+ \I \wh{\alpha}_{a'}} \sqrt{\wh{R}_{a'} + (r^2 - \wh{s}^2_{a'})}
\, , &\ls
\wt{p}_{a'}
\ &= \ 
\frac{\I}{2^{1/4}} \, \e^{- \I \wh{\alpha}_{a'}} \frac{-(\vartheta - \wh{t}^4_{a'}) + \I (r^1 - \wh{s}^1_{a'})}{\sqrt{\wh{R}_{a'} + (r^2 - \wh{s}^2_{a'})}}
\, ,
\end{alignat}
\esubeq
where $\alpha_a$ and $\wh{\alpha}_{a'}$ are arbitrary phases interpreted as the gauge parameters of $U(1)^k \times U(1)^{\ell}$ gauge symmetries.
Since the above configuration constrains the fluctuation modes of the charged scalar fields, we can rewrite their kinetic terms in terms of scalar fields of the neutral hypermultiplet:
\bsubeq \label{kin-qqpp}
\begin{align}
- \sum_a \Big\{
|D_m q_a|^2 + |D_m \wt{q}_a|^2
\Big\}
\ &= \ 
- \sum_a \frac{1}{2 \sqrt{2} R_a} \Big\{ (\del_m r^1)^2 + (\del_m r^2)^2 + (\del_m r^3)^2 \Big\}
\nn \\
\ & \ \ \ \ 
- \sum_a \sqrt{2} R_a \Big\{ 
\del_m \alpha_a 
- A_{m,a} 
+ \frac{1}{\sqrt{2}} \Omega_{i,a} (\del_m r^i) 
\Big\}^2
\, , \\
- \sum_{a'} \Big\{
|D_m p_{a'}|^2 + |D_m \wt{p}_{a'}|^2
\Big\}
\ &= \ 
- \sum_{a'} \frac{1}{2 \sqrt{2} \wh{R}_{a'}} \Big\{ (\del_m r^1)^2 + (\del_m r^2)^2 + (\del_m \vartheta)^2 \Big\}
\nn \\
\ & \ \ \ \ 
- \sum_{a'} \sqrt{2} \wh{R}_{a'} \Big\{
\del_m \wh{\alpha}_{a'}
- \wh{A}_{m,a'} 
+ \frac{1}{\sqrt{2}} \wh{\Omega}_{j',a'} (\del_m r^{j'})
\Big\}^2
\, ,
\end{align}
\esubeq
where we have introduced the following variables:
\bsubeq \label{Omega}
\begin{gather}
\begin{align}
\Omega_{i,a} (\del_m r^i)
\ &:= \ 
\frac{(r^1 - s^1_a) \del_m r^3 - (r^3 - t^3_a) \del_m r^1}{\sqrt{2} R_a (R_a + (r^2 - s^2_a))}
\, , \\
\wh{\Omega}_{j',a'} (\del_m r^{j'})
\ &:= \ 
\frac{- (r^1 - \wh{s}^1_{a'}) \del_m \vartheta + (\vartheta - \wh{t}^4_{a'}) \del_m r^1}{\sqrt{2} \wh{R}_{a'} (\wh{R}_{a'} + (r^2 - \wh{s}^2_{a'}))}
\, ,
\end{align}
\\
\begin{alignat}{3}
\Omega_{1,a} 
\ &= \ 
- \frac{r^3 - t^3_a}{\sqrt{2} R_a (R_a + (r^2 - s^2_a))}
\, , &\ls
\Omega_{2,a} 
\ &= \ 0
\, , &\ls
\Omega_{3,a} 
\ &= \ 
+ \frac{r^1 - s^1_a}{\sqrt{2} R_a (R_a + (r^2 - s^2_a))}
\, , \\
\wh{\Omega}_{1,a'} 
\ &= \ 
+ \frac{\vartheta - \wh{t}^4_{a'}}{\sqrt{2} \wh{R}_{a'} (\wh{R}_{a'} + (r^2 - \wh{s}^2_{a'}))}
\, , &\ls
\wh{\Omega}_{2,a'} 
\ &= \ 0 
\, , &\ls
\wh{\Omega}_{4,a'} 
\ &= \ 
- \frac{r^1 - \wh{s}^1_{a'}}{\sqrt{2} \wh{R}_{a'} (\wh{R}_{a'} + (r^2 - \wh{s}^2_{a'}))}
\, . 
\end{alignat}
\end{gather}
\esubeq
We can interpret that $\Omega_{a} = \Omega_{i,a} \d r^j$ of each index $a$ is a one-form in the space $(r^1,r^2,r^3)$, while
$\wh{\Omega}_{a'} = \wh{\Omega}_{j',a'} \d r^{j'}$ of each $a'$ is another one-form in the space $(r^1, r^2, \vartheta)$.
They play a significant role in the IR effective theory.
In particular, their explicit expressions (\ref{Omega}) are very important when we discuss the smearing procedure as in \cite{Kimura:2013fda}.

Substituting the supersymmetric vacuum configuration (\ref{Higgs-vacuum-SDL}) and the forms of the charged fields (\ref{kin-qqpp}) into the Lagrangian (\ref{SD-GLSM}), we obtain the effective gauge theory 
\begin{align}
\Scr{L}
\ &= \ 
\sum_a \frac{1}{2 e_a^2} (F_{01,a})^2 
+ \sum_{a'} \frac{1}{2 \wh{e}_{a'}^2} (\wh{F}_{01,a'})^2 
\nn \\
\ & \ \ \ \ 
- \frac{1}{2} H_0 \Big\{ (\del_m r^1)^2 + (\del_m r^2)^2 \Big\}
- \frac{1}{2} H (\del_m r^3)^2 
- \frac{1}{2} \wh{H} (\del_m \vartheta)^2 
\nn \\
\ & \ \ \ \ 
+ \eps^{mn} (\del_m r^3) (D_n \varphi_{\text{R}-})
+ \eps^{mn} (\del_m \vartheta) (D_n \varphi_{\text{I}+})
\nn \\
\ & \ \ \ \ 
- \sum_a \sqrt{2} R_a \Big\{
\del_m \alpha_a 
- A_{m,a} 
+ \frac{1}{\sqrt{2}} \Omega_{i,a} (\del_m r^i) 
\Big\}^2
\nn \\
\ & \ \ \ \ 
- \sum_{a'} \sqrt{2} \wh{R}_{a'} \Big\{
\del_m \wh{\alpha}_{a'}
- \wh{A}_{m,a'} 
+ \frac{1}{\sqrt{2}} \wh{\Omega}_{j',a'} (\del_m r^{j'})
\Big\}^2
\nn \\
\ & \ \ \ \ 
+ \sum_a \sqrt{2} \, \eps^{mn} \del_m \big( (\vartheta - t^4_a) A_{n,a} \big)
+ \sum_{a'} \sqrt{2} \, \eps^{mn} \del_m \big( (r^3 - \wh{t}^3_{a'}) \wh{A}_{n,a'} \big) 
\, , \label{SDL-SU2}
\end{align}
where we have omitted the fermionic field contents.
The functions $H_0$, $H$ and $\wh{H}$ are defined as
\bsubeq \label{Harm_H}
\begin{align}
H_0 
\ &:= \
\frac{1}{g^2} 
+ \sum_a \frac{1}{\sqrt{2} R_a}
+ \sum_{a'} \frac{1}{\sqrt{2} \wh{R}_{a'}}
\, , \\
H
\ &:= \
\frac{1}{g^2} 
+ \sum_a \frac{1}{\sqrt{2} R_a}
\, , \ls
\wh{H}
\ := \ 
\frac{1}{g^2} 
+ \sum_{a'} \frac{1}{\sqrt{2} \wh{R}_{a'}}
\, . 
\end{align}
\esubeq
Note that the pairs $(\Omega_{a}, H)$ and $(\wh{\Omega}_{a'}, \wh{H})$ satisfy the monopole equations designed as
\begin{align}
\big( \nabla \times \sum_a \Omega_{a} \big)_i
\ &= \ 
\nabla_i H
\, , \ls
\big( \nabla \times \sum_{a'} \wh{\Omega}_{a'} \big)_{j'}
\ = \ 
\nabla_{j'} \wh{H}
\, . \label{mono-eqs_2}
\end{align}
Here we have comments on the $SU(2)_R \times \wh{SU(2)}_R$ rotation (\ref{SU2-rotation}).
The equations (\ref{qq-SDL}) and (\ref{pp-SDL}) are transformed under the rotation, while the variables $R_a$ and $\wh{R}_{a'}$ in (\ref{qqRppR-SDL}) are invariant. 
Since (\ref{qq-SDL}) and (\ref{pp-SDL}) are transformed,
the solutions (\ref{qqpp-SU2}) are also transformed, 
hence the components of $(\Omega_{i,a}, \wh{\Omega}_{j',a'})$ themselves are also transformed.
On the other hand, the functions $(H, \wh{H})$ are invariant,
and the monopole equations (\ref{mono-eqs_2}) are covariant under the rotation.
Then we interpret that the algebraic equations (\ref{qq-SDL}) and (\ref{pp-SDL}) represent a point on the supersymmetry vacuum manifold given by (\ref{qqRppR-SDL}), and the $SU(2)_R \times \wh{SU(2)}_R$ rotation moves a vacuum to a different one in the same supersymmetry vacuum manifold.

\subsection{IR limit}

We take the IR limit and consider the low energy effective theory.
Since the gauge coupling constants have mass dimension one,
they go to infinity $e_a, \wh{e}_{a'} \to \infty$ in the IR limit.
Their kinetic terms are frozen out, and the gauge fields are no longer dynamical. 
Then we integrate them out from the system.
It is easy to solve the equations of motion:
\bsubeq \label{sol:A-SU2}
\begin{align}
A_{n,a} \ &= \ 
\del_n \alpha_a 
+ \frac{1}{\sqrt{2}} \Omega_{i,a} (\del_n r^i)
- \frac{1}{2 R_a} \eps_{mn} (\del^m \vartheta)
\, , \\
\wh{A}_{n,a}
\ &= \
\del_n \wh{\alpha}_{a'} 
+ \frac{1}{\sqrt{2}} \wh{\Omega}_{j',a'} (\del_n r^{j'})
- \frac{1}{2 \wh{R}_{a'}} \eps_{mn} (\del^m r^3)
\, .  
\end{align}
\esubeq
Substituting them into the Lagrangian (\ref{SDL-SU2}) under the IR limit, 
we obtain the NLSM 
\begin{align}
\Scr{L}^{\text{IR}}
\ &= \ 
- \frac{1}{2} H_0 \Big\{ (\del_m r^1)^2 + (\del_m r^2)^2 + (\del_m r^3)^2 + (\del_m \vartheta)^2 \Big\}
\nn \\
\ & \ \ \ \ 
+ \eps^{mn} (\del_m r^3) \Big\{ \del_n \wt{\varphi}_{\text{R}-} - \wh{\Omega}_{j'} (\del_n r^{j'}) \Big\} 
+ \eps^{mn} (\del_m \vartheta) \Big\{ \del_n \wt{\varphi}_{\text{I}+} - \Omega_i (\del_n r^i) \Big\} 
\nn \\
\ & \ \ \ \ 
+ \sum_a \sqrt{2} \, \eps^{mn} \del_m \big( (\vartheta - t^4_a) \mr{A}_{n,a} \big)
+ \sum_{a'} \sqrt{2} \, \eps^{mn} \del_m \big( (r^3 - \wh{t}^3_{a'}) \mr{\wh{A}}_{n,a'} \big) 
\, , \label{SDL-SU2-2}
\end{align}
where $\mr{A}_{n,a}$ and $\mr{\wh{A}}{}_{n,a'}$ are governed by the equations of motion (\ref{sol:A-SU2}).
As appeared in \cite{Tong:2002rq}, we have used the following expressions: 
\bsubeq \label{sums}
\begin{alignat}{2}
\Omega_i \ &:= \
\sum_a \Omega_{i,a}
\, , &\LS
\wh{\Omega}_{j'} \ &:= \
\sum_{a'} \wh{\Omega}_{j',a'}
\, , \label{sum_Omega} \\
\wt{\varphi}_{\text{I}+}
\ &:= \ 
\varphi_{\text{I}+} - \sqrt{2} \sum_a \alpha_a
\, , &\ls
\wt{\varphi}_{\text{R}-}
\ &:= \
\varphi_{\text{R}-} - \sqrt{2} \sum_{a'} \wh{\alpha}_{a'}
\, . \label{phi-alpha}
\end{alignat}
\esubeq
The target space of this NLSM describes the $k$ five-branes of codimension three living in the three-dimensional space $(r^1, r^2, r^3)$ and $\ell$ five-branes of codimension three in $(r^1,r^2,\vartheta)$.
But this is not a careful expression because there exist antisymmetric terms which represent the dual space of the two directions $(r^3, \vartheta)$.
We also notice that the target space of the NLSM (\ref{SDL-SU2}) has not obtained isometry along the $\vartheta$- and $r^3$-directions,
i.e., the function $H_0$ and the variables $(\Omega_i, \wh{\Omega}_{j'})$ are not invariant under the shifts $\vartheta \to \vartheta + a$ and $r^3 \to r^3 + b$, where $a$ and $b$ are arbitrary.
In order to generate the isometry, we perform the smearing procedure discussed in \cite{deBoer:2010ud, deBoer:2012ma} and realized in the NLSM framework \cite{Kimura:2013fda}.

\subsection{Smearing procedure}

In this subsection, we discuss the smearing procedure which generates the isometry along certain directions in the target space of the NLSM.
This procedure will be utilized in various sections.

First, for simplicity, we set the FI parameters $(s^i_a, \wh{s}^j_{a'})$ to zero.
We also set $(t^a_a, \wh{t}^4_{a'})$ to the following way:
\bsubeq \label{set-SP}
\begin{alignat}{3}
s^1_a \ &= \ 0 \ = \ s^2_a
\, , &\ls
t^3_a \ &= \ 2 \pi {\cal R}_3 \, a
\, , &\ls
a &\in {\mathbb Z}
\, , \\
\wh{s}^1_{a'} \ &= \ 0 \ = \ \wh{s}^2_{a'}
\, , &\ls
\wh{t}^4_{a'} \ &= \ 2 \pi {\cal R}_4 \, a'
\, , &\ls
a' &\in {\mathbb Z}
\, . 
\end{alignat}
\esubeq
From the geometrical viewpoint, the vanishing $(s^i_a, \wh{s}^j_{a'})$
implies that two kinds of the five-branes of codimension three are located at the origin of the two-plane $(r^1,r^2)$,
while $k$ five-branes are arrayed along the $r^3$-directions with period $2 \pi {\cal R}_3$, and $\ell$ different five-branes are arrayed along the $\vartheta$-direction with period $2 \pi {\cal R}_4$, respectively. 
Locally, this is equivalent that the two-directions are compactified on a two-torus with radii $({\cal R}_3, {\cal R}_4)$.

Second, we take the number of the gauge symmetries $k$ and $\ell$ to infinity.
In this limit, various functions which appeared in the previous subsection become divergent. 
We extract several terms which have physical (or geometrical) meanings \cite{deBoer:2010ud, deBoer:2012ma, Kimura:2013fda}: 
\bsubeq \label{smeared-HOmega}
\begin{align}
H_0 \ &\to \ 
\frac{1}{g^2} + \sigma_3 \log \frac{\Lambda_3}{\rho}
+ \sigma_4 \log \frac{\Lambda_4}{\rho}
\, , \\
\Omega_{1} 
\ &\to \ 0
\, , \ls
\Omega_{3} 
\ \to \
- \sigma_3 \Big\{ \arctan \Big( \frac{r^2}{r^1} \Big) - \frac{\pi}{2} \Big\}
\, , \label{smear-Omega}\\
\wh{\Omega}_{1} 
\ &\to \ 0
\, , \ls
\wh{\Omega}_{4} 
\ \to \ 
+ \sigma_4 \Big\{ \arctan \Big( \frac{r^2}{r^1} \Big) - \frac{\pi}{2} \Big\}
\, , \label{smear-whOmega} \\
\sigma_3 \ &:= \ \frac{1}{\sqrt{2} \, \pi {\cal R}_3}
\, , \ls 
\sigma_4 \ := \ \frac{1}{\sqrt{2} \, \pi {\cal R}_4}
\, ,
\end{align}
\esubeq
where $\Lambda_3$ and $\Lambda_4$ are IR cutoffs.
We have introduced a new variable $\rho^2 := (r^1)^2 + (r^2)^2$.
This measures the distance between an arbitrary point in the
two-dimensional plane $(r^1, r^2)$ and the core of the five-brane located at the origin.
The detailed computation to derive the above expressions can be seen in appendix \ref{A:Omega}.
Now, we emphasize the true reason why we introduced the $SU(2)_R \times \wh{SU(2)}_R$ rotation (\ref{SU2-rotation}).
If we did not adapt (\ref{SU2-rotation}) and straightforwardly described the one-forms $\Omega_a$ and $\wh{\Omega}_{a'}$ from (\ref{SDL}),
the smearing limit corresponding to (\ref{smear-Omega}) and (\ref{smear-whOmega}) become divergent and we lost any physical information (again, see appendix \ref{A:Omega}).
This was caused by inappropriate polarizations of these one-forms, even though they satisfy the monopole equations (\ref{mono-eqs_2}).
In order to obtain finite values of $(\Omega, \wh{\Omega})$ after the smearing procedure, we have to modify the polarizations of them.
Indeed, the rotation (\ref{SU2-rotation}) is the answer to keep the physical information after the smearing procedure.

Under the smearing procedure, the Lagrangian (\ref{SDL-SU2-2}) is reduced to
\begin{align}
\Scr{L}^{\text{IR}}
\ &= \
- \frac{1}{2} H_0 \Big\{ (\del_m r^1)^2 + (\del_m r^2)^2 + (\del_m r^3)^2 + (\del_m \vartheta)^2 \Big\}
\nn \\
\ & \ \ \ \ 
+ \eps^{mn} (\del_m r^3) (\del_n \wt{\varphi}_{\text{R}-}) 
+ \eps^{mn} (\del_m \vartheta) (\del_n \wt{\varphi}_{\text{I}+})
+ \eps^{mn} \Omega_0 (\del_m r^3) (\del_n \vartheta) 
\nn \\
\ & \ \ \ \ 
+ \sum_a \sqrt{2} \, \eps^{mn} \del_m \big( (\vartheta - t^4_a) \mr{A}_{n,a} \big)
+ \sum_{a'} \sqrt{2} \, \eps^{mn} \del_m \big( (r^3 - \wh{t}^3_{a'}) \mr{\wh{A}}_{n,a'} \big)
\, . \label{SDL-SU2-3}
\end{align}
This is the semi-doubled NLSM for defect five-branes. 
For convenience, we have introduced a new expression $\Omega_0 := \Omega_3 - \wh{\Omega}_4$.
Since this is still {\it semi-doubled}, we have to further reduce this model to standard ones.
By integrating out the original or dual scalar fields, we will surely obtain the standard NLSMs for defect five-branes.

\subsection{Reduction to standard NLSMs}

In order to regard various NLSMs reduced from the semi-doubled NLSM (\ref{SDL-SU2-3}) as the string worldsheet sigma models,
we introduce the following normalization:
\bsubeq \label{string-NLSM-bosons}
\begin{gather}
S \ = \ 
\frac{1}{2 \pi \alpha'} \int \d^2 x \sqrt{- g} \, \Scr{L}
\, , \\
\Scr{L}
\ = \ 
- \half G_{MN} \, g^{mn} \, \del_m X^M \del_n X^N
+ \half B_{MN} \, \ve^{mn} \, \del_m X^M \del_n X^N
\, . 
\end{gather}
\esubeq
Here $\alpha'$ is the Regge slope.
$G_{MN}$ and $B_{MN}$ are the metric and the NS-NS B-field in the considering background spacetime, while $g_{mn}$ and $\ve_{mn}$ are two-dimensional worldsheet metric and the Levi-Civita antisymmetric tensor, respectively.
In the flat two-dimensional case, the tensor $\ve_{mn}$ is simply given by the antisymmetric symbol $\eps_{mn}$.
$X^M$ are scalar fields in the sigma model. They represent the spacetime coordinates.

Now, we are ready to discuss the configurations of various defect five-branes.
The final task in this section is to derive the standard NLSM via the dualization procedure discussed in section \ref{S:twistedlinear}.

\subsubsection*{Standard NLSM for defect NS5-brane}

Let us first integrate out two scalar fields $\wt{\varphi}_{\text{R}-}$ and $\wt{\varphi}_{\text{I}+}$ in the semi-doubled NLSM (\ref{SDL-SU2-3}).
They are dual fields of the original scalars $r^3$ and $\vartheta$, respectively.
It is easy to find that the equations of motion are trivial:
\begin{alignat}{2}
\wt{\varphi}_{\text{R}-}: \ \ \ 
0 \ &= \ 
\del^n \Big\{
- \eps_{mn} (\del^m r^3)
\Big\}
\, , &\ls
\wt{\varphi}_{\text{I}+}: \ \ \ 
0 \ &= \ 
\del^n \Big\{
- \eps_{mn} (\del^m \vartheta)
\Big\}
\, .
\end{alignat}
Substituting this result into the Lagrangian, we immediately obtain 
\begin{align}
\Scr{L}_{\text{NS5}}^{\text{IR}}
\ &= \ 
- \frac{1}{2} H_0 \Big\{ (\del_m r^1)^2 + (\del_m r^2)^2 + (\del_m r^3)^2 + (\del_m \vartheta)^2 \Big\}
+ \Omega_0 \, \eps^{mn} (\del_m r^3) (\del_n \vartheta) 
\nn \\
\ & \ \ \ \ 
+ \sum_a \sqrt{2} \, \eps^{mn} \del_m \big( (\vartheta - t^4_a) \mr{A}_{n,a} \big)
+ \sum_{a'} \sqrt{2} \, \eps^{mn} \del_m \big( (r^3 - \wh{t}^3_{a'}) \mr{\wh{A}}_{n,a'} \big)
\, . \label{SDL-SU2-dNS5}
\end{align}
Compared with the string sigma model (\ref{string-NLSM-bosons}), we can read off the background configuration 
\begin{align}
G_{MN} \ &= \ H_0 \, \delta_{MN}
\, , \ls
B_{34} \ = \ \Omega_0
\, . \label{NLSM-dNS5}
\end{align}
Hence, we conclude that the NLSM (\ref{SDL-SU2-dNS5}) is nothing but the sigma model for a defect NS5-brane \cite{Kimura:2015yla, Kimura:2018hph}.

\subsubsection*{Standard NLSM for KK-vortex}

Next, we integrate out the dual scalar $\wt{\varphi}_{\text{R}-}$ and the original scalar $\vartheta$.
The equation of motion for $\wt{\varphi}_{\text{R}-}$ is again trivial:
\bsubeq
\begin{align}
0 \ &= \ 
\del^n \Big\{
- \eps_{mn} (\del^m r^3)
\Big\}
\, , 
\end{align}
but that for $\vartheta$ provides a non-trivial equation:
\begin{align}
0 \ &= \ 
\del^m \Big\{
H_0 (\del_m \vartheta)
- \eps_{mn} (\del^n \wt{\varphi}_{\text{I}+})
+ \Omega_0 \, \eps_{mn} (\del^n r^3)
\Big\}
\, , \\
\therefore \ \ \
\del_m \vartheta
\ &= \ 
\frac{1}{H_0} \eps_{mn} \Big\{
\del^n \wt{\varphi}_{\text{I}+}
- \Omega_0 (\del^n r^3) \Big\}
+ \frac{1}{H_0} (\del^n f_{mn}^{\text{sd}}) 
\, . \label{SDL-SU2-KKeq} 
\end{align}
\esubeq
Here $f_{mn}^{\text{sd}}$ is an arbitrary antisymmetric tensor.
Without loss of generality, we can set this to zero.
Substituting the solutions into the Lagrangian (\ref{SDL-SU2-3}), we obtain
\begin{align}
\Scr{L}_{\text{KK}}^{\text{IR}}
\ &= \ 
- \frac{1}{2} H_0 \Big\{ (\del_m r^1)^2 + (\del_m r^2)^2 + (\del_m r^3)^2 \Big\}
- \frac{1}{2 H_0} \Big\{ \del_m \wt{\varphi}_{\text{I}+} - \Omega_0 (\del_m r^3) \Big\}^2
\nn \\
\ & \ \ \ \ 
+ \sum_a \sqrt{2} \, \eps^{mn} \del_m \big( (\vartheta - t^4_a) \mr{A}_{n,a} \big)
+ \sum_{a'} \sqrt{2} \, \eps^{mn} \del_m \big( (r^3 - \wh{t}^3_{a'}) \mr{\wh{A}}_{n,a'} \big)
\, . \label{SDL-SU2-dKK5}
\end{align}
This NLSM gives the background configuration of a single KK-vortex, a smeared KK-monopole \cite{Okada:2014wma}, or called the periodic KK-monopole \cite{Cherkis:2001gm}:
\bsubeq \label{NLSM-dKK}
\begin{gather}
G_{ij} \ = \ H_0 \, \delta_{ij}
\, , \ls 
G_{33} \ = \ \frac{K}{H_0}
\, , \ls
G_{34} \ = \ - \frac{\Omega_0}{H_0}
\, , \ls 
G_{44} \ = \ \frac{1}{H_0}
\, , \ls i,j \ = \ 1,2
\, , \\
B_{MN} \ = \ 0
\, , \\
K \ := \ H_0^2 + \Omega_0^2
\, . \label{def-K}
\end{gather}
\esubeq

\subsubsection*{Standard NLSM for exotic $5^2_2$-brane}

Finally we consider the model obtained by integrating out the original scalars $r^3$ and $\vartheta$.
Their equations of motion are non-trivial and give a new set of equations:
\bsubeq 
\begin{alignat}{2}
r^3: \ \ \ &&
0 \ &= \ 
\del^m \Big\{
H_0 (\del_m r^3)
- \eps_{mn} (\del^n \wt{\varphi}_{\text{R}-})
- \Omega_0 \, \eps_{mn} (\del^n \vartheta)
\Big\}
\, , \\
&&&\to \ \ 
0 \ = \ 
H_0 (\del_m r^3)
- \eps_{mn} (\del^n \wt{\varphi}_{\text{R}-})
- \Omega_0 \, \eps_{mn} (\del^n \vartheta)
+ \del^n {\sf a}_{mn} 
\, , \\
\vartheta: \ \ \ &&
0 \ &= \ 
\del^m \Big\{
H_0 (\del_m \vartheta)
- \eps_{mn} (\del^n \wt{\varphi}_{\text{I}+})
+ \Omega_0 \, \eps_{mn} (\del^n r^3)
\Big\}
\, , \\
&&&\to \ \ 
0 \ = \ 
H_0 (\del_m \vartheta)
- \eps_{mn} (\del^n \wt{\varphi}_{\text{I}+})
+ \Omega_0 \, \eps_{mn} (\del^n r^3)
+ \del^n {\sf b}_{mn} 
\, .
\end{alignat}
\esubeq
Here we have introduced two arbitrary antisymmetric tensors
 ${\sf a}_{mn} = - {\sf a}_{nm}$ and ${\sf b}_{mn} = - {\sf b}_{nm}$.
The equations give the solutions:
\bsubeq \label{onshell-duality-E}
\begin{align}
\del_m r^3
\ &= \ 
+ \frac{1}{K} \Big\{
H_0 \Big( \eps_{mn} \del^n \wt{\varphi}_{\text{R}-} - \del^n {\sf a}_{mn} \Big)
+ \Omega_0 \Big( \del_m \wt{\varphi}_{\text{I}+} - \eps_{mn} \del_p {\sf b}^{np} \Big)
\Big\}
\, , \\
\del_m \vartheta
\ &= \ 
- \frac{1}{K} \Big\{
\Omega_0 \Big(\del_m \wt{\varphi}_{\text{R}-} - \eps_{mn} \del_p {\sf a}^{np} \Big)
- H_0 \Big( \eps_{mn} \del^n \wt{\varphi}_{\text{I}+} - \del^n {\sf b}_{mn} \Big)
\Big\}
\, .
\end{align}
\esubeq
We substitute them into (\ref{SDL-SU2-3}).
In this configuration, we can set ${\sf a}_{mn}$ and ${\sf b}_{mn}$ to zero, without loss of generality.
The result is 
\begin{align}
\Scr{L}_{\text{E}}^{\text{IR}}
\ &= \ 
- \frac{1}{2} H_0 \Big\{ (\del_m r^1)^2 + (\del_m r^2)^2 \Big\}
- \frac{H_0}{2 K} \Big\{ (\del_m \wt{\varphi}_{\text{R}-})^2 + (\del_m \wt{\varphi}_{\text{I}+})^2 \Big\}
- \frac{\Omega_0}{K} \eps_{mn} (\del^m \wt{\varphi}_{\text{R}-}) (\del^n \wt{\varphi}_{\text{I}+}) 
\nn \\
\ & \ \ \ \ 
+ \sum_a \sqrt{2} \, \eps^{mn} \del_m \big( (\vartheta - t^4_a) \mr{A}_{n,a} \big)
+ \sum_{a'} \sqrt{2} \, \eps^{mn} \del_m \big( (r^3 - \wh{t}^3_{a'}) \mr{\wh{A}}_{n,a'} \big)
\, . \label{SDL-SU2-522}
\end{align}
Applying this form to the string sigma model (\ref{string-NLSM-bosons}),
we obtain the configuration of a single exotic $5^2_2$-brane \cite{Kimura:2013fda}:
\bsubeq \label{NLSM-E}
\begin{gather}
G_{ij} \ = \ H_0 \, \delta_{ij}
\, , \ls 
G_{ab} \ = \ \frac{H_0}{K} \, \delta_{ab}
\, , \ls 
G_{ia} \ = \ 0 
\, , \ls i,j \ = \ 1,2, \ \ a,b \ = \ 3,4
\, , \\
B_{34} \ = \ - \frac{\Omega_0}{K}
\, .
\end{gather}
\esubeq

Hence, we conclude that the semi-doubled NLSM (\ref{SDL-SU2-3}) truly contains the three different standard NLSMs whose target spaces are defect five-branes.
This result also indicates that the semi-doubled GLSM (\ref{SD-GLSM}) has properly describes the UV completion of the standard NLSMs.
We note that,
in this section, we performed the duality transformations at the final stage of the analysis.
We would like to investigate whether the dualization procedure is also applicable at the UV level.
If this is also true, we will find a powerful T-duality transformation procedure for defect branes in terms of a complex twisted linear superfield.
This procedure can be applied to UV gauge theories as well as various IR
effective theories, even if the IR target spaces do not possess isometry.

\section{Standard GLSMs}
\label{S:standard}

In this section, we reduce the semi-doubled GLSM (\ref{SD-GLSM}) to various standard GLSMs. 
In each GLSM, we take the IR limit.
We will find its IR effective theory corresponds to the NLSM discussed in section \ref{S:SDNLSM}.
This result will conclude that the duality transformations introduced in terms of the complex twisted linear superfield discussed in section \ref{S:twistedlinear} is applicable to both the IR theory and its UV theory.
In other words, the T-duality transformation procedure given in section \ref{S:twistedlinear} can be extended away from the IR limit, as demonstrated by Hori and Vafa \cite{Hori:2000kt}. 
This feature is very strong and helpful to investigate the string worldsheet instanton corrections to the five-branes configurations via the gauge theory vortex corrections in section \ref{S:instantons}.

\subsection{Standard GLSM for defect NS5-brane}

We integrate out the dual scalar fields $\varphi_{\text{R}-}$ and $\varphi_{\text{I}+}$ in the gauge theory (\ref{SD-GLSM}). 
Their equations of motion are
\begin{alignat}{2} 
\varphi_{\text{R}-}: \ \ \ 
0 \ &= \ 
\del^n \Big\{
- \eps_{mn} (\del^m r^3)
\Big\}
\, , &\ls
\varphi_{\text{I}+}: \ \ \ 
0 \ &= \ 
\del^n \Big\{
- \eps_{mn} (\del^m \vartheta)
\Big\}
\, . \label{eom-phiRI-GLSM-dNS5}
\end{alignat}
The solutions of these equations are trivial. 
Substituting the equations of motion into the semi-doubled GLSM (\ref{SD-GLSM}), we find that the covariant derivatives $D_m \varphi_{\text{R}-} = \del_m \varphi_{\text{R}-} - \sqrt{2} \sum_{a'} \wh{A}_{m,a'}$ and $D_m \varphi_{\text{I}+} = \del_m \varphi_{\text{I}+} - \sqrt{2} \sum_a A_{m,a}$ are decomposed into two parts. 
The ordinary derivative part vanishes due to the equations of motion (\ref{eom-phiRI-GLSM-dNS5}),
while the gauge fields part is combined with the total derivative term.
Then we obtain the standard GLSM: 
\begin{align}
\Scr{L}_{\text{NS5}}
\ &= \ 
\sum_{a=1}^k \frac{1}{2 e_a^2} (F_{01,a})^2 
+ \sum_{a'=1}^{\l} \frac{1}{2 \wh{e}_{a'}^2} (\wh{F}_{01,a'})^2 
+ \sqrt{2} \sum_a (\vartheta - t^4_a) F_{01,a} 
+ \sqrt{2} \sum_{a'} (r^3 - \wh{t}^3_{a'}) \wh{F}_{01,a'} 
\nn \\
\ & \ \ \ \ 
- \sum_a \frac{1}{e_a^2} \Big\{ |\del_m \phi_a|^2 + |\del_m \sigma_a|^2 \Big\}
- \sum_{a'} \frac{1}{\wh{e}_{a'}^2} \Big\{ |\del_m \wh{\phi}_{a'}|^2 + |\del_m \wh{\sigma}_{a'}|^2 \Big\}
\nn \\
\ & \ \ \ \ 
- \sum_a 2 \big( |q_a|^2 + |\wt{q}_a|^2 \big) \big( |\phi_a|^2 + |\sigma_a|^2 \big) 
- \sum_{a'} 2 \big( |p_{a'}|^2 + |\wt{p}_{a'}|^2 \big) \big( |\wh{\phi}_{a'}|^2 + |\wh{\sigma}_{a'}|^2 \big)
\nn \\
\ & \ \ \ \ 
- 2 g^2 \Big| \sum_a \phi_a + \I \sum_{a'} \wh{\phi}_{a'} \Big|^2
- 2 g^2 \Big| \sum_a \sigma_a + \I \sum_{a'} \wh{\sigma}_{a'} \Big|^2
\nn \\
\ & \ \ \ \ 
- \sum_a \Big\{ |D_m q_a|^2 + |D_m \wt{q}_a|^2 \Big\}
- \sum_{a'} \Big\{ |D_m p_{a'}|^2 + |D_m \wt{p}_{a'}|^2 \Big\}
\nn \\
\ & \ \ \ \ 
- \frac{1}{2 g^2} \Big\{
(\del_m r^1)^2
+ (\del_m r^2)^2
+ (\del_m r^3)^2 
+ (\del_m \vartheta)^2 
\Big\}
\nn \\
\ & \ \ \ \ 
- \sum_a \frac{e_a^2}{2} \Big\{ (|q_a|^2 - |\wt{q}_a|^2) - \sqrt{2} \, (r^2 - s^2_a) \Big\}^2
- \sum_a e_a^2 \Big| \sqrt{2} \, q_a \wt{q}_a + \big( (r^1 - s^1_a) - \I (r^3 - t^3_a) \big) \Big|^2
\nn \\
\ & \ \ \ \ 
- \sum_{a'} \frac{\wh{e}_{a'}^2}{2} \Big\{ (|p_{a'}|^2 - |\wt{p}_{a'}|^2) - \sqrt{2} \, (r^2 - \wh{s}^2_{a'}) \Big\}^2
- \sum_{a'} \wh{e}_{a'}^2 \Big| \sqrt{2} \, p_{a'} \wt{p}_{a'} + \I \big( (r^1 - \wh{s}^1_{a'}) + \I (\vartheta - \wh{t}^4_{a'}) \big) \Big|^2
\, . \label{L-dNS5}
\end{align}
Note that, if the gauge coupling constants $\wh{e}_{a'}$ shrink to zero, the system (\ref{L-dNS5}) is reduced to the standard GLSM for H-monopoles (\ref{Tong-L}) discussed in \cite{Tong:2002rq}.
We comment that this system again acquire the topological terms $\vartheta F_{01,a}$ and $r^3 \wh{F}_{01,a'}$.
We have emphasized the role of these two terms in section \ref{S:introduction}. 
We will investigate their contributions to the non-perturbative corrections in section \ref{S:instantons}.

Analogous to the previous section, we investigate the supersymmetry vacuum of this system.
This can be evaluated by the vanishing condition of the scalar potentials given by the same equations as (\ref{Higgs-vacuum-SDL}).
On this vacuum, the configuration of the charged scalar fields $(q_a,
\wt{q}_a)$ and $(p_{a'}, \wt{p}_{a'})$ is also the same as (\ref{qqpp-SU2}).
Then the effective gauge theory is given by
\begin{align}
\Scr{L}_{\text{NS5}}
\ &= \ 
\sum_{a=1}^k \frac{1}{2 e_a^2} (F_{01,a})^2 
+ \sum_{a'=1}^{\l} \frac{1}{2 \wh{e}_{a'}^2} (\wh{F}_{01,a'})^2 
\nn \\
\ & \ \ \ \ 
+ \sqrt{2} \sum_a \eps^{mn} \del_m \big( (\vartheta - t^4_a) A_{n,a} \big) 
+ \sqrt{2} \sum_{a'} \eps^{mn} \del_m \big( (r^3 - \wh{t}^3_{a'}) \wh{A}_{n,a'} \big)
\nn \\
\ & \ \ \ \ 
- \frac{H_0}{2} \Big\{ (\del_m r^1)^2 + (\del_m r^2)^2 \Big\}
- \frac{H}{2} (\del_m r^3)^2 
- \frac{\wh{H}}{2} (\del_m \vartheta)^2 
\nn \\
\ & \ \ \ \ 
- \sum_a \sqrt{2} \, R_a \Big\{
\del_m \alpha_a 
- A_{m,a} 
+ \frac{1}{\sqrt{2}} \Omega_{i,a} (\del_m r^i)
\Big\}^2
- \sum_a \eps^{mn} (\del_m \vartheta) (\sqrt{2} A_{n,a}) 
\nn \\
\ & \ \ \ \ 
- \sum_{a'} \sqrt{2} \, \wh{R}_{a'} \Big\{
\del_m \wh{\alpha}_{a'} 
- \wh{A}_{m,a'} 
+ \frac{1}{\sqrt{2}} \wh{\Omega}_{j',a'} (\del_m r^{j'})
\Big\}^2
- \sum_{a'} \eps^{mn} (\del_m r^3) (\sqrt{2} \wh{A}_{n,a'}) 
\, . \label{L-dNS5_2}
\end{align}
Here the variables $(R_a, \wh{R}_{a'})$, $(\Omega_{i,a}, \wh{\Omega}_{j',a'})$ and $(H_0, H, \wh{H})$ are also exactly the same as those in (\ref{qqRppR-SDL}), (\ref{Omega}) and (\ref{Harm_H}), respectively.

Now we study the IR limit $e_a, \wh{e}_{a'} \to \infty$. 
In this limit, the kinetic terms of the gauge fields are frozen, and the gauge fields become non-dynamical.
Then we integrate them out via the equations of motion.
The solutions are easily obtained:
\bsubeq \label{sol:Am_dNS5}
\begin{align}
A_{n,a} 
\ &= \ 
- \frac{1}{2 R_a} \eps_{mn} (\del^m \vartheta)
+ \del_n \alpha_a
+ \frac{1}{\sqrt{2}} \Omega_{i,a} (\del_n r^i)
\, , \\
\wh{A}_{n,a} 
\ &= \ 
- \frac{1}{2 \wh{R}_{a'}} \eps_{mn} (\del^m r^3)
+ \del_n \wh{\alpha}_{a'}
+ \frac{1}{\sqrt{2}} \wh{\Omega}_{j',a'} (\del_n r^{j'})
\, .
\end{align}
\esubeq
Substituting the solutions into the effective Lagrangian (\ref{L-dNS5_2}) in the IR limit, we find the standard NLSM
\begin{align}
\Scr{L}_{\text{NS5}}^{\text{IR}}
\ &= \ 
- \frac{H_0}{2} \Big\{ (\del_m r^1)^2 + (\del_m r^2)^2 + (\del_m r^3)^2 + (\del_m \vartheta)^2 \Big\}
+ \Omega_0 \, \eps^{mn} (\del_m r^3) (\del_n \vartheta) 
\nn \\
\ & \ \ \ \ 
+ \sum_a \sqrt{2} \, \eps^{mn} \del_m \big( (\vartheta - t^4_a) \mr{A}_{n,a} \big)
+ \sum_{a'} \sqrt{2} \, \eps^{mn} \del_m \big( (r^3 - \wh{t}^3_{a'}) \mr{\wh{A}}_{n,a'} \big)
\nn \\
\ & \ \ \ \ 
+ \eps^{mn} \, \Omega_{1} (\del_m r^1) (\del_n \vartheta) 
+ \eps^{mn} \, \wh{\Omega}_{1} (\del_m r^1) (\del_n r^3) 
\nn \\
\ & \ \ \ \ 
- \sqrt{2} \sum_a \eps^{mn} (\del_m \vartheta) (\del_n \alpha_a)
- \sqrt{2} \sum_{a'} \eps^{mn} (\del_m r^3) (\del_n \wh{\alpha}_{a'})
\, . \label{L-dNS5_3}
\end{align}
This configuration has not possessed isometry along the $r^3$- and $\vartheta$-directions yet. 
Performing the smearing procedure expressed by (\ref{smeared-HOmega}), 
we find that the third line disappears and obtain properly the same NLSM as (\ref{SDL-SU2-dNS5}), i.e., the standard NLSM for a single defect NS5-brane, up to the terms containing the gauge parameters.
These terms can be set to zero.
This result is merely expected, because we started the GLSM for H-monopoles (\ref{Tong-L}) and extended it to the model for a defect NS5-brane.
Indeed, the dualization procedure was not seriously utilized in this system.

\subsection{Standard GLSM for KK-vortex}

We go back to the semi-doubled GLSM (\ref{SD-GLSM}), and construct the standard GLSM for a KK-vortex.
The KK-vortex is a gravitational object via T-dualization of a defect NS5-brane.
In this subsection, we confirm that the GLSM for the KK-vortex is dual to that for the defect NS5-brane at the UV level, as well as at the IR level.

First, we evaluate the equation of motion for the dual scalar field $\varphi_{\text{R}-}$.
\begin{alignat}{2}
{\varphi}_{\text{R}-}: \ \ \ &&
0 \ &= \
\del^n \Big\{
- \eps_{mn} (\del^m r^3)
\Big\} 
\, . \label{EOM:phiR_KK}
\end{alignat}
Again this is trivially satisfied.
On the other hand, the equation of motion for the original scalar $\vartheta$ is non-trivial:
\begin{alignat}{2}
\vartheta: \ \ \ &&
0 \ &= \
\sum_{a'} \wh{e}_{a'}^2 \Big\{ \sqrt{2} \, p_{a'} \wt{p}_{a'} + \I (r^1 - \wh{s}^1_{a'}) - (\vartheta - \wh{t}^4_{a'}) \Big\}
+ \sum_{a'} \wh{e}_{a'}^2 \Big\{ \sqrt{2} \, \ol{p}{}_{a'} \ol{\wt{p}}{}_{a'} - \I (r^1 - \wh{s}^1_{a'}) - (\vartheta - \wh{t}^4_{a'}) \Big\}
\nn \\
\ &&& \ \ \ \  
- \del^m \Big\{
\frac{1}{g^2} (\del_m \vartheta)
- \eps_{mn} (D^n \varphi_{\text{I}+})
\Big\}
\, . \label{EOM-theta_KK_original}
\end{alignat}
For convenience, we abbreviate the first two terms in the right-hand side:
\begin{align}
\mathscr{A}_{a'} 
\ &:= \
\sqrt{2} \, p_{a'} \wt{p}_{a'} + \I (r^1 - \wh{s}^1_{a'}) - (\vartheta - \wh{t}^4_{a'}) 
\, , \ls
\mathscr{A}^m
\ := \ 
\int \d x^m \sum_{a'} \wh{e}_{a'}^2 (\mathscr{A}_{a'} + \ol{\mathscr{A}}{}_{a'})
\, . \label{Am_KK}
\end{align}
Note that $\mathscr{A}_{a'}$ is a building block of the scalar potentials.
The equation of motion (\ref{EOM-theta_KK_original}) can be formally solved as
\begin{align}
0 \ &= \ 
\mathscr{A}^m
- \frac{1}{g^2} (\del^m \vartheta) 
+ \eps^{mn} (D_n \varphi_{\text{I}+})
+ \del_n f^{mn} (x)
\, . \label{EOM:theta_KK}
\end{align}
We note that $f^{mn}(x)$ is an arbitrary antisymmetric tensor.
Since we now study the model in two dimensions, this arbitrary tensor can be given in terms of the Levi-Civita symbol:
\begin{align}
f^{mn} (x) \ &= \ \eps^{mn} f(x)
\, , \label{f_KKG}
\end{align}
where $f(x)$ is an arbitrary function.
In later discussion, this function will play a central role in confirming the dual configuration at the UV level.
Now, we obtain the dual form of the original scalar field $\vartheta$ as follows:
\begin{align}
\frac{1}{g^2} \del_m \mr{\vartheta}
\ &= \ 
\eps_{mn} \Big\{ D^n \varphi_{\text{I}+} + \del^n f(x) \Big\}
+ \mathscr{A}_m
\, . \label{Kin:theta_KK}
\end{align}
Here $\mr{\vartheta}$ means that $\vartheta$ is no longer independent field and governed by the equation of motion (\ref{EOM:theta_KK}).
Substituting the solutions (\ref{EOM:phiR_KK}) and (\ref{Kin:theta_KK}) into the Lagrangian (\ref{SD-GLSM}), we obtain a gauge theory:
\begin{align}
\Scr{L}_{\text{KK}}
\ &= \ 
\sum_{a=1}^k \frac{1}{2 e_a^2} (F_{01,a})^2 
+ \sum_{a'=1}^{\l} \frac{1}{2 \wh{e}_{a'}^2} (\wh{F}_{01,a'})^2 
\nn \\
\ & \ \ \ \ 
+ \sum_a \sqrt{2} \, \eps^{mn} \del_m \big( (\mr{\vartheta} - t^4_a) A_{n,a} \big)
+ \sqrt{2} \sum_{a'} (r^3 - \wh{t}^3_{a'}) \eps^{mn} \del_m \wh{A}_{n,a'} 
\nn \\
\ & \ \ \ \ 
- \sum_a \frac{1}{e_a^2} \Big\{ |\del_m \phi_a|^2 + |\del_m \sigma_a|^2 \Big\}
- \sum_{a'} \frac{1}{\wh{e}_{a'}^2} \Big\{ |\del_m \wh{\phi}_{a'}|^2 + |\del_m \wh{\sigma}_{a'}|^2 \Big\}
\nn \\
\ & \ \ \ \ 
- \sum_a 2 \big( |q_a|^2 + |\wt{q}_a|^2 \big) \big( |\phi_a|^2 + |\sigma_a|^2 \big) 
- \sum_{a'} 2 \big( |p_{a'}|^2 + |\wt{p}_{a'}|^2 \big) \big( |\wh{\phi}_{a'}|^2 + |\wh{\sigma}_{a'}|^2 \big)
\nn \\
\ & \ \ \ \ 
- 2 g^2 \Big| \sum_a \phi_a + \I \sum_{a'} \wh{\phi}_{a'} \Big|^2
- 2 g^2 \Big| \sum_a \sigma_a + \I \sum_{a'} \wh{\sigma}_{a'} \Big|^2
\nn \\
\ & \ \ \ \ 
- \sum_a \Big\{ |D_m q_a|^2 + |D_m \wt{q}_a|^2 \Big\}
- \sum_{a'} \Big\{ |D_m p_{a'}|^2 + |D_m \wt{p}_{a'}|^2 \Big\}
\nn \\
\ & \ \ \ \ 
- \frac{1}{2 g^2} \Big\{
(\del_m r^1)^2
+ (\del_m r^2)^2
+ (\del_m r^3)^2 
+ (\del_m \mr{\vartheta})^2 
\Big\}
+ \eps^{mn} (\del_m \mr{\vartheta}) (D_n \varphi_{\text{I}+})
\nn \\
\ & \ \ \ \ 
- \sum_a \frac{e_a^2}{2} \Big\{ (|q_a|^2 - |\wt{q}_a|^2) - \sqrt{2} \, (r^2 - s^2_a) \Big\}^2
- \sum_a e_a^2 \Big| \sqrt{2} \, q_a \wt{q}_a + \big( (r^1 - s^1_a) - \I (r^3 - t^3_a) \big) \Big|^2
\nn \\
\ & \ \ \ \ 
- \sum_{a'} \frac{\wh{e}_{a'}^2}{2} \Big\{ (|p_{a'}|^2 - |\wt{p}_{a'}|^2) - \sqrt{2} \, (r^2 - \wh{s}^2_{a'}) \Big\}^2
- \sum_{a'} \wh{e}_{a'}^2 |\mathscr{A}_{a'}|^2 
\, . \label{L-KK}
\end{align}
We next consider the supersymmetric vacuum on which the scalar potentials vanish. 
The equations are exactly the same as (\ref{Higgs-vacuum-SDL}), except that the scalar $\vartheta$ is replaced with $\mr{\vartheta}$.
We also find the solution of the charged fields $(q_a, \wt{q}_a)$ and $(p_{a'}, \wt{p}_{a'})$ as (\ref{qqpp-SU2}). 
Then we obtain the gauge theory on the supersymmetric vacuum:
\begin{align}
\Scr{L}_{\text{KK}}
\ &= \ 
\sum_{a=1}^k \frac{1}{2 e_a^2} (F_{01,a})^2 
+ \sum_{a'=1}^{\l} \frac{1}{2 \wh{e}_{a'}^2} (\wh{F}_{01,a'})^2 
\nn \\
\ & \ \ \ \ 
+ \sum_a \sqrt{2} \, \eps^{mn} \del_m \big( (\mr{\vartheta} - t^4_a) A_{n,a} \big)
+ \sum_{a'} \sqrt{2} \, \eps^{mn} \del_m \big( (r^3 - \wh{t}^3_{a'}) \wh{A}_{n,a'} \big)
- \sum_{a'} \eps^{mn} \del_n \big( (\del_m r^3) (\sqrt{2} \wh{\alpha}_{a'}) \big)
\nn \\
\ & \ \ \ \ 
- \frac{H_0}{2} \Big\{ (\del_m r^1)^2 + (\del_m r^2)^2 \Big\}
- \frac{H}{2} (\del_m r^3)^2 
\nn \\
\ & \ \ \ \ 
+ \frac{g^4 \wh{H}}{2} \big\{ D_m \varphi_{\text{I}+} + \del_m f \big\}^2
- g^2 \big\{ D^m \varphi_{\text{I}+} + \del^m f \big\} (D_m \varphi_{\text{I}+})
+ \sum_{a'} \eps^{mn} (\del_m r^3) \sqrt{2} (\del_n \wh{\alpha}_{a'} - \wh{A}_{n,a'}) 
\nn \\
\ & \ \ \ \ 
- \sum_a \sqrt{2} R_a \Big\{ 
\del_m \alpha_a - A_{m,a}
+ \frac{1}{\sqrt{2}} \Omega_{i,a} (\del_m r^i)
\Big\}^2
\nn \\
\ & \ \ \ \ 
+ \sum_{a'} \sqrt{2} \wh{R}_{a'} \Big\{ 
\eps^{np} (\del_p \wh{\alpha}_{a'} - \wh{A}_{p,a'}) 
+ \frac{1}{\sqrt{2}} \wh{\Omega}_{1,a'} \eps^{np} (\del_p r^1)
+ \frac{g^2}{\sqrt{2}} \wh{\Omega}_{4,a'} \big( D^n \varphi_{\text{I}+} + \del^n f \big)
\Big\}^2
\, . \label{L-KK_2}
\end{align}

We analyze the IR limit $e_a, \wh{e}_{a'} \to \infty$.
To simplify our computations, we introduce the following expressions:
\bsubeq \label{simpleA}
\begin{alignat}{2}
{\sf A}_{m,a} 
\ &:= \ 
\sqrt{2} (\del_m \alpha_a - A_{m,a})
\, , &\ls
\wh{\sf A}_{m,a'} 
\ &:= \ 
\sqrt{2} (\del_m \wh{\alpha}_{a'} - \wh{A}_{m,a'})
\, , \\
\wt{\varphi}_{\text{I}+}
\ &:= \ 
\varphi_{\text{I}+} - \sum_{a} \sqrt{2} \alpha_a
\, , &\ls
D_m \varphi_{\text{I}+}
\ &\hphantom{:}= \ 
\del_m \wt{\varphi}_{\text{I}+} + \sum_a {\sf A}_{m,a}
\, .
\end{alignat}
\esubeq
In the IR limit, the gauge fields ${\sf A}_{m,a}$ and $\wh{\sf A}_{m,a'}$ are no longer dynamical, and they should be integrated out from the system.
We evaluate their equations of motion:
\bsubeq
\begin{align}
0 
\ &= \ 
g^4 \wh{H} (D_n \varphi_{\text{I}+} + \del_n f)
- g^2 (D_n \varphi_{\text{I}+})
- g^2 (D_n \varphi_{\text{I}+} + \del_n f)
\nn \\
\ & \ \ \ \ 
- \sqrt{2} R_a \Big\{ {\sf A}_{n,a} + \Omega_{1,a} (\del_n r^1) + \Omega_{3,a} (\del_n r^3) \Big\}
\nn \\
\ & \ \ \ \ 
+ \sum_{a'} \sqrt{2} \wh{R}_{a'} \Big\{ \eps_{np} \wh{\sf A}^p_{a'} + \eps_{np} \wh{\Omega}_{1,a'} (\del^p r^1) + g^2 \wh{\Omega}_{4,a'} (D_n \varphi_{\text{I}+} + \del_n f) \Big\} (g^2 \wh{\Omega}_{4,a'})
\, , \label{EOM-A_180818} \\
0 
\ &= \ 
\del_n r^3
+ \sqrt{2} \wh{R}_{a'} \Big\{ \eps_{nq} \wh{\sf A}^q_{a'} + \eps_{nq} \wh{\Omega}_{1,a'} (\del^q r^1) + g^2 \wh{\Omega}_{4,a'} (D_n \varphi_{\text{I}+} + \del_n f) \Big\}
\, . \label{EOM-At_180818}
\end{align}
\esubeq
Via the same technique as in \cite{Kimura:2013fda}, we first obtain the sum of the gauge fields $\sum_a {\sf A}_{n,a}$. 
This is enough to understand the IR behavior of the system.
\begin{align}
\sum_b {\sf A}_{n,b}
\ &= \ 
- \frac{g^2 \wh{H}_2}{H_3} \sum_b \frac{1}{\sqrt{2} R_b} (\del_n \wt{\varphi}_{\text{I}+})
- \frac{g^2 \wh{H}_2 - 1}{H_3} \sum_b \frac{1}{\sqrt{2} R_b} (\del_n f)
\nn \\
\ & \ \ \ \ 
- \frac{\Omega_1}{g^2 H_3} (\del_n r^1)
- \frac{1}{H_3} \Big( \frac{\Omega_3}{g^2} + \wh{\Omega}_{4} \sum_b \frac{1}{\sqrt{2} R_b} \Big) (\del_n r^3)
\, , \\
D_n \varphi_{\text{I}+} + \del_n f
\ &= \ 
\del_n \wt{\varphi}_{\text{I}+} + \del_n f + \sum_b {\sf A}_{n,b}
\nn \\
\ &= \ 
\frac{1}{g^2 H_3} \Big\{ (\del_n \wt{\varphi}_{\text{I}+})
- \Big( \Omega_3 + \sum_b \frac{g^2 \wh{\Omega}_{4}}{\sqrt{2} R_b} \Big) (\del_n r^3)
+ g^2 H (\del_n f)
- \Omega_1 (\del_n r^1)
\Big\}
\, . \label{onshell-covphiI+}
\end{align}
To make the expressions simple, we have introduced the following functions:
\begin{align}
\wh{H}_2 
\ &:= \ 
\frac{1}{g^2} - \sum_{a'} \frac{1}{\sqrt{2} \wh{R}_{a'}}
\, , \ls
H_3 
\ := \ 
\frac{1}{g^2} + \sum_a \frac{g^2 \wh{H}_2}{\sqrt{2} R_a}
\, .
\end{align}

We notice that the arbitrary function $f(x)$ (\ref{f_KKG}) is not constrained by any physical condition.
This can be merely regarded as a diffeomorphism parameter from the target space viewpoint.
However, in order to confirm that the gauge theory (\ref{L-KK}) is truly the UV completion of the NLSM for a single KK-vortex (\ref{SDL-SU2-dKK5}), 
it is better to fix this function to an appropriate form.
In the IR limit, we have already known the dual form of $\del_m \mr{\vartheta}$ in terms of (\ref{SDL-SU2-KKeq}), which we call the ``on-shell'' form because the gauge fields are on-shell and integrated out.
On the other hand, $\del_m \mr{\vartheta}$ is also given by (\ref{onshell-covphiI+}) via (\ref{Kin:theta_KK}).
Combining them, we obtain the ``on-shell'' expression of the arbitrary function $f(x)$:
\begin{align}
\del_n f
\ &= \ 
\frac{1}{H_0} \Big\{ 
- \sum_{a'} \frac{1}{\sqrt{2} \wh{R}_{a'}} (\del_n \wt{\varphi}_{\text{I}+}) 
+ \Big( 
\sum_{a'} \frac{\Omega_3}{\sqrt{2} \wh{R}_{a'}} 
+ \wh{\Omega}_4 H
\Big) (\del_n r^3)
+ \sum_{a'} \frac{\Omega_1}{\sqrt{2} \wh{R}_{a'}} (\del_n r^1)
\Big\}
\, . \label{solf_180818}
\end{align}
However, the function $f(x)$ which want to know is the ``off-shell'' form before the gauge fields are integrated out.
We construct it in the following way. 
Substitute (\ref{solf_180818}) into (\ref{onshell-covphiI+}), and describe the scalar field $\wt{\varphi}_{\text{I}+}$ in terms of the covariant derivative $D_m \varphi_{\text{I}+}$ and other functions:
\begin{align}
\del_n \wt{\varphi}_{\text{I}+}
\ &= \ 
\frac{H_0}{\wh{H}} D_n \varphi_{\text{I}+} 
+ \frac{1}{\wh{H}} \Big(
\Omega_3 \wh{H}
+ \sum_a \frac{\wh{\Omega}_4}{\sqrt{2} R_a} 
\Big) (\del_n r^3)
+ \Omega_1 (\del_n r^1)
\, .
\end{align}
This is a big hint to obtain the ``off-shell'' form of $f(x)$.
Substitute it into (\ref{solf_180818}) and remove $\wt{\varphi}_{\text{I}+}$.
Then the derivative $\del_n f$ is expressed in terms of the covariant derivative $D_m \varphi_{\text{I}+}$ which carries the off-shell gauge fields.
Then we obtain the ``off-shell'' form: 
\begin{align}
\del_n f
\ &= \ 
- \sum_{a'} \frac{1}{\sqrt{2} \wh{R}_{a'} \wh{H}} D_n \varphi_{\text{I}+} 
+ \frac{\wh{\Omega}_4}{g^2 \wh{H}} (\del_n r^3)
\, . \label{offshell-f_KK}
\end{align}
By using this, we expect that the following is the duality relation at the UV level: 
\begin{align}
\frac{1}{g^2} \del_m \mr{\vartheta}
\ &= \ 
\eps_{mn} \Big\{ D^n \varphi_{\text{I}+} + \del^n f \Big\}
\ = \
\frac{1}{g^2 \wh{H}} \eps_{mn} \Big\{ D^n \varphi_{\text{I}+} + \wh{\Omega}_4 (\del^n r^3) \Big\}
\, . \label{offshell-dualK_180818}
\end{align}
We go back to the gauge theory (\ref{L-KK}) where the original field $\mr{\vartheta}$ is now governed by (\ref{offshell-dualK_180818}) (with $\mathscr{A}^m$ (\ref{Am_KK}) before setting them to zero):
\bsubeq \label{Complete-GLSMK}
\begin{align}
\Scr{L}_{\text{KK}}
\ &= \ 
\sum_{a=1}^k \frac{1}{2 e_a^2} (F_{01,a})^2 
+ \sum_{a'=1}^{\l} \frac{1}{2 \wh{e}_{a'}^2} (\wh{F}_{01,a'})^2 
\nn \\
\ & \ \ \ \ 
+ \sum_a \sqrt{2} \, \eps^{mn} \del_m \big( (\mr{\vartheta} - t^4_a) A_{n,a} \big)
+ \sqrt{2} \sum_{a'} (r^3 - \wh{t}^3_{a'}) \eps^{mn} \del_m \wh{A}_{n,a'} 
\nn \\
\ & \ \ \ \ 
- \sum_a \frac{1}{e_a^2} \Big\{ |\del_m \phi_a|^2 + |\del_m \sigma_a|^2 \Big\}
- \sum_{a'} \frac{1}{\wh{e}_{a'}^2} \Big\{ |\del_m \wh{\phi}_{a'}|^2 + |\del_m \wh{\sigma}_{a'}|^2 \Big\}
\nn \\
\ & \ \ \ \ 
- \sum_a 2 \big( |q_a|^2 + |\wt{q}_a|^2 \big) \big( |\phi_a|^2 + |\sigma_a|^2 \big) 
- \sum_{a'} 2 \big( |p_{a'}|^2 + |\wt{p}_{a'}|^2 \big) \big( |\wh{\phi}_{a'}|^2 + |\wh{\sigma}_{a'}|^2 \big)
\nn \\
\ & \ \ \ \ 
- 2 g^2 \Big| \sum_a \phi_a + \I \sum_{a'} \wh{\phi}_{a'} \Big|^2
- 2 g^2 \Big| \sum_a \sigma_a + \I \sum_{a'} \wh{\sigma}_{a'} \Big|^2
\nn \\
\ & \ \ \ \ 
- \sum_a \Big\{ |D_m q_a|^2 + |D_m \wt{q}_a|^2 \Big\}
- \sum_{a'} \Big\{ |D_m p_{a'}|^2 + |D_m \wt{p}_{a'}|^2 \Big\}
\nn \\
\ & \ \ \ \ 
- \frac{1}{2 g^2} \Big\{
(\del_m r^1)^2
+ (\del_m r^2)^2
+ (\del_m r^3)^2 
+ (\del_m \mr{\vartheta})^2 
\Big\}
+ \eps^{mn} (\del_m \mr{\vartheta}) (D_n \varphi_{\text{I}+})
\nn \\
\ & \ \ \ \ 
- \sum_a \frac{e_a^2}{2} \Big\{ (|q_a|^2 - |\wt{q}_a|^2) - \sqrt{2} \, (r^2 - s^2_a) \Big\}^2
- \sum_a e_a^2 \Big| \sqrt{2} \, q_a \wt{q}_a + \big( (r^1 - s^1_a) - \I (r^3 - t^3_a) \big) \Big|^2
\nn \\
\ & \ \ \ \ 
- \sum_{a'} \frac{\wh{e}_{a'}^2}{2} \Big\{ (|p_{a'}|^2 - |\wt{p}_{a'}|^2) - \sqrt{2} \, (r^2 - \wh{s}^2_{a'}) \Big\}^2
- \sum_{a'} \wh{e}_{a'}^2 |\mathscr{A}_{a'}|^2 
\, , \\
\frac{1}{g^2} \del_m \mr{\vartheta}
\ &= \ 
\frac{1}{g^2 \wh{H}} \eps_{mn} \Big\{ D^n \varphi_{\text{I}+} + \wh{\Omega}_4 (\del^n r^3) \Big\}
+ \mathscr{A}_m
\, . \label{off-theta_KK}
\end{align}
\esubeq
This should lead us to the NLSM for the KK-vortex. 
We explore the IR regime. 
In the IR limit $e_a, \wh{e}_{a'} \to \infty$, 
we again evaluate the equations of motion for the gauge fields ${\sf A}_{m,a}$ and $\wh{\sf A}_{m,a'}$.
After the straightforward calculations, we can solve them and substitute the solutions into the Lagrangian. 
Finally we obtain the IR NLSM in following form:
\begin{align}
\Scr{L}_{\text{KK}}^{\text{IR}}
\ &= \ 
- \frac{H_0}{2} \Big\{ (\del_m r^1)^2 + (\del_m r^2)^2 + (\del_m r^3)^2 \Big\}
- \frac{1}{2 H_0} \Big\{ \del_m \wt{\varphi}_{\text{I}+} - \Omega_0 (\del_m r^3) \Big\}^2
\nn \\
\ & \ \ \ \ 
+ \sum_a \sqrt{2} \, \eps^{mn} \del_m \big( (\mr{\vartheta} - t^4_a) \mr{A}_{n,a} \big)
+ \sum_{a'} \sqrt{2} \, \eps^{mn} \del_m \big( (r^3 - \wh{t}^3_{a'}) \mr{\wh{A}}{}_{n,a'} \big)
\nn \\
\ & \ \ \ \ 
- \frac{\Omega_1^2}{2 H_0} (\del_m r^1)^2 
+ \frac{\Omega_1}{H_0} \Big\{ \del_m \wt{\varphi}_{\text{I}+} - \Omega_0 (\del_m r^3) \Big\} (\del^m r^1) 
+ \wh{\Omega}_{1} \eps^{pm} (\del_p r^1) (\del_m r^3) 
\nn \\
\ & \ \ \ \ 
- \sum_{a'} \eps^{mn} \del_n \big( (\del_m r^3) (\sqrt{2} \wh{\alpha}_{a'}) \big)
\, . \label{L-KK_4_180819}
\end{align}
This is the NLSM which we have already described as the NLSM for the single KK-vortex (\ref{SDL-SU2-dKK5}), expect for the third and fourth lines.
But this is not a problem.
In the smearing procedure (\ref{smeared-HOmega}), the third lines disappear.
The fourth line does not contribute to the system because it is a total derivative.
Hence, we conclude that we successfully confirmed the gauge theory (\ref{Complete-GLSMK}) as the GLSM the single KK-vortex.
Then the duality relation between the original field $\vartheta$ and the dual field $\varphi_{\text{I}+}$ is given as (\ref{offshell-dualK_180818}) (or more precisely (\ref{off-theta_KK})), which is reduced to (\ref{SDL-SU2-KKeq}) in the IR limit.

\subsection{Standard GLSM for exotic $5^2_2$-brane}

We will also find that the semi-doubled GLSM (\ref{SD-GLSM}) contains the standard GLSM for the single $5^2_2$-brane.
To confirm this, we integrate out the original scalar fields $\vartheta$ and $r^3$.
The equation of motion for the former has been analyzed in (\ref{EOM-theta_KK_original}),
and its final solution is given by (\ref{off-theta_KK}).  
Here we focus on the equation of motion for the latter:
\begin{alignat}{2}
r^3: \ \ \ &&
0 \ &= \ 
- \I \sum_a e_a^2 \Big\{ \sqrt{2} \, q_a \wt{q}_a + (r^1 - s^1_a) - \I (r^3 - t^3_a) \Big\}
+ \I \sum_a e_a^2 \Big\{ \sqrt{2} \, \ol{q}_a \ol{\wt{q}}{}_a + (r^1 - s^1_a) + \I (r^3 - t^3_a) \Big\}
\nn \\
&&& \ \ \ \
- \del^m \Big\{
\frac{1}{g^2} (\del_m r^3)
- \eps_{mn} (D^n \varphi_{\text{R}-})
\Big\}
\, . \label{EOM-r3_original}
\end{alignat}
For later convenience, this is also abbreviated by introducing the following forms:
\begin{align}
\mathscr{B}_a 
\ &:= \ 
\sqrt{2} \, q_a \wt{q}_a + (r^1 - s^1_a) - \I (r^3 - t^3_a) 
\, , \ls
\mathscr{B}^m
\ := \ 
\int \d x^m \Big\{ - \I \sum_a e_a^2 (\mathscr{B}_a - \ol{\mathscr{B}}{}_a) \Big\}
\, . \label{Bm_E}
\end{align}
We note that $\mathscr{B}_a$ is also the building block of the scalar potentials.
Now (\ref{EOM-r3_original}) is described by
\begin{align}
\frac{1}{g^2} \del_m \mr{r}^3
\ &= \ 
\eps_{mn} \Big\{ D^n \varphi_{\text{R}-} + \del^n \wh{f}(x) \Big\} + \mathscr{B}_m
\, . \label{Kin:r3_E}
\end{align}
Here we notice that $\wh{f}(x)$ is an arbitrary function similar to $f(x)$ (\ref{f_KKG}).
In the current stage, we have already understood how this function is treated.
Actually, this function is also interpreted as a gauge parameter of the diffeomorphism and the B-field gauge transformation on the target space configuration.
Analogous to (\ref{off-theta_KK}), the form (\ref{Kin:r3_E}) can be obtained as
\begin{align}
\frac{1}{g^2} \del_m \mr{r}^3
\ &= \ 
\frac{1}{g^2 H} \eps_{mn} \Big\{ D^n \varphi_{\text{R}-} + \Omega_3 (\del^n \vartheta) \Big\} + \mathscr{B}_m
\, . \label{off-r3_E}
\end{align}
By using the expressions (\ref{off-theta_KK}) and (\ref{off-r3_E}), we find the off-shell duality relations at the UV level:
\bsubeq \label{rtheta-dualE_1_180820}
\begin{align}
\frac{1}{g^2} \del_m \mr{\vartheta}
\ &= \ 
\frac{1}{g^2 {\cal M}} 
\Big\{
H \eps_{mn} D^n \varphi_{\text{I}+}
+ \wh{\Omega}_4 D_m \varphi_{\text{R}-}
+ \Big( \mathscr{A}_m + \frac{\wh{\Omega}_4}{\wh{H}} \eps_{mn} \mathscr{B}^n \Big) 
\Big\}
\, , \\
\frac{1}{g^2} \del_m \mr{r}^3
\ &= \ 
\frac{1}{g^2 {\cal M}} \Big\{
\wh{H} \eps_{mn} D^n \varphi_{\text{R}-}
+ \Omega_3 D_m \varphi_{\text{I}+}
+ \Big( \frac{\Omega_3}{H} \eps_{mn} \mathscr{A}^n + \mathscr{B}_m \Big)
\Big\}
\, , \\
{\cal M}
\ &:= \ 
H \wh{H} - \Omega_3 \wh{\Omega}_4
\, . 
\vphantom{\int}
\end{align}
\esubeq
We emphasize that $(\mr{\vartheta}, \mr{r}^3)$ are no longer dynamical but governed by the dual fields $(\varphi_{\text{I}+}, \varphi_{\text{R}-})$.
Substituting the duality relations into the semi-doubled GLSM (\ref{SD-GLSM}), we obtain a standard gauge theory:
\begin{align}
\Scr{L}_{\text{E}}
\ &= \ 
\sum_{a=1}^k \frac{1}{2 e_a^2} (F_{01,a})^2 
+ \sum_{a'=1}^{\l} \frac{1}{2 \wh{e}_{a'}^2} (\wh{F}_{01,a'})^2 
\nn \\
\ & \ \ \ \ 
+ \sum_a \sqrt{2} \, \eps^{mn} \del_m \big( (\mr{\vartheta} - t^4_a) A_{n,a} \big)
+ \sum_{a'} \sqrt{2} \, \eps^{mn} \del_m \big( (\mr{r}^3 - \wh{t}^3_{a'}) \wh{A}_{n,a'} \big) 
\nn \\
\ & \ \ \ \ 
- \sum_a \frac{1}{e_a^2} \Big\{ |\del_m \phi_a|^2 + |\del_m \sigma_a|^2 \Big\}
- \sum_{a'} \frac{1}{\wh{e}_{a'}^2} \Big\{ |\del_m \wh{\phi}_{a'}|^2 + |\del_m \wh{\sigma}_{a'}|^2 \Big\}
\nn \\
\ & \ \ \ \ 
- \sum_a 2 \big( |q_a|^2 + |\wt{q}_a|^2 \big) \big( |\phi_a|^2 + |\sigma_a|^2 \big) 
- \sum_{a'} 2 \big( |p_{a'}|^2 + |\wt{p}_{a'}|^2 \big) \big( |\wh{\phi}_{a'}|^2 + |\wh{\sigma}_{a'}|^2 \big)
\nn \\
\ & \ \ \ \ 
- 2 g^2 \Big| \sum_a \phi_a + \I \sum_{a'} \wh{\phi}_{a'} \Big|^2
- 2 g^2 \Big| \sum_a \sigma_a + \I \sum_{a'} \wh{\sigma}_{a'} \Big|^2
\nn \\
\ & \ \ \ \ 
- \sum_a \Big\{ |D_m q_a|^2 + |D_m \wt{q}_a|^2 \Big\}
- \sum_{a'} \Big\{ |D_m p_{a'}|^2 + |D_m \wt{p}_{a'}|^2 \Big\}
\nn \\
\ & \ \ \ \ 
- \frac{1}{2 g^2} \Big\{
(\del_m r^1)^2
+ (\del_m r^2)^2
+ (\del_m \mr{r}^3)^2
+ (\del_m \mr{\vartheta})^2
\Big\}
+ \eps^{mn} (\del_m \mr{r}^3) (D_n \varphi_{\text{R}-})
+ \eps^{mn} (\del_m \mr{\vartheta}) (D_n \varphi_{\text{I}+})
\nn \\
\ & \ \ \ \ 
- \sum_a \frac{e_a^2}{2} \Big\{ (|q_a|^2 - |\wt{q}_a|^2) - \sqrt{2} \, (r^2 - s^2_a) \Big\}^2
- \sum_a e_a^2 |\mathscr{B}_a|^2 
\nn \\
\ & \ \ \ \ 
- \sum_{a'} \frac{\wh{e}_{a'}^2}{2} \Big\{ (|p_{a'}|^2 - |\wt{p}_{a'}|^2) - \sqrt{2} \, (r^2 - \wh{s}^2_{a'}) \Big\}^2
- \sum_{a'} \wh{e}_{a'}^2 |\mathscr{A}_{a'}|^2 
\, . \label{L-E}
\end{align}
We expect that this is also the GLSM for the single $5^2_2$-brane which correctly flows to the IR NLSM (\ref{SDL-SU2-522}).

Now we investigate the IR behavior.
Since the form of the scalar potentials in (\ref{L-E}) is still the same
as in (\ref{SD-GLSM}), we again use the equations
(\ref{Higgs-vacuum-SDL}) to describe the supersymmetric vacuum.
On this vacuum the charged fields also have the same configuration as (\ref{qqpp-SU2}).
Then, replacing $(\vartheta, r^3)$ with $(\mr{\vartheta}, \mr{r}^3)$, 
we reduce the Lagrangian (\ref{L-E}) to the following form:
\begin{align}
\Scr{L}_{\text{E}}
\ &= \
\sum_{a=1}^k \frac{1}{2 e_a^2} (F_{01,a})^2 
+ \sum_{a'=1}^{\l} \frac{1}{2 \wh{e}_{a'}^2} (\wh{F}_{01,a'})^2 
\nn \\
\ & \ \ \ \ 
+ \sum_a \sqrt{2} \, \eps^{mn} \del_m \big( (\mr{\vartheta} - t^4_a) A_{n,a} \big)
+ \sum_{a'} \sqrt{2} \, \eps^{mn} \del_m \big( (\mr{r}^3 - \wh{t}^3_{a'}) \wh{A}_{n,a'} \big) 
\nn \\
\ & \ \ \ \ 
- \frac{H_0}{2} \Big\{
(\del_m r^1)^2
+ (\del_m r^2)^2
\Big\}
\nn \\
\ & \ \ \ \ 
- \frac{H}{2 {\cal M}^2} \Big\{ \wh{H} \eps_{mn} D^n \varphi_{\text{R}-} + \Omega_3 D_m \varphi_{\text{I}+} \Big\}^2 
+ \frac{1}{{\cal M}} \Big\{ \wh{H} \eps_{mn} D^n \varphi_{\text{R}-} + \Omega_3 D_m \varphi_{\text{I}+} \Big\} (\eps^{mp} D_p \varphi_{\text{R}-})
\nn \\
\ & \ \ \ \ 
+ \frac{\wh{H}}{2 {\cal M}^2} \Big\{ H D_m \varphi_{\text{I}+} + \wh{\Omega}_4 \eps_{mn} D^n \varphi_{\text{R}-} \Big\}^2 
- \frac{1}{{\cal M}} \Big\{ H D_m \varphi_{\text{I}+} + \wh{\Omega}_4 \eps_{mn} D^n \varphi_{\text{R}-} \Big\} (D^m \varphi_{\text{I}+})
\nn \\
\ & \ \ \ \ 
- \sum_a \sqrt{2} R_a \Big\{ 
\del_m \alpha_a - {A}_{m,a}
+ \frac{\Omega_{1,a}}{\sqrt{2}} (\del_m r^1)
+ \frac{\Omega_{3,a}}{\sqrt{2} {\cal M}} \Big(
\wh{H} \eps_{mn} D^n \varphi_{\text{R}-}
+ \Omega_3 D_m \varphi_{\text{I}+}
\Big) 
\Big\}^2
\nn \\
\ & \ \ \ \ 
+ \sum_{a'} \sqrt{2} \wh{R}_{a'} \Big\{ 
\eps_{mn} (\del^n \wh{\alpha}_{a'} - \wh{A}^n_{a'})
+ \frac{\wh{\Omega}_{1,a'}}{\sqrt{2}} \eps_{mn} (\del^n r^1)
+ \frac{\wh{\Omega}_{4,a'}}{\sqrt{2} {\cal M}} \Big(
H D_m \varphi_{\text{I}+}
+ \wh{\Omega}_4 \eps_{mn} D^n \varphi_{\text{R}-}
\Big) 
\Big\}^2
\, . \label{L-E_180820}
\end{align}
As discussed in \cite{Tong:2002rq} and (\ref{simpleA}), we introduce the following description, for convenience:
\begin{align}
\wt{\varphi}_{\text{R}-} 
\ &:= \ 
\varphi_{\text{R}-} - \sqrt{2} \sum_{a'} \wh{\alpha}_{a'}
\, , \ \ \ 
D_m \varphi_{\text{R}-}
\ = \ 
\del_m \wt{\varphi}_{\text{R}-}
+ \sum_a \wh{\sf A}_{m,a'}
\, . 
\end{align}

Next, we solve the equations of motion for ${\sf A}_{m,a}$ and $\wh{\sf A}_{m,a'}$ in the IR limit $e_a, \wh{e}_{a'} \to \infty$.
After tedious but straightforward calculations, we first obtain the sum of the gauge potentials as follows:
\bsubeq \label{sol_sumA-At_180821}
\begin{align}
\sum_a {\sf A}_{n,a}
\ &= \ 
- \frac{1}{K} \Big\{
(K - X_4) \del_n \wt{\varphi}_{\text{I}+}
+ Y_4 \eps_{np} \del^p \wt{\varphi}_{\text{R}-}
\Big\}
- \frac{X_4 \Omega_1}{K} (\del_n r^1)
+ \frac{Y_4 \wh{\Omega}_1}{K} \eps_{np} (\del^p r^1)
\, , \label{solsum_A_180821} \\
\sum_{a'} \eps_{np} \wh{\sf A}^p_{a'}
\ &= \ 
- \frac{1}{K} \Big\{
Y_3 \del_n \wt{\varphi}_{\text{I}+}
+ (K - X_3) \eps_{np} \del^p \wt{\varphi}_{\text{R}-}
\Big\}
+ \frac{Y_3 \Omega_1}{K} (\del_n r^1)
- \frac{X_3 \wh{\Omega}_1}{K} \eps_{np} (\del^p r^1)
\, , \label{solsum_At_180821}
\end{align}
\esubeq
where, coefficients are given by the following functions:
\bsubeq
\begin{alignat}{2}
X_3
\ &:= \ 
H_0 H + \Omega_0 \Omega_3 
\, , &\ls
Y_4 
\ &:= \ 
\wh{\Omega}_4 H_0 + \Omega_0 \wh{H}
\, , \\
Y_3
\ &:= \
\Omega_3 H_0 - \Omega_0 H 
\, , &\ls
X_4 
\ &:= \
H_0 \wh{H} - \Omega_0 \wh{\Omega}_4 
\, .
\end{alignat}
\esubeq
Substituting them into the covariant derivatives in the equations of motion for the gauge potentials, we eventually obtain themselves:
\bsubeq \label{sol2_180821}
\begin{align}
{\sf A}_{n,a} 
\ &= \ 
\frac{1}{\sqrt{2} R_a {\cal M}} \Big\{ H D_n \varphi_{\text{I}+} + \wh{\Omega}_4 \eps_{np} D^p \varphi_{\text{R}-} \Big\}
\nn \\
\ & \ \ \ \
- \Omega_{1,a} (\del_n r^1) 
- \frac{\Omega_{3,a}}{{\cal M}} \Big( \wh{H} \eps_{np} D^p \varphi_{\text{R}-} + \Omega_3 D_n \varphi_{\text{I}+} \Big)
\, , \\
\eps_{np} \wh{\sf A}^p_{a'} 
\ &= \ 
\frac{1}{\sqrt{2} \wh{R}_{a'} {\cal M}} \Big( \wh{H} \eps_{np} D^p \varphi_{\text{R}-} + \Omega_3 D_n \varphi_{\text{I}+} \Big)
\nn \\
\ & \ \ \ \ 
- \wh{\Omega}_{1,a'} \eps_{np} (\del^p r^1) 
- \frac{\wh{\Omega}_{4,a'}}{{\cal M}} \Big\{ H D_n \varphi_{\text{I}+} + \wh{\Omega}_4 \eps_{np} D^p \varphi_{\text{R}-} \Big\}
\, .
\end{align}
\esubeq
Substituting the solutions into the IR limit of the Lagrangian (\ref{L-E}),
we finally obtain
\begin{align}
\Scr{L}_{\text{E}}^{\text{IR}}
\ &= \ 
- \frac{H_0}{2} \Big\{ (\del_m r^1)^2 + (\del_m r^2)^2 \Big\}
- \frac{H_0}{2 K} \Big\{ (\del_m \wt{\varphi}_{\text{R}-})^2 + (\del_m \wt{\varphi}_{\text{I}+})^2 \Big\}
- \frac{\Omega_0}{K} \eps^{mn} (\del_m \wt{\varphi}_{\text{R}-}) (\del_n \wt{\varphi}_{\text{I}+}) 
\nn \\
\ & \ \ \ \ 
+ \sum_a \sqrt{2} \, \eps^{mn} \del_m \big( (\mr{\vartheta} - t^4_a) \mr{A}_{n,a} \big)
+ \sum_{a'} \sqrt{2} \, \eps^{mn} \del_m \big( (\mr{r}^3 - \wh{t}^3_{a'}) \mr{\wh{A}}{}_{n,a'} \big) 
\nn \\
\ & \ \ \ \ 
+ \Scr{L} (\Omega_1, \wh{\Omega}_1)
\, . \label{NLSME_180821}
\end{align}
Here we have introduced $\Scr{L}(\Omega_1, \wh{\Omega}_1)$ which carries terms
containing $\Omega_{1,a}$ or $\wh{\Omega}_{1,a'}$. This disappears in the smearing procedure.
It turns out that the IR Lagrangian (\ref{NLSME_180821}) exactly coincides with the NLSM (\ref{SDL-SU2-522}) derived from the semi-doubled NLSM.
Hence, we understand that (\ref{L-E}) is also the GLSM for the single exotic $5^2_2$-brane.
We have a comment. 
This is different from the Lagrangian discussed in \cite{Kimura:2013fda}.
Even though the difference merely comes from difference of the isometry directions, this model is much simpler than that in \cite{Kimura:2013fda}.
This is because we need not introduce the prepotential of the chiral superfield $\Phi$ which contains a lot of redundant fields \cite{Kimura:2014aja, Kimura:2015cza}.

\subsection{From semi-doubled GLSM to standard NLSMs}

In this section, we confirmed that the semi-doubled GLSM (\ref{SD-GLSM}) correctly contains the standard GLSMs for the single defect NS5-brane (\ref{L-dNS5}), the single KK-vortex (\ref{Complete-GLSMK}), and the single exotic $5^2_2$-brane (\ref{L-E}).
In the IR limit, each gauge theory flows to the NLSM for (\ref{SDL-SU2-dNS5}), (\ref{SDL-SU2-dKK5}), and (\ref{SDL-SU2-522}), respectively.
We emphasize that the NLSMs are originally derived from the semi-doubled NLSM (\ref{SDL-SU2-3}) as the IR limit of the semi-doubled GLSM (\ref{SD-GLSM}).
Summarizing these discussions, we understand that 
the dualization procedure at the IR level is preserved even at the UV level where the gauge coupling constants are finite.
This is a very strong statement because we now understand that the complex twisted linear $\wt{L}$, a reducible superfield, is also quite useful to analyze the duality transformations among various different systems, as those in terms of irreducible superfields.
Originally, reducible superfields are introduced in order to discuss the T-duality transformations of NLSMs without global isometry on the target space geometries \cite{Gates:1984nk, Grisaru:1997ep}.
In our analysis, we also understand that, as in the same discussions 
in \cite{Hori:2000kt}, the duality transformations of them without global isometry can be traced in the language of gauge theory.
Here we illustrate this feature in Figure \ref{fig:IR-dual}.
\begin{center}
\begin{tikzpicture}
\def\w{15mm}
\def\h{15mm}
\tikzset{scirc/.style={scale=1, fill=white, rounded corners=.7mm, draw, line width=.5pt, minimum width=25mm}};
\matrix (M) [matrix of math nodes, nodes = {scirc},,column sep=\w,row sep=\h] 
{
\text{
\renewcommand{\arraystretch}{.85}
\begin{tabular}{c}
semi-doubled GLSM
\cr
(\ref{SD-GLSM})
\end{tabular}
} 
&
&
\text{
\renewcommand{\arraystretch}{.85}
\begin{tabular}{c}
standard GLSMs
\cr
(\ref{L-dNS5}), (\ref{Complete-GLSMK}), (\ref{L-E})
\end{tabular}
} 
\\
\text{
\renewcommand{\arraystretch}{.85}
\begin{tabular}{c}
semi-doubled NLSM
\cr
(\ref{SDL-SU2-3})
\end{tabular}
}
&
&
\text{
\renewcommand{\arraystretch}{.85}
\begin{tabular}{c}
standard NLSMs
\cr
(\ref{SDL-SU2-dNS5}), (\ref{SDL-SU2-dKK5}), (\ref{SDL-SU2-522})
\end{tabular}
}
\\
};
\tikzset{LW/.style={-latex, line width=1.0pt, color=black}}
\draw[LW] ([xshift=3mm]M-1-1.east) -- ([xshift=-3mm]M-1-3.west) node[midway,above] {reduction};
\draw[LW] ([xshift=3mm]M-2-1.east) -- ([xshift=-3mm]M-2-3.west) node[midway,below] {reduction};
\tikzset{LW2/.style={-latex, line width=1.0pt, color=black}}
\draw[LW2] ([yshift=-1.5mm]M-1-1.south) -- ([yshift=1.5mm]M-2-1.north) node[midway,left] {IR limit};
\draw[LW2] ([yshift=-1.5mm]M-1-3.south) -- ([yshift=1.5mm]M-2-3.north) node[midway,right] {IR limit};
\end{tikzpicture}
\figcaption{\footnotesize The flows from the semi-doubled GLSM (\ref{SD-GLSM}) to various standard NLSMs (\ref{SDL-SU2-dNS5}), (\ref{SDL-SU2-dKK5}) and (\ref{SDL-SU2-522}). The terminology ``reduction'' implies that we solve the equations of motion for original/dual scalar fields and integrate them out. The closure of the flows represents applicability of the duality transformations by using (ir)reducible superfields.} 
\label{fig:IR-dual}
\end{center}

We have understood that the semi-doubled GLSM is not only a linearized simple model, but also the UV completion of the defect five-branes.
Our next study is to understand quantum phenomena in the gauge theory, and its stringy effects on the defect branes. 
In next section, we focus on non-perturbative vortex corrections in gauge theory.

\section{Vortices in GLSM and worldsheet instantons}
\label{S:instantons}

We now established a GLSM framework to implement the isometry in the torus fibration. 
By exploiting this model, we study the instanton effects in the GLSMs for each five-brane of codimension two.
As discussed in \cite{Witten:1993yc}, the gauge instantons in
a GLSM can be interpreted as the worldsheet instantons in the IR.
It was shown that the geometry of the H-monopole receives gauge instanton corrections \cite{Tong:2002rq}.
One finds that the isometry of the geometry is broken and the H-monopole localizes in the compactified direction.
Equivalently, the geometry becomes that of the NS5-brane on $S^1$ due to the instanton corrections.
The $n$-instanton contributions to the geometry are identified with the KK $n$-modes in $S^1$.
Things get more involved in the T-dualized KK-monopole picture.
The analogous calculations of the instanton effects were performed in the GLSM for the single centered KK-monopole \cite{Harvey:2005ab}.
Similar to the H-monopole case, the instantons modify the Taub-NUT geometry.
However, the $n$-instanton contributions to the Taub-NUT geometry should
be identified with the $n$-winding modes since the T-duality symmetry
exchanges the KK- and the string winding modes.
Since any geometries related by T-duality symmetry is physically equivalent in string theory, we expect that there are winding modes corrections to branes in the T-duality orbit that include NS5-brane.
Indeed, several geometries that receive winding corrections have been found in the context of DFT \cite{Berman:2014jsa,Bakhmatov:2016kfn,Lust:2017jox,Kimura:2018hph}.

The worldsheet instantons are configurations of wrapped fundamental strings on spacetime two-cycles.
Since there are no two-cycles in the H-monopole and Taub-NUT geometries, the instantons
in the GLSM are interpreted as a kind of singular (point-like)
instantons. The calculations of these point-like instantons are justified in a certain limit of 
parameters of the geometry.
For example, the non-trivial two-cycle in the single centered Taub-NUT 
geometry is realized as a limit of the open cigar in the geometry.
This reflect the fact that the worldsheet instantons in the single
centered Taub-NUT space can be interpreted as a limit of the disk
instantons \cite{Okuyama:2005gx}.
This notion is also carried over to the analysis of instanton effects
for the $5^2_2$-brane \cite{Kimura:2013zva}. 
In this limit of the parameter, one finds that GLSMs are reduced to truncated models where instanton calculus are legitimately performed.
In the following, we write down the truncated models in a limit of parameters for the GLSMs and discuss instanton effects in each geometries.

\subsection{Instanton corrections to defect NS5-brane}

We begin with the GLSM (\ref{L-dNS5}) for the defect NS5-brane.
Since the procedure in this section does not depend on $k$, 
we consider the $k = 1$ single centered model in the following. 
The isometries along the transverse directions to the single defect NS5-brane are realized as the $U(1) \times U(1)$ gauge symmetry in the GLSM.
In order to elucidate the instanton effects, we look for a field configuration that solves the equation of motion.
We first require that the scalar fields in the gauge multiplets stay in the supersymmetric vacuum $\phi = \wh{\phi} = \sigma = \wh{\sigma} = 0$. 
By utilizing the $SU(2)_R$ and $\widehat{SU(2)}_R$ symmetries, the charged scalar fields in the hypermultiplets can be taken to be $\wt{q} = \wt{p} = 0$ without loss of generality. 

What kind of limit of the parameters should we consider for the instanton calculus is a subtle issue since the parameters in the GLSMs for the five-branes of codimension two do not have obvious geometrical meanings.
We therefore employ the natural limit $g \to 0$ that appears in the GLSM for the H-, KK-monopoles and the $5^2_2$-brane \cite{Tong:2002rq, Harvey:2005ab, Kimura:2013zva}.
In the limit $g \to 0$, the fields $(r^1, r^2, r^3, \vartheta)$ whose kinetic terms have the overall factor ${1}/{g^2}$ are frozen. 
We then set these fields to constants.
Using again the $SU(2)_R$ symmetry, we choose $r^1 = s^1$, $r^3 = t^3$.
Collecting all together, and performing the Wick rotation $x^0 \to - \I x^2$, the GLSM (\ref{L-dNS5}) is reduced to the following truncated model in
the Euclidean space:
\begin{align}
\Scr{L}_{\text{NS5}} 
\ &= \ 
\frac{1}{2e^2} F_{12}^2 
+ \frac{1}{2\wh{e}^2} \wh{F}_{12}^2 
+ \I \sqrt{2} \, \vartheta_4 F_{12} 
+ \I \sqrt{2} \, \vartheta_3 \wh{F}_{12}
\nn \\
\ & \ \ \ \ 
+ |D_m q|^2 + |D_m p|^2 
+ \frac{e^2}{2} \Big( |q|^2 - \sqrt{2} \zeta \Big)^2
+ \frac{\wh{e}^2}{2} \Big( |p|^2 - \sqrt{2} \wh{\zeta} \Big)^2
\, ,
\label{eq:truncated_NS5}
\end{align}
where we have ignored the irrelevant constant terms and defined the following quantities:
\begin{align}
r^2 - s^2 
\ = \ 
\zeta
\, , \ls
r^2 - \wh{s}^2 
\ = \ \wh{\zeta}
\, , \ls
\vartheta - t^4 
\ = \ \vartheta_4
\, , \ls 
r^3 - \wh{t}^3 
\ = \ \vartheta_3
\, .
\end{align}
The Lagrangian (\ref{eq:truncated_NS5}) represents a pair of the Abelian-Higgs model in the critical couplings. 
This is a natural extension of the Abelian-Higgs model 
appeared in the GLSM for the H-monopole \cite{Tong:2002rq}.
Indeed, if one of the gauge multiplet $(\wh{A}_m, p)$ is dropped, the
model is reduced to the one studied in \cite{Tong:2002rq} where 
the instantons in the $(A_m, q)$ sector breaks the isometry of the
H-monopole.

In (\ref{eq:truncated_NS5}), one observes that when the gauge fields
$(A_m, \wh{A}_m)$ take values in the non-trivial topological sectors, the isometries along the
$\vartheta$- and $r^3$-directions are broken.
The Bogomol'nyi completion of the Lagrangian reads
\begin{align}
\Scr{L}_{\text{NS5}} 
\ &= \ 
\frac{1}{2 e^2} \Big\{ F_{12} \pm e^2 (|q|^2 - \sqrt{2} \zeta) \Big\}^2 
+ \big| (D_1 \pm \I D_2) q \big|^2
\nn \\
\ & \ \ \ \ 
+ \frac{1}{2 \wh{e}^2} \Big\{ \wh{F}_{12} \pm e^2 (|p|^2 - \sqrt{2} \wh{\zeta}) \Big\}^2 
+ \big| (D_1 \pm \I D_2) p \big|^2
\nn \\
\ & \ \ \ \ 
\pm \sqrt{2} \, \zeta F_{12} 
+ \I \sqrt{2} \, \vartheta_4 F_{12}
\pm \sqrt{2} \, \wh{\zeta} \wh{F}_{12} 
+ \I \sqrt{2} \, \vartheta_3 \wh{F}_{12}
\, .
\end{align}
Thus the Euclidean action 
$S_{\text{E}} = \frac{1}{2 \pi} \int \d^2 x \Scr{L}_{\text{NS5}}$ 
is bounded from below: 
\begin{align}
S_{\text{E}} 
\ &\geq \ 
\sqrt{2} \, \zeta \, |n| 
- \I \sqrt{2} \, \vartheta_4 \, n
+ \sqrt{2} \, \wh{\zeta} \, |\wh{n}| 
- \I \sqrt{2} \, \vartheta_3 \, \wh{n}
\, .
\end{align}
Here the integers $(n, \wh{n})$ are the topological numbers defined by
the first Chern number:
\begin{alignat}{2}
n 
\ &= \ 
- \frac{1}{2 \pi} \int \! \d^2 x \, F_{12}
\, , &\ls
\wh{n} 
\ &= \ 
- \frac{1}{2 \pi} \int \! \d^2 x \, \wh{F}_{12}
\, .
\end{alignat}
The Euclidean action is most stringently bounded by
\begin{align}
S_{\text{E}} 
\ &\geq \ 
\sqrt{2} \, |\vec{\zeta}| \, \sqrt{\vec{n}^2} 
- \I \sqrt{2} \, (\vec{\vartheta} \cdot \vec{n})
\ = \ S_{\text{inst}}
\, , \label{eq:inst_action}
\end{align}
Here we have defined 
$\vec{\zeta} = (\zeta, \wh{\zeta})$, $\vec{n} = (n,\wh{n})$ and
$\vec{\vartheta} = (\vartheta_4, \vartheta_3)$.
The inequality is saturated when $\vec{\zeta} \parallel \vec{n}$ and the
following BPS equations are satisfied:
\bsubeq \label{eq:ANO_vortices}
\begin{alignat}{2}
F_{12} \ &= \ 
\mp e^2 \Big( |q|^2 - \sqrt{2} \zeta \Big)
\, , &\ls
(D_1 \pm \I D_2) q \ &= \ 0
\, , \\
\wh{F}_{12} 
\ &= \ 
\mp \wh{e}^2 \Big( |p|^2 - \sqrt{2} \wh{\zeta} \Big)
\, , &\ls
(D_1 \pm \I D_2) p \ &= \ 0
\, .
\end{alignat}
\esubeq
The solutions $(A_m, q), (\wh{A}_m, p)$ to the above equations are the Abrikosov-Nielsen-Olesen (ANO) vortices which have non-trivial topological numbers.
Vortices in the Euclidean theory are understood as instantons in two dimensions.
By applying the standard technique of the instanton calculus in gauge theories, 
one can compute the instanton corrections to the four point functions of the fermions 
$(\psi_{\pm,q}, \wt{\psi}_{\pm,q})$, $(\psi_{\pm,p}, \wt{\psi}_{\pm,p})$ 
in the charged hypermultiplets. They are related to the geometric fermions $(\chi_{\pm}, \wt{\chi}_{\pm})$ 
through the (SUSY partners of the) constraints (\ref{qqR-SDL}), (\ref{ppR-SDL}).
From the four-point function of the geometric fermions, one can read off
the corrections to the Riemann tensor, hence, the spacetime metric and
the harmonic functions.
The technical details are well established but since it goes beyond the
scope of the present study, we provide a qualitative discussions in the
following.

In \cite{Kimura:2013zva}, two of the present authors studied the instanton effects to the $5^2_2$-brane.
In order to realize the five-brane of codimension two, we have prepared 
the multi-centered GLSM to represent an infinite array of five-branes.
As discussed in \cite{Okuyama:2005gx}, we have assumed that instantons
in each single centered five-brane sector are completely decoupled and
they can be superimposed in the path-integral.
The situation is the same with the defect NS5-brane.
The path-integral of the GLSM for the defect NS5-brane around the classical solutions to
(\ref{eq:ANO_vortices}) apparently results in the broken isometries in the IR geometry.
One expects that the instanton-corrected geometry is a localized defect NS5-brane in the two-torus $T^2$.
If the instanton numbers $(n,\wh{n})$ are identified with the KK-modes in $T^2$, the instanton corrections to the harmonic function should be proportional to the factor $\e^{-S_{\text{inst}}}$ where the instanton action is given in (\ref{eq:inst_action}).
This is indeed plausible.
The harmonic function of the NS5-brane in $T^2$ is calculated as 
\begin{align}
H (\rho) 
\ &= \ 
\sigma \left[
\log \frac{\mu}{\rho} + \sum_{(n_3,n_4) \not= (0,0)} 
\e^{\I n_3 \frac{x_3}{{\cal R}_3} + \I n_4 \frac{x_4}{{\cal R}_4}} 
K_0 
\left(
\rho 
\sqrt{
\Big( \frac{n_3}{{\cal R}_3} \Big)^2
+ \Big( \frac{n_4}{{\cal R}_4} \Big)^2
}
\right)
\right]
\, ,
\label{eq:harmonic_func_inst}
\end{align}
where we simply set $\sigma := \sigma_3 = \sigma_4$. 
$K_0 (x)$ is the modified Bessel function of the second kind and 
the integers $(n_3, n_4)$ label the KK-modes in $T^2$.
The large-$\rho$ expansion of $K_0 (x)$ reveals the exponential
behavior of the harmonic function:
\begin{align}
\e^{\I n_3 \frac{x_3}{{\cal R}_3} + \I n_4 \frac{x_4}{{\cal R}_4}} 
K_0 
\left(
\rho 
\sqrt{
\Big( \frac{n_3}{{\cal R}_3} \Big)^2
+ \Big( \frac{n_4}{{\cal R}_4} \Big)^2
}
\right) 
\ &\sim \
\exp
\left[
- \rho 
\sqrt{
\Big( \frac{n_3}{{\cal R}_3} \Big)^2
+ \Big( \frac{n_4}{{\cal R}_4} \Big)^2
}
+ \I n_3 \frac{x_3}{{\cal R}_3} + \I n_4 \frac{x_4}{{\cal R}_4}
\right]
\, .
\end{align}
After a suitable redefinition of the parameters 
$(\zeta, \wh{\zeta}) \to (\zeta/{\cal R}_4, \wh{\zeta}/ {\cal R}_3)$,
$\vartheta_{3,4} \to x_{3,4} / {\cal R}_{3,4}$,
and an overall rescaling of the action, this is nothing but the expected factor $\e^{-S_{\text{inst}}}$.

\subsection{Instanton corrections to KK-vortex}

We next consider the GLSM (\ref{Complete-GLSMK}) for the KK-vortex.
We follow the same steps discussed in the defect NS5-brane.
The fields $(\phi, \sigma, \wh{\phi}, \wh{\sigma})$ are required to stay in the
supersymmetric vacuum.
We set $\wt{q} = \wt{p} = 0$ by the $SU(2)_R$ and
$\widehat{SU(2)}_R$ symmetries.
Again, the instanton calculus is justified in the limit $g \to 0$
\cite{Harvey:2005ab}. In this limit the fields $r^1, r^2, r^3$ and
$\mathring{\vartheta}$ are frozen and become constants.
We set $r^1 = s^1$ and $r^3 = t^3$ by $SU(2)_R$ symmetry.
Then, the truncated model becomes
\begin{align}
\Scr{L}_{\text{KK}} 
\ &= \ 
\frac{1}{2e^2} F_{12}^2 
+ \frac{1}{2\wh{e}^2} \wh{F}_{12}^2 
+ \I \sqrt{2} \, \wt{\vartheta}_4 F_{12} 
+ \I \sqrt{2} \, \vartheta_3 \wh{F}_{12}
\nn \\
\ & \ \ \ \ 
+ |D_m q|^2 + |D_m p|^2 
+ \frac{e^2}{2} \Big( |q|^2 - \sqrt{2} \zeta \Big)^2
+ \frac{\wh{e}^2}{2} \Big( |p|^2 - \sqrt{2} \wh{\zeta} \Big)^2
\, , \label{eq:truncated_KK5}
\end{align}
where we have defined
\begin{align}
r^2 - s^2 
\ &= \ \zeta
\, , \ls
r^2 - \wh{s}^2 
\ = \ \wh{\zeta}
\, , \ls
\mathring{\vartheta} - t^4 
\ = \ \wt{\vartheta}_4
\, , \ls
r^3 - \wh{t}^3 \ = \ \vartheta_3
\, .
\end{align}
This is a pair of the Abelian-Higgs models and it is essentially the same with (\ref{eq:truncated_NS5}).
However, the most crucial difference from the defect NS5-brane case is that the theta parameter $\vartheta_4$ in the (\ref{eq:truncated_NS5}) is replaced by
its dual $\wt{\vartheta}_4$ corresponding to the winding coordinate associated with one of the compact direction in the two-torus $T^2$.
This means the instantons in (\ref{eq:truncated_KK5}) break the isometry along the $\vartheta_3$ and the {\it winding direction} $\wt{\vartheta}_4$. 
This is the torus generalization of the discussion in \cite{Harvey:2005ab} where the only one isometry is realized in the KK-monopole.
The resulting geometry represents a localized KK-monopole in the winding space \cite{Gregory:1997te}.
This is obtained by formally applying the Buscher rule to the localized H-monopole without assuming the isometry.
Although this geometry is not a solution to supergravity anymore, it is a solution to the double field theory \cite{Kimura:2018hph}.

\subsection{Instanton corrections to exotic $5^2_2$-brane}

Finally, we discuss the instanton effects in the GLSM (\ref{L-E}) for the exotic $5^2_2$-brane.
Again, the fields $(\phi, \sigma, \wh{\phi}, \wh{\sigma})$ stay in the
supersymmetric vacuum and we set $\wt{q} = \wt{p} = 0$ without
loss of generality. In the limit $g \to 0$, the fields 
$(r^1, r^2, \mathring{r}^3, \mathring{\vartheta})$ are frozen and they become constants.
If we set $r^1 = s^1$, $r^3 = t^3$, the GLSM is reduced to 
\begin{align}
\Scr{L}_{\text{E}}
\ &= \ 
\frac{1}{2e^2} F_{12}^2 
+ \frac{1}{2\wh{e}^2} \wh{F}_{12}^2 
+ \I \sqrt{2} \, \wt{\vartheta}_4 F_{12} 
+ \I \sqrt{2} \, \wt{\vartheta}_3 \wh{F}_{12}
\nn \\
\ & \ \ \ \ 
+ |D_m q|^2 + |D_m p|^2 
+ \frac{e^2}{2} \Big( |q|^2 - \sqrt{2} \zeta \Big)^2
+ \frac{\wh{e}^2}{2} \Big( |p|^2 - \sqrt{2} \wh{\zeta} \Big)^2
\, , \label{eq:truncated_522}
\end{align}
where we have defined
\begin{align}
r^2 - s^2 
\ &= \ \zeta
\, , \ls
r^2 - \wh{s}^2 
\ = \ \wh{\zeta}
\, , \ls
\mathring{\vartheta} - t^4 
\ = \ \wt{\vartheta}_4
\, , \ls
\mathring{r}^3 - \wh{t}^3 
\ = \ \wt{\vartheta}_3
\, .
\end{align}
Once again, this is essentially the same with (\ref{eq:truncated_NS5})
but the theta parameters $(\vartheta_3, \vartheta_4)$ in the (\ref{eq:truncated_NS5}) are replaced by their duals $(\wt{\vartheta}_3, \wt{\vartheta}_4)$.
Therefore instantons in this model breaks the isometries along the two winding directions $\wt{\vartheta}_3$ and $\wt{\vartheta}_4$ in $T^2$.
The IR geometry is that of the $5^2_2$-brane corresponding to the first logarithmic term in (\ref{eq:harmonic_func_inst}) together with the correction terms.
Therefore, the resulting geometry is the localized $5^2_2$-brane in the winding space.
This is what expected.
Since the geometry of the localized defect NS5-brane in $T^2$ contains the KK-modes, it is conceivable that the geometry in the T-dualized $5^2_2$-brane is characterized by the winding modes.
We have shown that the $5^2_2$-brane localized in the two winding
directions is not a solution to supergravity but a solution to DFT \cite{Kimura:2018hph}.
The model (\ref{eq:truncated_522}) is nothing but the one we discussed in the previous work \cite{Kimura:2018hph} and it is a natural extension of the Abelian-Higgs model discussed in \cite{Kimura:2013zva}.

\section{Conclusion and discussions}
\label{S:conclusion}

In this paper, we established the double dualization of a twisted chiral superfield in terms of a complex twisted linear superfield.
This is necessary when we study the physics behind the geometry of the two-torus fibration and T-duality transformations along that directions.
The situation typically involves exotic branes of codimension two.
The formalism developed in this paper is very powerful when we try to dualize both the real and imaginary parts of the twisted chiral superfield. 
Such a dualization cannot be realized in the ordinary dualization in terms of irreducible (twisted) chiral superfields.
Furthermore, we also established the dualization procedure in a systematic way, which was not clear in the previous work \cite{Kimura:2015qze}.

By using this technique, we constructed the semi-doubled GLSM for five-branes of codimension two.
This model is a natural and powerful extension of the models proposed by Tong and the present authors \cite{Kimura:2013fda, Kimura:2013khz}.
The UV descriptions of the defect NS5-brane, KK-vortex and the $5^2_2$-brane, all of them are related by T-duality transformations, are included together in the GLSM (\ref{SD-GLSM}).
This situation is similar to the double field theory where T-duality symmetry is manifestly realized and a DFT solution represents a family of objects related by T-duality.
This is a reason of the name ``semi-doubled''.
In the IR limit of this gauge theory, we obtained a sigma model.
This is still semi-doubled. 
Performing the dualization procedure, we successfully obtained the string sigma models for the single defect NS5-brane, the single KK-vortex, and the single exotic $5^2_2$-brane. 
The corresponding IR sigma models are also derived from the standard GLSMs which appears via the dualization procedure at the UV level.
Throughout the analysis, we confirmed that the dualization procedure is applicable at any level where the gauge couplings are finite. 
This is quite a powerful statement because we can perform the T-duality transformations without global isometry at any energy level in terms of reducible superfields.
This is completely parallel to the discussion by Hori and Vafa \cite{Hori:2000kt} where the duality transformations in the presence of isometry are performed via the exchange between a chiral and a twisted chiral superfields.  

The most efficient applications of the GLSM we constructed are instanton effects in IR geometries.
By utilizing the GLSMs that incorporate the torus fibration, 
we discuss the worldsheet instanton corrections to five-branes of codimension two.
Although, there are some technical issues to implement the rigorous calculations, we can elucidate the essential consequences of the instanton effects.
By writing down the truncated models in a suitable limit of parameters, we found that the expected winding corrections to the geometries appear.
This is due to the specific structures of the topological terms in the GLSM.
The result is quite natural and is consistent with the calculations in \cite{Tong:2002rq} and subsequent works \cite{Harvey:2005ab, Kimura:2013zva}.
We also confirmed that this completely agrees with the result in DFT analysis discussed in our previous work \cite{Kimura:2018hph}.
Since our new GLSMs are much simpler than the previous one \cite{Kimura:2013fda, Kimura:2015qze},
then it would be possible to evaluate the instanton calculus by the localization techniques (see, for instance, the pioneer works
\cite{Benini:2012ui, Doroud:2012xw, Gomis:2012wy, Gadde:2013ftv, Benini:2013nda, Benini:2013xpa, Harvey:2014nha, Yoshida:2011au} and many others).
The localization with semi-chiral superfields \cite{Nian:2014fma, Benini:2015isa} would be also helpful to analyze our semi-doubled GLSM.

The models we obtained from the semi-doubled GLSM are alternative GLSMs
for respective five-branes of codimension two.
Even though they are not described in the superfield formalism,
the constituents are much simpler than the formulation in \cite{Kimura:2013fda} proposed by the present authors. 
This is because we replaced the complex twisted linear superfield with the sum of irreducible superfields. This replacement is completely equivalent.
On the other hand, in the previous formulation, we introduced the prepotential of the chiral superfield $\Psi$.
This is also reducible. We have to remove many redundant degrees of freedom by performing the gauge-fixing \cite{Kimura:2015qze}. This is quite troublesome when we consider quantum corrections.

As discussed in section \ref{S:instantons} and in the previous work \cite{Kimura:2018hph}, one expects that there are winding corrections even in the branes of codimension less than two.
In the present framework, 
the chiral superfield $\Psi$, which carries the remaining two transverse directions of the five-branes, never couples to the gauge field strength, i.e., the twisted chiral $\Sigma$.
Then, it is impossible to generate further topological terms to trace the worldsheet instanton corrections along $\Psi$.
In other words, as far as we employ the GLSM scheme proposed by Tong and its extension, there are no way to analyze the quantum corrections to five-branes of codimension one (a domain wall) or codimension zero (a space-filling brane) as discussed in \cite{Kimura:2018hph}.
However, if we apply the present technique to DFT, it might be possible to trace them. 
The gauged DFT \cite{Grana:2012rr} and the doubled-yet-gauged DFT
\cite{Park:2013mpa, Lee:2013hma} would be good candidates to apply it.
It would be interesting to proceed to these directions.

\section*{Acknowledgments}

The authors would like to thank
Akitsugu Miwa
for valuable discussions and comments.
The authors would also like to thank the Yukawa Institute for Theoretical Physics at Kyoto University. Discussions during the workshop {\tt YITP-T-18-04} {``New Frontiers in String Theory 2018''} were useful to complete this work.
The work of TK is supported by the Iwanami-Fujukai Foundation.
The work of SS is supported by the Japan Society for the Promotion of
Science (JSPS) KAKENHI Grant Number {\tt JP17K14294}.

\begin{appendix}

\section*{Appendix}

\section{$\N=(2,2)$ superfields in two dimensions}
\label{A:SF}

In this appendix, we introduce two-dimensional $\N=(2,2)$ superfields and their component fields.
This appendix is prepared to support detailed computations for readers.

\subsection{Convention}

First of all, we introduce the supercovariant derivative with respect to anti-commuting Grassmann coordinates $(\theta^{\pm}, \ol{\theta}{}^{\pm})$ which are given as one-component Weyl spinors in two-dimensional spacetime:
\begin{align}
D_{\pm} 
\ &:= \ 
\frac{\del}{\del \theta^{\pm}} 
- \I \ol{\theta}{}^{\pm} \big( \del_0 \pm \del_1 \big)
\, , \ls
\ol{D}{}_{\pm} 
\ := \ 
- \frac{\del}{\del \ol{\theta}{}^{\pm}} 
+ \I \theta^{\pm} \big( \del_0 \pm \del_1 \big)
\, . \label{covD} 
\end{align}
Here $\del_m$ are the coordinate derivatives of two-dimensional spacetime.
For later convenience, we often use the light-cone coordinates $\del_{\pm} = \del_0 \pm \del_1$.
We note that the one-component Weyl spinor $\theta_{\pm}$ has chirality $\pm$, whose index is raised in terms of $\ve^{+-} = - \ve^{-+} = -1$ in such a way that
\begin{align}
\theta^+ \ &= \ 
\ve^{+-} \theta_- \ = \ 
- \theta_-
\, , \ls
\theta^- \ = \ 
\ve^{-+} \theta_+ \ = \ 
+ \theta_+
\, .
\end{align}
We also note that the above two derivatives are related via the hermitian conjugate which is defined as $(\eta_+ \lambda_-)^{\dagger} = + \ol{\lambda}{}_- \ol{\eta}{}_+$.
Chirality is not flipped when we study two-dimensional spacetime with Lorentz signature.
Here we also introduce the measure of the Grassmann coordinates $(\theta^{\pm}, \ol{\theta}{}^{\pm})$:
\bsubeq \label{f-measure-22} 
\begin{gather}
\d^2 \theta 
\ := \ 
- \half \, \d \theta^+ \, \d \theta^- 
\, , \ls
\d^2 \wt{\theta} 
\ := \ 
- \half \, \d \theta^+ \, \d \ol{\theta}{}^- 
\, , \ls 
\d^4 \theta 
\ := \ 
- \frac{1}{4} \d \theta^+ \, \d \theta^- \, \d \ol{\theta}{}^+ \, \d \ol{\theta}{}^- 
\, , \\
\int \! \d^2 \theta \, \theta \theta 
\ = \ 
1 
\, , \ls
\int \! \d^2 \wt{\theta} \, \theta^+ \ol{\theta}{}^- 
\ = \ 
\half 
\, , \ls
\int \! \d^4 \theta \, \theta^+ \theta^- \ol{\theta}{}^+ \ol{\theta}{}^- 
\ = \ 
- \frac{1}{4}
\, . 
\end{gather}
\esubeq
By using these measure, we construct supersymmetric Lagrangians.

\subsection{Superfields}

Here we write down various superfields which appear in the main part of this paper.

\begin{itemize}
\item {Chiral superfield}:

The definition of a chiral superfield is 
\begin{align}
0 \ &= \ 
\ol{D}{}_{\pm} \Phi
\, . 
\end{align}
$\Phi$ is expanded in terms of a complex scalar $\phi$, 
two complex Weyl spinors $\psi_{\pm}$ and a complex scalar ${\sf F}$ in the following way:
\begin{align}
\Phi \ &= \ 
\phi
+ \I \sqrt{2} \, \theta^+ \psi_{+} 
+ \I \sqrt{2} \, \theta^- \psi_{-}
+ 2 \I \, \theta^+ \theta^- {\sf F}
\nn \\
\ & \ \ \ \ 
- \I \, \theta^+ \ol{\theta}{}^+ \del_+ \phi
- \I \, \theta^- \ol{\theta}{}^- \del_- \phi
+ \sqrt{2} \, \theta^+ \ol{\theta}{}^+ \theta^- \del_+ \psi_{-}
+ \sqrt{2} \, \theta^- \ol{\theta}{}^- \theta^+ \del_- \psi_{+}
\nn \\
\ & \ \ \ \ 
+ \theta^+ \theta^- \ol{\theta}{}^+ \ol{\theta}{}^- \del_+ \del_- \phi
\, .
\end{align}
We sometimes express the superfield to $\Phi(\phi, \psi_{\pm}, {\sf F})$ to show the component fields. If an auxiliary field ${\sf F}$ is not important to discuss, we also express it to $\Phi(\phi,\psi_{\pm})$.

\item {Vector superfield}:

A vector superfield is given as a hermitian superfield: 
\begin{align}
\ol{V} \ &= \ V
\, . 
\end{align}
The expansion is given in the following way as 
\begin{align}
V \ &= \ 
- \theta^+ \ol{\theta}{}^+ (A_{0} + A_{1})
- \theta^- \ol{\theta}{}^- (A_{0} - A_{1})
- \sqrt{2} \, \theta^- \ol{\theta}{}^+ \sigma
- \sqrt{2} \, \theta^+ \ol{\theta}{}^- \ol{\sigma}
\nn \\
\ & \ \ \ \ 
- 2 \I \, \theta^+ \theta^-
\big( \ol{\theta}{}^+ \ol{\lambda}{}_{+} 
+ \ol{\theta}{}^- \ol{\lambda}{}_{-} \big)
+ 2 \I \, \ol{\theta}{}^+ \ol{\theta}{}^-
\big( \theta^+ \lambda_{+} + \theta^- \lambda_{-} \big)
+ 2 \, \theta^+ \theta^- \ol{\theta}{}^+ \ol{\theta}{}^- {\sf D}_{V}
\, ,
\end{align}
where $A_m$ is a real vector,
$\sigma$ is a complex scalar and $\ol{\sigma}$ is its conjugate,
$\lambda_{\pm}$ and $\ol{\lambda}{}_{\pm}$ are complex Weyl spinors,
and ${\sf D}_V$ is a real scalar.
Here we have already applied the Wess-Zumino gauge.

\item {Twisted chiral superfield}:

A twisted chiral superfield is defined as
\begin{align}
0 \ &= \ 
\ol{D}{}_+ Y
\ = \ 
D_- Y
\, . 
\end{align}
The explicit expansion of $Y(\sigma, \ol{\chi}{}_+, \chi_-, \wt{\sf G})$ is given as
\begin{align}
Y
\ &= \ 
\sigma
+ \I \sqrt{2} \, \theta^+ \ol{\chi}{}_{+}
- \I \sqrt{2} \, \ol{\theta}{}^- \chi_{-}
+ 2 \I \theta^+ \ol{\theta}{}^- \wt{\sf G}
\nn \\
\ & \ \ \ \ 
- \I \theta^+ \ol{\theta}{}^+ \del_+ \sigma
+ \I \theta^- \ol{\theta}{}^- \del_- \sigma
- \sqrt{2} \, \theta^+ \theta^- \ol{\theta}{}^- \del_- \ol{\chi}{}_{+}
- \sqrt{2} \, \theta^+ \ol{\theta}{}^+ \ol{\theta}{}^- \del_+ \chi_{-}
\nn \\
\ & \ \ \ \ 
- \theta^+ \theta^- \ol{\theta}{}^+ \ol{\theta}{}^- \del_+ \del_- \sigma
\, ,
\end{align}
where component fields are a complex scalar $\sigma$,
complex Weyl spinors $\chi_-$ and $\ol{\chi}{}_+$,
and a complex scalar $\wt{\sf G}$.
It is useful to introduce a twisted chiral superfield $\Sigma$ for the (Abelian) gauge field strength $F_{mn} = \del_m A_n - \del_n A_m$ in the following form:
\begin{align}
\Sigma
\ = \ 
\frac{1}{\sqrt{2}} \ol{D}{}_+ D_- V
\ &= \ 
\sigma
+ \I \sqrt{2} \, \theta^+ \ol{\lambda}{}_{+} 
- \I \sqrt{2} \, \ol{\theta}{}^- \lambda_{-}
- \sqrt{2} \, \theta^+ \ol{\theta}{}^- ({\sf D}_{V} - \I F_{01})
\nn \\
\ & \ \ \ \ 
- \I \, \ol{\theta}{}^- \theta{}^- \, \del_- \sigma
- \I \, \theta{}^+ \ol{\theta}{}^+ \, \del_+ \sigma
- \sqrt{2} \, \ol{\theta}{}^- \theta{}^+ \theta{}^- \, \del_- \ol{\lambda}{}_{+}
+ \sqrt{2} \, \theta{}^+ \ol{\theta}{}^- \ol{\theta}{}^+ \, \del_+ \lambda_{-} 
\nn \\
\ & \ \ \ \ 
- \theta{}^+ \ol{\theta}{}^- \theta{}^- \ol{\theta}{}^+ \, \del_+ \del_- \sigma
\, .
\end{align}

\item {Complex twisted linear superfield}:

The above three superfields are irreducible, which means that the number of the degrees of freedom is minimum.
On the other hand, we can also define reducible superfields.
Here, a complex twisted linear superfield is a typical one:
\begin{align}
0 \ &= \ 
\ol{D}{}_+ D_- \wt{L}
\ = \ 
- D_+ \ol{D}{}_+ \wt{L}
\, .
\end{align}
The expansion is introduced as 
\begin{align}
\wt{L} \ &= \ 
\wt{\phi}_{L} 
+ \I \sqrt{2} \, \theta^+ \wt{\psi}_{L+} 
+ \I \sqrt{2} \, \theta^- \wt{\psi}_{L-} 
+ \I \sqrt{2} \, \ol{\theta}{}^+ \wt{\chi}_{L+} 
+ \I \sqrt{2} \, \ol{\theta}{}^- \wt{\chi}_{L-}
\nn \\
\ & \ \ \ \ 
+ \I \, \theta^+ \theta^- \wt{\sf F}_{L} 
+ \I \, \ol{\theta}{}^+ \ol{\theta}{}^- \wt{\sf M}_{L}
+ \theta^+ \ol{\theta}{}^- \wt{\sf G}_{L} 
+ \theta^- \ol{\theta}{}^- \wt{\sf A}_{L=}
+ \theta^+ \ol{\theta}{}^+ \wt{\sf B}_{L\+}
\nn \\
\ & \ \ \ \ 
- \sqrt{2} \, \theta^+ \theta^- \ol{\theta}{}^+ \del_+ \wt{\psi}{}_{L-}
- \sqrt{2} \, \theta^+ \theta^- \ol{\theta}{}^- \wt{\zeta}_{L-}
- \sqrt{2} \, \theta^+ \ol{\theta}{}^+ \ol{\theta}{}^- \wt{\lambda}_{L+}
+ \sqrt{2} \, \theta^- \ol{\theta}{}^+ \ol{\theta}{}^- \del_- \wt{\chi}{}_{L+}
\nn \\
\ & \ \ \ \ 
- \theta^+ \theta^- \ol{\theta}{}^+ \ol{\theta}{}^- 
\Big(
- \del_+ \del_- \wt{\phi}_{L} 
- \I \del_+ \wt{\sf A}_{L=} + \I \del_- \wt{\sf B}_{L\+} 
\Big)
\, . 
\end{align}
The component fields are as follows:
complex scalars $\wt{\phi}_L$, $\wt{\sf F}_L$, $\wt{\sf M}_L$ and $\wt{\sf G}_L$,
complex Weyl spinors $\wt{\psi}_{L\pm}$, $\wt{\chi}_{L \pm}$, 
$\wt{\lambda}_{L+}$ and $\wt{\zeta}_{L-}$,
and certain components of complex vectors $\wt{\sf A}_{L=}$ and $\wt{\sf B}_{L\+}$.
The degrees of freedom of this reducible superfield has three times as many as that of irreducible superfields. 
Indeed, this is described in terms of three irreducible superfields in such a way that
\begin{align}
\wt{L} \ &= \ 
X + \ol{W} + Y 
\, , \ls
\left\{
\begin{array}{rl}
X : & \text{chiral} \, , 
\\
\ol{W} : & \text{anti-chiral} \, ,
\\
Y : & \text{twisted chiral} \, .
\end{array}
\right.
\label{tL=XYW}
\end{align}
The relation between the component fields of 
the left-hand side $\wt{L}$ and those of the right-hand side
$X(\phi_X, \psi_{X\pm}, {\sf F}_X)$, 
$\ol{W}(\ol{\phi}{}_W, \ol{\psi}{}_{W\pm}, \ol{\sf F}{}_W)$ and 
$Y(\sigma_Y, \ol{\wt{\chi}}{}_{Y+}, \wt{\chi}_{Y-}, \wt{\sf G}_Y)$ 
are determined as 
\bsubeq
\begin{alignat}{2}
\wt{\phi}_L
\ &= \
\phi_X + \ol{\phi}{}_W + \sigma_Y
\, , &\ls
\wt{\sf F}_L
\ &= \ 
2 {\sf F}_X
\, , \\
\wt{\sf M}_L
\ &= \
2 \ol{\sf F}{}_W 
\, , &\ls
\wt{\sf G}_L
\ &= \ 
2 \I \wt{\sf G}_Y
\, , \\
\wt{\sf A}_{L=}
\ &= \ 
- \I \del_- \big( \phi_X - \ol{\phi}{}_W - \sigma_Y \big)
\, , &\ls
\wt{\sf B}{}_{L\+}
\ &= \ 
- \I \del_+ \big( \phi_X - \ol{\phi}{}_W + \sigma_Y \big)
\, , \\
\wt{\psi}_{L+}
\ &= \ 
\psi_{X+} + \ol{\wt{\chi}}{}_{Y+}
\, , &\ls
\wt{\psi}_{L-}
\ &= \ 
\psi_{X-}
\, , \\
\wt{\chi}_{L+}
\ &= \ 
\ol{\psi}{}_{W+}
\, , &\ls
\wt{\chi}_{L-}
\ &= \ 
\ol{\psi}{}_{W-} - \wt{\chi}_{Y-}
\, , \\
\wt{\lambda}{}_{L+}
\ &= \ 
\del_+ \big( \ol{\psi}{}_{W-} + \wt{\chi}_{Y-} \big)
\, , &\ls
\wt{\zeta}{}_{L-}
\ &= \ 
- \del_- \big( \psi_{X+} - \ol{\wt{\chi}}{}_{Y+} \big)
\, .
\end{alignat}
\esubeq

\end{itemize}

\section{One-form $\Omega$}
\label{A:Omega}

In section \ref{S:SDNLSM}, we introduced two one-forms $\Omega_a$ and $\wh{\Omega}_{a'}$ whose components are described in (\ref{Omega}).
In this appendix, we consider the smearing procedure. 
At the infinity limit $k, \ell \to \infty$, 
the FI parameters can be replaced with continuous variables 
$t^3_a \to t$ and $\wh{t}^4_a \to \wh{t}$.
Furthermore, 
$R_a$ and $\wh{R}_{a'}$ becomes functions of $t$ and $\wh{t}$, respectively.
The sums $\sum_a \Omega_{i,a}$ and $\sum_{a'} \wh{\Omega}_{j',a'}$ are
also replaced with integrals with respect to $t$ and $\wh{t}$:
\bsubeq
\begin{align}
H_0 \ &= \ 
\frac{1}{g^2} 
+ \lim_{L_3 \to \infty} \frac{1}{2 \pi {\cal R}_3} \int_{-L_3}^{L_3} \d t \, \frac{1}{\sqrt{2} R}
+ \lim_{L_4 \to \infty} \frac{1}{2 \pi {\cal R}_4} \int_{-L_4}^{L_4} \d \wh{t} \, \frac{1}{\sqrt{2} \wh{R}}
\nn \\
\ &= \ 
\frac{1}{g^2}
+ \lim_{L_3 \to \infty} \frac{1}{2 \sqrt{2} \, \pi {\cal R}_3} \log \frac{L_3 + \sqrt{\rho^2 + L_3^2}}{\rho}
+ \lim_{L_4 \to \infty} \frac{1}{2 \sqrt{2} \, \pi {\cal R}_4} \log \frac{L_4 + \sqrt{\rho^2 + L_4^2}}{\rho}
\nn \\
\ &= \ 
\frac{1}{g^2} 
+ \sigma_3 \log \frac{\Lambda_3}{\rho}
+ \sigma_4 \log \frac{\Lambda_4}{\rho}
\, , \\
\Omega_1 \ &= \ 
- \lim_{L_3 \to \infty} \frac{1}{2 \pi {\cal R}_3} \int_{-L_3}^{L_3} \d t \, \frac{r^3 - t}{\sqrt{2} R (R + r^2)}
\ = \ 0
\, , \\
\Omega_3 \ &= \ 
+ \lim_{L_3 \to \infty} \frac{1}{2 \pi {\cal R}_3} \int_{-L_3}^{L_3} \d t \, \frac{r^1}{\sqrt{2} \, R (R + r^2)}
\ = \ 
- \sigma_3 \Big\{ \arctan \Big( \frac{r^2}{r^1} \Big) - \frac{\pi}{2} \Big\}
\, , \\
\wh{\Omega}_1 \ &= \ 
+ \lim_{L_4 \to \infty} \frac{1}{2 \pi {\cal R}_4} \int_{-L_4}^{L_4} \d \wh{t} \, \frac{\vartheta - \wh{t}}{\sqrt{2} \wh{R} (\wh{R} + r^2)}
\ = \ 0
\, , \\
\wh{\Omega}_4 \ &= \ 
- \lim_{L_4 \to \infty} \frac{1}{2 \pi {\cal R}_4} \int_{-L_4}^{L_4} \d \wh{t} \, \frac{r^1}{\sqrt{2} \wh{R} (\wh{R} + r^2)}
\ = \
+ \sigma_4 \Big\{ \arctan \Big( \frac{r^2}{r^1} \Big) - \frac{\pi}{2} \Big\}
\, .
\end{align}
\esubeq
Here we have introduced the IR cutoffs $\Lambda_3$ and $\Lambda_4$.
As defined in the main part, we used $\rho^2 = (r^1)^2 + (r^2)^2$.
The values $(\sigma_3, \sigma_4)$ are also defined in (\ref{smeared-HOmega}).

If we investigate the supersymmetric effective theory of (\ref{SDL}) without the $SU(2)_R \times \wh{SU(2)}_R$ rotation (\ref{SU2-rotation}), we obtain two one-forms $\Omega_a$ and $\wh{\Omega}_{a'}$ whose components are explicitly expressed as
\bsubeq \label{Omega-wo-SU2}
\begin{alignat}{3}
\Omega_{1,a} 
\ &= \ 
+ \frac{r^2 - s^2_a}{\sqrt{2} R_a (R_a + (r^3 - t^3_a))}
\, , &\ls
\Omega_{2,a} 
\ &= \ 
- \frac{r^1 - s^1_a}{\sqrt{2} R_a (R_a + (r^3 - t^3_a))}
\, , &\ls
\Omega_{3,a} 
\ &= \ 0
\, , \\
\wh{\Omega}_{1,a'} 
\ &= \ 
+ \frac{r^2 - \wh{s}^2_{a'}}{\sqrt{2} \wh{R}_{a'} (\wh{R}_{a'} - (\vartheta - \wh{t}^4_{a'}))}
\, , &\ls
\wh{\Omega}_{2,a'} 
\ &= \ 
- \frac{r^1 - \wh{s}^1_{a'}}{\sqrt{2} \wh{R}_{a'} (\wh{R}_{a'} - (\vartheta - \wh{t}^4_{a'}))}
\, , &\ls
\wh{\Omega}_{4,a'} 
\ &= \ 0
\, .
\end{alignat}
\esubeq
These one-forms also satisfy the monopole equations (\ref{mono-eqs_2}).
But the polarizations are different from those in (\ref{Omega}).

We perform the smearing procedure in this configuration.
Setting the FI parameters to (\ref{set-SP}), and taking the infinity limit $k, \ell \to \infty$, we replace the sums $\sum_a \Omega_{i,a}$ and $\sum_{a'} \wh{\Omega}_{j',a'}$ to integrals:  
\bsubeq
\begin{align}
\Omega_1 \ &= \ 
\lim_{L_3 \to \infty} \frac{1}{2 \pi {\cal R}_3} \int_{-L_3}^{L_3} \d t \, \frac{r^2}{\sqrt{2} R (R + (r^3 - t))}
\ = \ 
+ \lim_{L_3 \to \infty} \frac{\sigma_3 r^2}{\rho^2} L_3
\, , \\
\Omega_2 \ &= \ 
\lim_{L_3 \to \infty} \frac{1}{2 \pi {\cal R}_3} \int_{-L_3}^{L_3} \d t \, \frac{-r^1}{\sqrt{2} R (R + (r^3 - t))}
\ = \ 
- \lim_{L_3 \to \infty} \frac{\sigma_3 r^1}{\rho^2} L_3
\, , \\
\wh{\Omega}_1 \ &= \ 
\lim_{L_4 \to \infty} \frac{1}{2 \pi {\cal R}_4} \int_{-L_4}^{L_4} \d \wh{t} \, \frac{r^2}{\sqrt{2} \wh{R} (\wh{R} - (\vartheta - \wh{t}))}
\ = \ 
+ \lim_{L_4 \to \infty} \frac{\sigma_4 r^2}{\rho^2} L_4
\, , \\
\wh{\Omega}_2 \ &= \ 
\lim_{L_4 \to \infty} \frac{1}{2 \pi {\cal R}_4} \int_{-L_4}^{L_4} \d \wh{t} \frac{- r^1}{\sqrt{2} \, \wh{R} (\wh{R} - (\vartheta - \wh{t}))}
\ = \ 
- \lim_{L_4 \to \infty} \frac{\sigma_4 r^1}{\rho^2} L_4
\, .
\end{align}
\esubeq
Unfortunately, we cannot extract any physical information which should be described as finite values. 
Since the one-forms play a central role in investigating the background configurations of defect five-branes as discussed in \cite{Kimura:2013fda}, 
we have to conclude that the polarizations (\ref{Omega-wo-SU2}) is useless.
This is the primary reason why we performed the $SU(2)_R \times \wh{SU(2)}_R$ rotation (\ref{SU2-rotation}) in the construction of the semi-doubled GLSM.

\end{appendix}
\phantomsection
\addcontentsline{toc}{section}{References}
{

}

}
\end{document}